%% file: main.tex
\documentclass[fleqn,usenatbib]{mnras}

\usepackage{newtxtext,newtxmath}

\usepackage[T1]{fontenc}
\usepackage{ae,aecompl}
\usepackage{lineno}

\usepackage{graphicx}	
\usepackage{amsmath}	
\usepackage{amssymb}	
\usepackage{multicol}        
\usepackage{bm}		
\usepackage{pdflscape}	
\usepackage{xspace}
\usepackage{comment}
\usepackage{booktabs,caption}
\usepackage[flushleft]{threeparttable}
\usepackage{todonotes}
\usepackage{float}
\usepackage{lastpage}

\newcommand{\sne}{SNe Ia\xspace}
\newcommand{\sn}{SN Ia\xspace}
\newcommand{\Ha}{H$\alpha$\xspace}
\newcommand{\Hb}{H$\beta$\xspace}
\newcommand{\Hg}{H$\gamma$\xspace}

\newcommand{\dd}{DD\xspace}
\newcommand{\sd}{SD\xspace}
\newcommand{\dm}{\ensuremath{\Delta m_{15}}\xspace}
\newcommand{\NaID}{Na\texttt{I}D\xspace}
\input{InputFiles/FinalNumbers.tex}
\input{InputFiles/InstrCounts.tex}

\title[111 Nebular SNe Ia]{Nebular Spectra of 111 Type Ia Supernovae Disfavor Single Degenerate Progenitors}

\author[M. A. Tucker et al.]{
M. A. Tucker$^{1,}$\thanks{E-mail: tuckerma@hawaii.edu}\thanks{DOE CSGF Fellow},
B. J. Shappee$^{1}$,
P. J. Vallely$^2$,
K. Z. Stanek$^{2,3}$,
J. L. Prieto$^{4,5}$, 
\newauthor
J. Botyanszki$^{6}$, 
C.~S.~Kochanek$^{2,3}$,
J. P. Anderson$^{7}$,
J. Brown$^2$,
L. Galbany$^{8}$,
\newauthor
T. W.-S. Holoien$^9$,
E. Y. Hsiao$^{10}$, 
S. Kumar$^{10}$,
H. Kuncarayakti$^{11,12}$,
N. Morrell$^{13}$,
\newauthor
M. M. Phillips$^{13}$, 
M. D. Stritzinger$^{14}$, and
Todd A. Thompson$^{2,3}$
\\
$^{1}$ Institute for Astronomy, University of Hawai`i at Manoa, 2680 Woodlawn Dr., Honolulu, Hi 96822\\
$^2$Department of Astronomy, The Ohio State University, 140 West 18th Avenue, Columbus, OH 43210, USA \\
$^3$Center for Cosmology and AstroParticle Physics (CCAPP), The Ohio State University, 191 W. Woodruff Avenue, Columbus, OH 43210, USA \\
$^{4}$N\'ucleo de Astronom\'ia de la Facultad de Ingenier\'ia y Ciencias, Universidad Diego Portales, Av. Ej\'ercito 441, Santiago, Chile \\
$^{5}$Millennium Institute of Astrophysics, Santiago, Chile \\
$^{6}$Physics Department, University of California, Berkeley, CA 94720, USA\\
$^{7}$European Southern Observatory, Alonso de Cordova 3107 Casilla 19001, Vitacura, Santiago, Chile \\
$^{8}$PITT PACC, Department of Physics and Astronomy, University of Pittsburgh, Pittsburgh, PA 15260, USA\\
$^9$Carnegie Observatories, 813 Santa Barbara Street, Pasadena, CA 91101, USA \\
$^{10}$Department of Physics, Florida State University, 77 Chieftan Way, Tallahassee, FL 32306, USA  \\
$^{11}$Finnish Centre for Astronomy with ESO (FINCA), FI-20014 University of Turku, Finland\\
$^{12}$Tuorla Observatory, Department of Physics and Astronomy, FI-20014 University of Turku, Finland\\
$^{13}$Las Campanas Observatory, Carnegie Observatories, Casilla 601, La Serena, Chile \\
$^{14}$Department of Physics and Astronomy, Aarhus University, Ny Munkegade 120, DK-8000 Aarhus C, Denmark 
}

\date{Accepted XXX. Received YYY; in original form ZZZ}

\pubyear{2019}

\begin{document}
\label{firstpage}
\pagerange{\pageref{firstpage}--\pageref{lastpage}}
\maketitle

\begin{abstract}

We place statistical constraints on Type Ia supernova (SN Ia) progenitors using \GoodSpeccount nebular phase spectra of \Nsn SNe Ia. We find no evidence of stripped companion emission in any of the nebular phase spectra. Upper limits are placed on the amount of mass that could go undetected in each spectrum using recent hydrodynamic simulations. With these null detections, we place an observational $3\sigma$ upper limit on the fraction of SNe Ia that are produced through the classical H-rich non-degenerate companion scenario of $ < \NCnormTbgSCpercThree\%$. Additionally, we set a tentative $3\sigma$ upper limit on He star progenitor scenarios of $< \HenormTbgSCpercThree\%$, although further theoretical modelling is required. These limits refer to our most representative sample including normal, 91bg-like, 91T-like, and ``Super Chandrasekhar'' \sne but excluding SNe Iax and SNe Ia-CSM. As part of our analysis, we also derive a Nebular Phase Phillips Relation, which approximates the brightness of a SN Ia from $150-500$~days after maximum using the peak magnitude and decline rate parameter $\Delta m_{15} (B)$.
\end{abstract}


\section{Introduction}

Type Ia supernovae (\sne) are utilised across many astronomical disciplines, including the extragalactic distance scale, dark energy studies, and Galactic chemical evolution. Despite their prevalence, the origins of \sne are still unclear even after decades of study. The general consensus is that they are explosions of carbon/oxygen (C/O) white dwarfs \citep{hoyle60} with fairly homogeneous properties. For example, the magnitude of \sne at peak is well constrained \citep[$M_{\rm{max}}\sim -19$, e.g.; ][]{folatelli10}, and, after correcting for light curve decline and color, they have an intrinsic scatter of $\sim 0.1$ mag \citep[e.g., Fig. 19, ][]{folatelli10}. Many formation mechanisms for \sne have been proposed to reproduce this level of uniformity, which can be grouped into two main categories: the double degenerate (DD) and single degenerate (SD) scenarios \citep[see ][ for reviews on \sne and their progenitors]{maoz14, livio18, jha19}.

The \dd scenario consists of two degenerate stars, usually C/O white dwarfs, which induce a \sne through accretion, collision, or merger. This can occur due to gravitational wave emission \citep{tutukov79, iben84, webbink84}, collision/violent merger due to perturbations by external bodies \citep{thompson11, katz12, shappeekozai, pejcha13, antogini14}, accretion from a low-mass white dwarf onto a smaller, higher-mass white dwarf \citep{taam80, livne90, tutukov96, pakmor12}, or a "double detonation" where an accreted helium layer detonates and drives the core to detonate \citep{woosley94, fink10, kromer10}. Due to the intrinsic faintness of both components in these systems, observational confirmation of \dd systems is exceptionally difficult \citep[e.g., ][]{rebassa18}. Some progress has been made on this front, such as bimodal emission in the nebular phase \citep{dong15, vallely} and possible hyper-velocity remnants \citep{shen18, ruffini19}. However, most of the evidence for \dd systems comes from the exclusion of \sd progenitors \citep[e.g., ][]{shappee17}.

The SD scenario involves a WD with a nearby non-degenerate companion \citep{whelan73, nomoto82, yoon03}, usually undergoing Roche Lobe overflow (RLOF). The WD accumulates material until reaching critical mass and then explodes. This critical mass is typically considered the Chandrasekhar mass ($M_{ch}\sim 1.4~M_\odot$), although sub-$M_{ch}$ explosions, including double detonation scenarios, are also possible \citep[e.g., ][]{livne95}. There are several predicted observational signatures of the SD degenerate scenario due to the interaction of the ejecta/explosion and the donor star \citep{wheeler75}, including effects on the rising \sn light curve \citep{kasen10}, soft X-ray emission in the accretion phase \citep{lanz05, tutukov07, woods18}, surviving companions with anomalous characteristics \citep[e.g., ][]{canal01, shappee_overlum}, and the amount of $^{56}$Ni decay products synthesized in the explosion \citep[e.g., ][]{ropke12, shappee17}. 

One of the most promising signatures of a RLOF companion to an exploding WD are emission lines produced by material stripped/ablated from the non-degenerate companion \citep[e.g., ][]{wheeler75, chugai86, marietta00, mattila05, pan12}, observable in nebular-phase spectra once the \sn has faded considerably and become optically thin. For example, \citet{boehner17} simulated stripping from red giant (RG), main sequence (MS), and sub-giant (SG) stars, finding approximately $0.33$, $0.25$, and $0.17M_\odot$, respectively, of stripped mass. \citet{botyanszki18} converted these estimates into expected \Ha luminosities and found that the emitted \Ha luminosity does not vary linearly with amount of stripped companion mass, which had been the assumption of previous studies \citep[e.g. ][]{leonard07,shappee13}, but instead the relation is closer to logarithmic. Additionally, the \Ha emission is powered by gamma-ray deposition from the \sn ejecta and roughly follows the bolometric luminosity. 

In this work we compile a comprehensive sample of \sne nebular spectra spanning $200-500~\rm {days}$ after explosion ($181-481~\rm{days}$ after maximum assuming a rise time of $\sim 19~\rm{days}$ from \citealp{firth15}) to search for the expected emission from stripped/ablated material. We find no such emission in any spectrum in our sample, and place new or updated stripped/ablated mass constraints for each \sn. The entirety of similar work in the literature totals 33 \sne \citep{mattila05, leonard07, shappee13, lundquist13, lundquist15, maguire16, graham17,shappee18, sand18, holmbo18, dimitriadis19, tucker18, sand19}, a fraction of the sample analyzed in this work. All \sne included in this study are listed in Table \ref{tab:targets} and photometric parameters ($t_{max}$, $\Delta m_{15}$, $\mu$, $E(B-V)_{\rm{host}}$) are provided in Table \ref{tab:snparams}. 

We outline our data sources and reduction techniques, including absolute flux calibration, in \S\ref{sec:data}. In \S\ref{sec:Hsearch}, we discuss our methodology in searching for and placing limits on material stripped from a RLOF companion. Our upper limits on stripped material are provided in \S\ref{sec:results}, and our findings are discussed in the context of \sne formation in \S\ref{sec:discuss}. Included in \S\ref{sec:discuss} are discussions about peculiar \sne and their role in our study including SNe Ia-CSM. Finally, in \S\ref{sec:conclusion}, we summarize our results.

\section{Data Sources and Reduction}\label{sec:data}

Our sample of \GoodSpeccount spectra of \Nsn \sne comes from the 40 instruments on 29 telescopes listed in Table \ref{tab:instruments}. All spectroscopically peculiar \sne are included except for those exhibiting signatures of circumstellar material (\sne-CSM). These \sne exhibit \Ha emission, but the velocity and magnitude of the emission is inconsistent with material stripped from a nearby companion; instead, these \sn appear to have exploded in a dense circumstellar environment \citep[e.g., SN~2002ic, ][]{wang04} and exhibit \Ha emission before the SN enters the nebular phase  \citep[e.g., ][]{silverman13}. A discussion of SNe Ia-CSM and their role in our results is provided in \S\ref{sec:CSM}. For non-CSM \sne we impose the following criteria when selecting nebular spectra:
\begin{itemize}
    \item Obtained between 200 and 500 days after explosion to maintain consistency with the models of \citet{botyanszki18}, assuming a typical rise time of $t_{\rm{rise}} \approx 19~\rm {days}$ \citep{firth15}.
    \item Cover $\pm1\,000~\rm{km}~\rm s^{-1}$ of at least one H or He line in Table \ref{tab:ModelInfo}.
    \item Have at least one method of absolute flux calibration, outlined in \S\ref{subsec:fluxcal}. 
    \item Published, posted, or observed by our team before the submission date of this article (15 March 2019). 
\end{itemize}

\noindent The complete list of new and archival spectra is provided in Table \ref{tab:spectra-params}. Additionally, we include new and archival photometry to supplement our spectral data and analysis. Early-phase photometry ($\lesssim 50$~days after maximum light) is used in deriving the photometric properties of each \sn using the photometric fitting code \texttt{SNooPy} \citep{burns11}, including time of maximum ($t_{max}$), the decline rate parameter $\Delta m_{15}$, extinction along line of sight, and the distance modulus. 
Late- and nebular-phase photometry are used for flux calibrating the nebular spectra and deriving a Nebular Phase Phillips Relation (NPPR). The NPPR approximates the nebular magnitude of a \sn given its peak magnitude and decline rate, calibrated to an extensive sample of new and archival \sne photometry. A complete description of the NPPR, its derivation and usage is provided in Appendix \ref{app:LTPR}.

\begin{table*}
\begin{threeparttable}
\caption{All telescopes and instruments utilised in this work. If a reference could not be found for a given instrument, the corresponding instrument website is provided in the table notes.}
\label{tab:instruments}
\input{InstrumentTable.tex}

\end{threeparttable}
\end{table*}

\subsection{New Spectra and Photometry}

We present \NewSpeccount new nebular-phase spectra of  \NewSNe \sne, of which \UniqueSNe have no prior published nebular spectra. These spectra were acquired in our ongoing study of \sne progenitors, taken with MagE and IMACS on Baade, MUSE on the VLT, and WFCCD on duPont (see Table \ref{tab:instruments} for telescope and instrument designations). For the new spectra presented here, each spectrum was reduced using telescope and instrument-specific pipelines, if available, otherwise typical IRAF\footnote{\url{http://iraf.noao.edu/}} tasks were used. The spectra acquired with MagE/Baade were reduced with a pipeline provided by the Carnegie Observatories\footnote{\url{http://code.obs.carnegiescience.edu/mage-pipeline}} \citep{kelson00, kelson03}, with the exception of standard star calibrations and stitching together each echellette spectrum, which was done with custom \texttt{Python} routines. For newly presented MUSE data acquired as part of the AMUSING survey \citep{galbany16}, spectra were extracted in a 1" circular aperture at the \sn location using the PyMUSE package \citep{pymuseref}, and corrected for host galaxy contributions using a background annulus extending from 2" to 3". New IMACS/Baade spectra were reduced with typical IRAF procedures including bias subtraction, flat-field correction, arc lamp exposures for wavelength calibration and standard star onbservations to correct for instrument and atmospheric response. 

For absolute flux calibrations, we also include nebular photometry for any \sne in our sample. This includes new observations and reproccessed archival images for which we could not find a published magnitude. New photometry includes $V$-band images taken with FORS2, $r$-band images from MODS1, and $BVRI$ images from WFCCD. Archival imaging includes $UBVRI$ imaging from FORS1/2 and $BVRgri$ imaging from EFOSC2 (Table \ref{tab:newphot}). All images are bias subtracted and flat-field corrected before performing aperture photometry with the \texttt{IRAF} \textit{apphot} task. For targets with $\delta \geq -30^{\circ}$, photometry from the Pan-STARRS Stack Object catalog\footnote{\url{http://archive.stsci.edu/panstarrs/stackobject/search.php}} \citep{chambers16,flewelling16} was used in calibrating the images, otherwise \textit{Gaia} DR2 photometry \citep{Gaiaref, GaiaDR2ref, riello18} was used. When transforming reported magnitudes to other photometric systems, \citet{tonry12} and \citet{evans18} were used for Pan-STARRS and \textit{Gaia}, respectively. The only exceptions to this procedure are the $B$-band FORS2/VLT images, which are calibrated using the reported photometric zeropoints\footnote{\url{https://www.eso.org/observing/dfo/quality/FORS2/qc/zeropoints/zeropoints.html}}.

\subsection{Archival Spectra and Photometry}

The primary sources of our archival spectra and photometry are the Berkeley SuperNova Ia Program\footnote{\url{http://heracles.astro.berkeley.edu/sndb/}} \citep[BSNIP, ][]{silverman12, silverman13}, the Center for Astrophysics (CfA) Supernova Data Archive\footnote{\url{https://www.cfa.harvard.edu/supernova/SNarchive.html}} \citep[][]{riess99, jha06, matheson08, blondin12}, the Carnegie Supernova Project\footnote{\url{http://csp.obs.carnegiescience.edu/}} \citep[CSP, ][]{CSPref, folatelli10, contreras10, stritzinger11, folatelli13, krisciunas17, phillips19}, the 100IAs project \citep{dong18}, the ANU WiFeS SuperNovA Program \citep[AWSNAP; ][]{childress16}, and the All-Sky Automated Survey for SuperNovae \citep[ASAS-SN, ][ Vallely et al., Chen et al., in prep]{shappee14, ASASSNcat14, ASASSNcat15, ASASSNcat16, ASASSNcat17}. The majority of the publicly available data were retrieved using the Open Supernova Catalog \citep[OSC, ][]{OSCref} and the Weizmann Interactive Supernova data REPository \citep[WISeREP, ][]{WRref}. All data provided by these sources are already reduced with the exception of precise spectral flux calibration, which we outline in \S\ref{subsec:fluxcal}. Additionally, we supplement these sources with archival data obtained from telescope databases, including the Keck Observatory Archive\footnote{\url{https://koa.ipac.caltech.edu/}} (KOA), the ESO Science Archive Facility \footnote{\url{http://archive.eso.org/cms.html}} (ESO SAF), the Isaac Newton Group Archive\footnote{\url{http://casu.ast.cam.ac.uk/casuadc/ingarch/query}} and the Gemini Observatory Archive\footnote{\url{https://archive.gemini.edu/}} (GOA).  Information on all the spectra in this study is presented in Table \ref{tab:spectra-params}.

Data reduction and calibration was performed as uniformly as possible across all sources of spectra. Data retrieved from public archives were already reduced, with the exception of absolute flux calibration. The reduction of data retrieved from telescope archives was generally less complete. All spectra retrieved from the ESO SAF were already reduced (excluding flux corrections) with the exception of FORS1/2 data. For any ESO SAF data reduction, both spectroscopy and photometry, we used the ESO SAF \texttt{esorex} data reduction pipeline \citep{esorexref}.

Spectra obtained from the KOA and GOA were not reduced prior to retrieval and had to be manually reduced. Recent LRIS spectra were reduced using Lpipe\footnote{\url{http://www.astro.caltech.edu/~dperley/programs/lpipe.html}}, while older LRIS and DEIMOS data were reduced using the LowRedux/XIDL pipeline\footnote{\url{http://www.ucolick.org/~xavier/LowRedux/}}. Gemini North/South GMOS spectra were reduced with the GMOS Data Reduction Cookbook\footnote{\url{http://ast.noao.edu/sites/default/files/GMOS_Cookbook/}}. 

We manually reduced any unreduced spectra for which no pipeline exists using standard IRAF procedures. Images were flat-fielded and bias-subtracted using archival calibration images taken near the epoch of observation, and wavelength calibrated with arc lamp exposures. Spectrophotometric standard star observations were used to correct for telescope/instrumental artefacts, atmospheric effects, and to place each spectrum on a reliable relative flux scale. 

\subsection{Accurate Flux Calibration}\label{subsec:fluxcal}

For our analysis in \S\ref{sec:Hsearch}, the spectra must be on a reliable absolute flux scale. While calibrating spectra with spectrophotometric standard stars places these spectra on a dependable relative flux scale, slit losses, atmospheric conditions, and other effects can cause the resulting spectra to deviate from an absolute flux scale. To scale a spectrum to the absolute scale, we employed Eq. 7 from \citet{fukugita96} to calculate synthetic photometry from the spectra. The spectra are then scaled so that the synthetic photometry matches the observed photometry. There were several different sources of photometry used to calibrate the spectra. In order of preference and reliability, with accuracy estimates in given parentheses:

\begin{enumerate}
    \item For spectra with acquisition images taken at the time of observation, we scale the entire spectrum to match these photometric observations, usually in the $V$ or $r$ filters ($\sim 5-10\%$). 
    
    \item If acquisition images are unavailable, we next tried to use photometry within $\pm 5~\rm d$ of the spectral observations. Photometry in all available filters within this temporal limit were used in the flux calibration ($\sim 10-15\%$).  
    
    \item If no photometric data was available within $\pm 5 ~\rm d$, we searched for photometry within $\pm 50~\rm d$. If at at least 3 photometric data points fell within this time span, we linearly interpolated to estimate the magnitude at the time of the spectral observation ($\sim 15-20\%$). 
    
    \item If none of these were available, the nebular $BVR$ magnitude was estimated with the NPPR and used to calibrate the spectrum (see Appendix \ref{app:LTPR}, $\sim 20\%$). 
    
\end{enumerate}

We required $> 90\%$ of the filter's transmission curve be covered by the observed spectrum for viable calibrations. If only a single filter was available, the entire spectrum was scaled to match the observation. If two filters were available for flux calibration, a simple linear fit was applied to the scale factors. If $>3$ filters were available, we use spline fits with fixed endpoints to ensure a robust flux correction across the entire spectrum. After placing the spectrum on an absolute flux scale, we correct for host galaxy and Milky Way reddening using the $E(B-V)_{\rm{host}}$ derived from the light curve fits. We implement a \citet{fitzpatrick99} extinction law and a \citet{schlegel98} Milky Way dust map for our reddening corrections. We assume $R_V = 3.1$ unless stated otherwise (see Appendix \ref{app:tables}).

\subsection{Sample Demographics}\label{subsec:demographics}

\begin{figure}
    \centering
    \includegraphics[width=\linewidth]{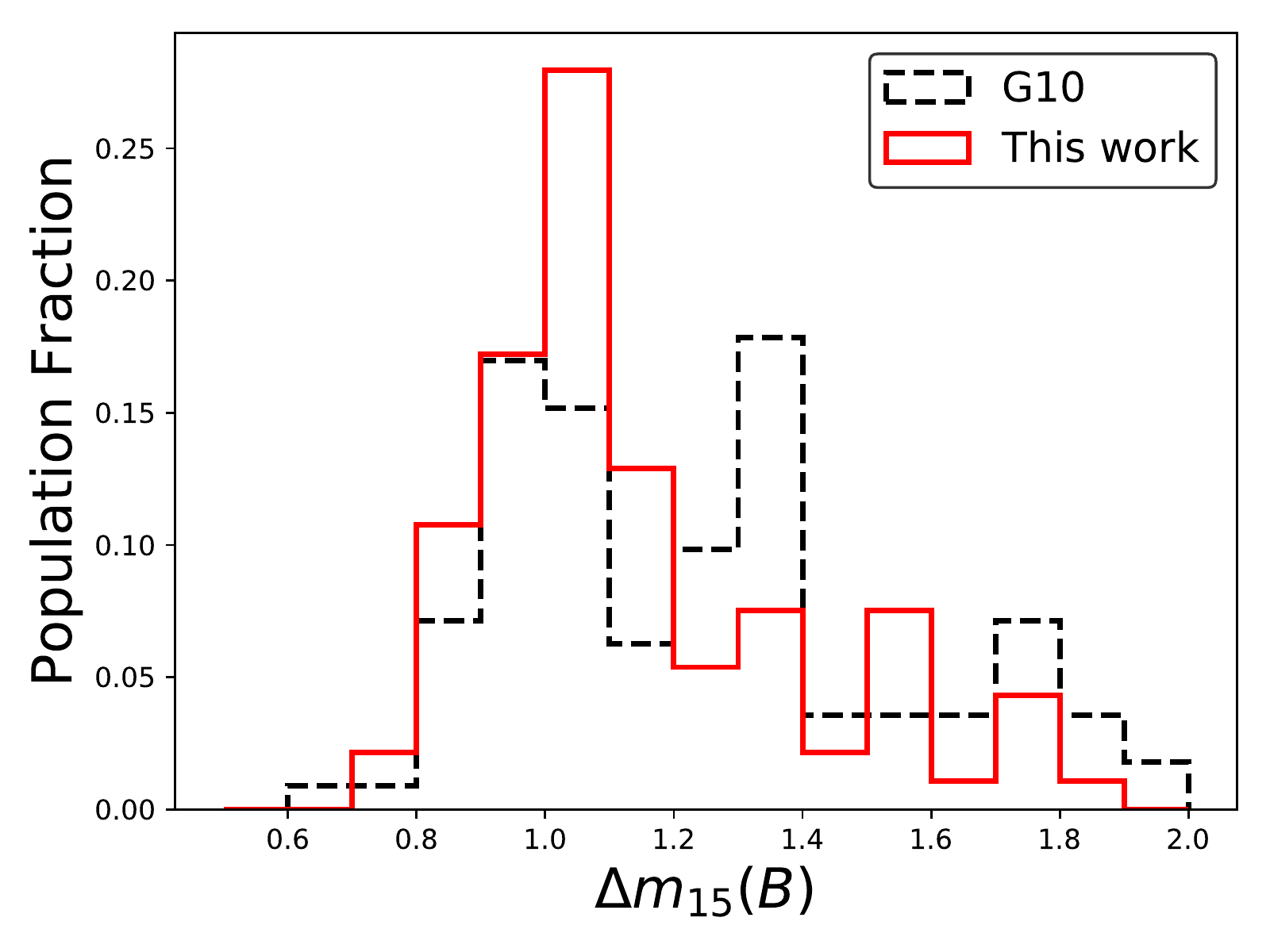}
    \caption{Distribution of $\Delta m_{15}(B)$ in our sample compared to the photometric sample from LOSS \citep{ganeshalingam10}. As expected, the nebular sample is biased towards brighter and broader \sne.}
    \label{fig:dmhist}
\end{figure}


\begin{figure}
    \centering
    \includegraphics[width=\linewidth]{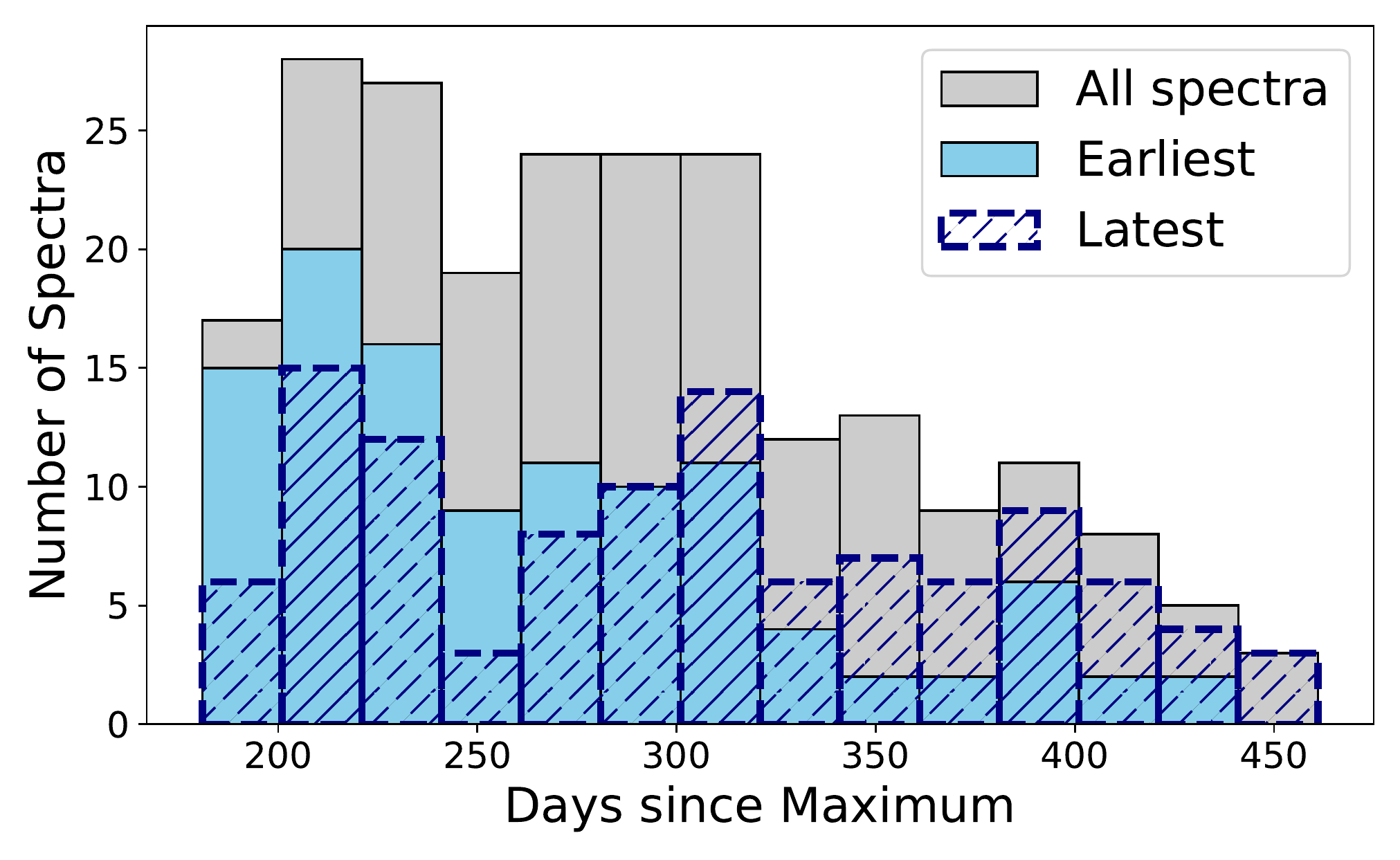}
    \caption{The number of spectra per days since maximum for 3 distributions, 1) all spectra in our sample (gray), 2) the earliest spectrum for each \sn (light blue), and 3) the latest spectrum for each \sn (dark blue, hatched). }
    \label{fig:phasedist}
\end{figure}

After collecting all nebular spectra that meet our temporal and calibration requirements, our sample consists of 111 \sne (Table \ref{tab:targets}). Due to the comprehensive nature of our search for nebular spectra, our sample is inherently biased towards brighter or peculiar \sne as these objects have a higher likelihood of being observed in the nebular phase. This effect is readily apparent in Fig. \ref{fig:dmhist} where the \dm values of \sne in our sample are compared to a purely photometric sample from the Lick Observatory Supernova Search \citep[LOSS, ][]{li00, ganeshalingam10}. Similar to the luminosity bias, we are biased to lower redshift \sne with a median redshift of $z_{\rm{med}}\approx 0.015$. We provide the temporal distribution for our set of nebular spectra in Fig. \ref{fig:phasedist}, including distributions for the earliest and latest spectra for each \sn. As expected, the number of spectra generally decline at later phases due to the supernovae fading from view. 

\section{Searching for Emission from Stripped Companion Material}\label{sec:Hsearch}

Prior to the work of \citet{botyanszki18}, the majority of unbound mass limits in the literature utilised the work of \citet{mattila05} and \citet{leonard07} to compute stripped mass limits from comparing observed spectra to expected \Ha luminosities. Several subsequent studies have adopted these methodologies in their work \citep[e.g., ][]{maguire16, graham17} with notable success in ruling out hydrogen-rich companions. Yet the models of \citet{mattila05} had several shortcomings in observational implementation. In particular, \citet{leonard07} assumed a linear scaling between the amount of unbound companion mass and the corresponding \Ha luminosity.

\citet{botyanszki18}, using the MS38 model (a main-sequence star undergoing RLOF) from \citet{boehner17}, instead found the emitted \Ha luminosity scales logarithmically with the amount of stripped mass. Additionally, \citet{botyanszki18} computed a simplified helium-star model, where all the stripped mass from the MS38 model is replaced with helium instead of Solar abundance material. This is not a true helium-star model, as helium-star companions are expected to have lower amounts of stripped mass than their hydrogen-rich counterparts and a modestly different velocity distribution \citep[e.g. ][]{pan12}, but it provides a starting point for calculating limits on the amount of unbound helium in a \sn spectrum.

While the models of \citet{botyanszki18} clarify the mass-luminosity scaling issue and expand to helium emission, they share two other shortcomings with the models of \citet{mattila05}: the requirement of using \Ha to constrain the amount of unbound mass, and only calculating the expected \Ha luminosity at a single epoch (200 days post-explosion for \citealp{botyanszki18} and 350 days post-maximum for \citealp{mattila05}). In the following subsections we discuss our stripped mass limits given these limitations.

\subsection{Expanding on these Models}\label{subsec:expansion}

\input{ModelInfoTable}

For \sne with star-forming host galaxies, the region around \Ha can be contaminated by narrow host galaxy \Ha and N\texttt{II} emission lines which complicates setting limits on \Ha emission. However, the unbound material has emission lines besides \Ha, including \Hb, \Hg, and the Paschen series. Assuming roughly Solar metallicity, the stripped material will also exhibit prominent He\texttt{I} lines in the optical and NIR \citep{botyanszki18}. We provide the luminosities for each of these lines in Table \ref{tab:ModelInfo} at $(t-t_0) = 200$ days from explosion for the hydrogen-rich (H-rich) model using the same MS38 model as \citet{botyanszki18}. Additionally, we supply similar data for the simplified helium star model from \citet{botyanszki18}, which we refer to as the He-rich model.  

\begin{figure}
    \centering
    \includegraphics[width=\linewidth]{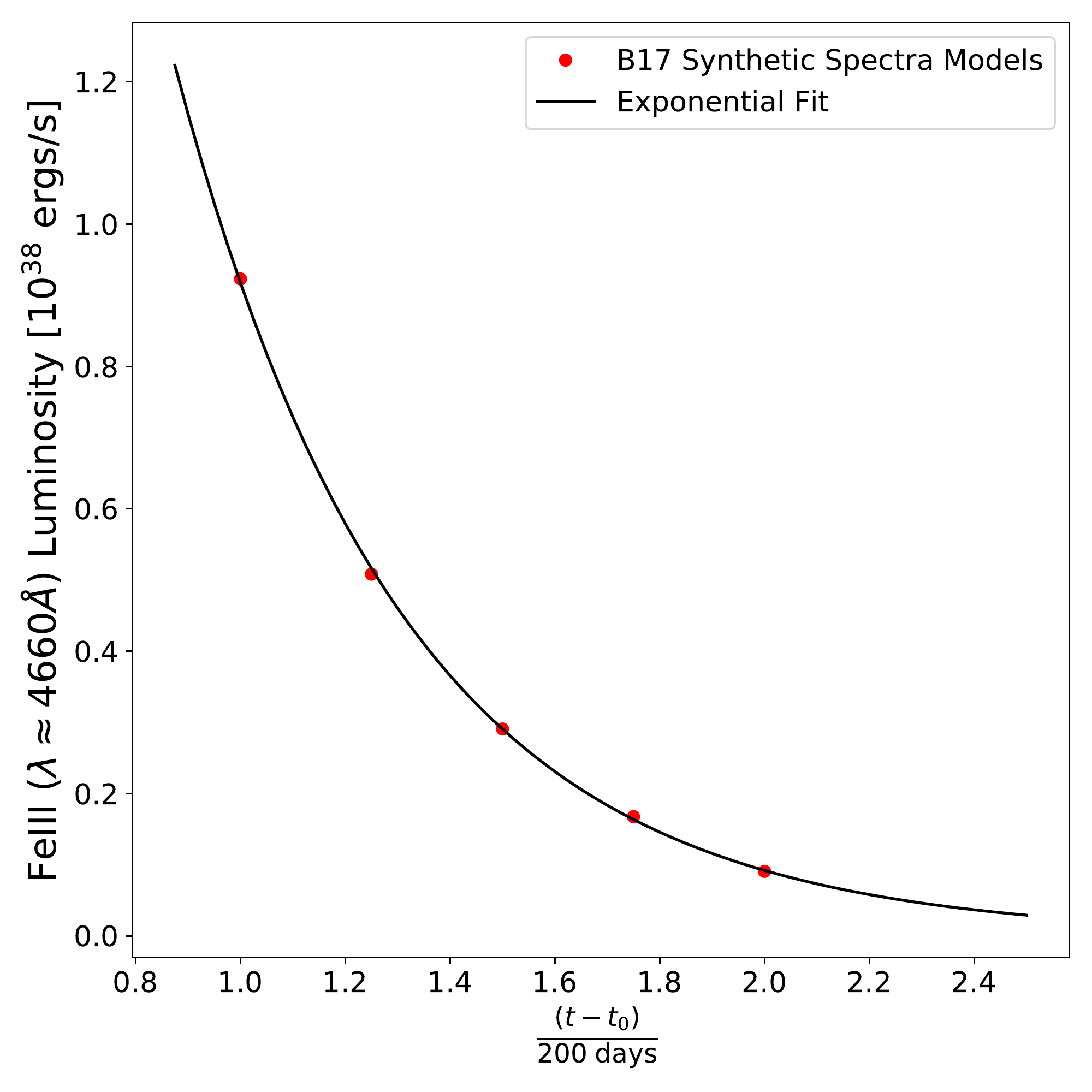}
    \caption{Peak luminosity of the Fe\texttt{III} line (red points) versus days since explosion $(t-t_0)$ from the time-dependent \sn spectral models of \citet{botyanszki17} and the exponential fit (black line).}
    \label{fig:FeIIIevolution}
\end{figure}

\begin{figure*}
    \centering
    \includegraphics[width=\linewidth]{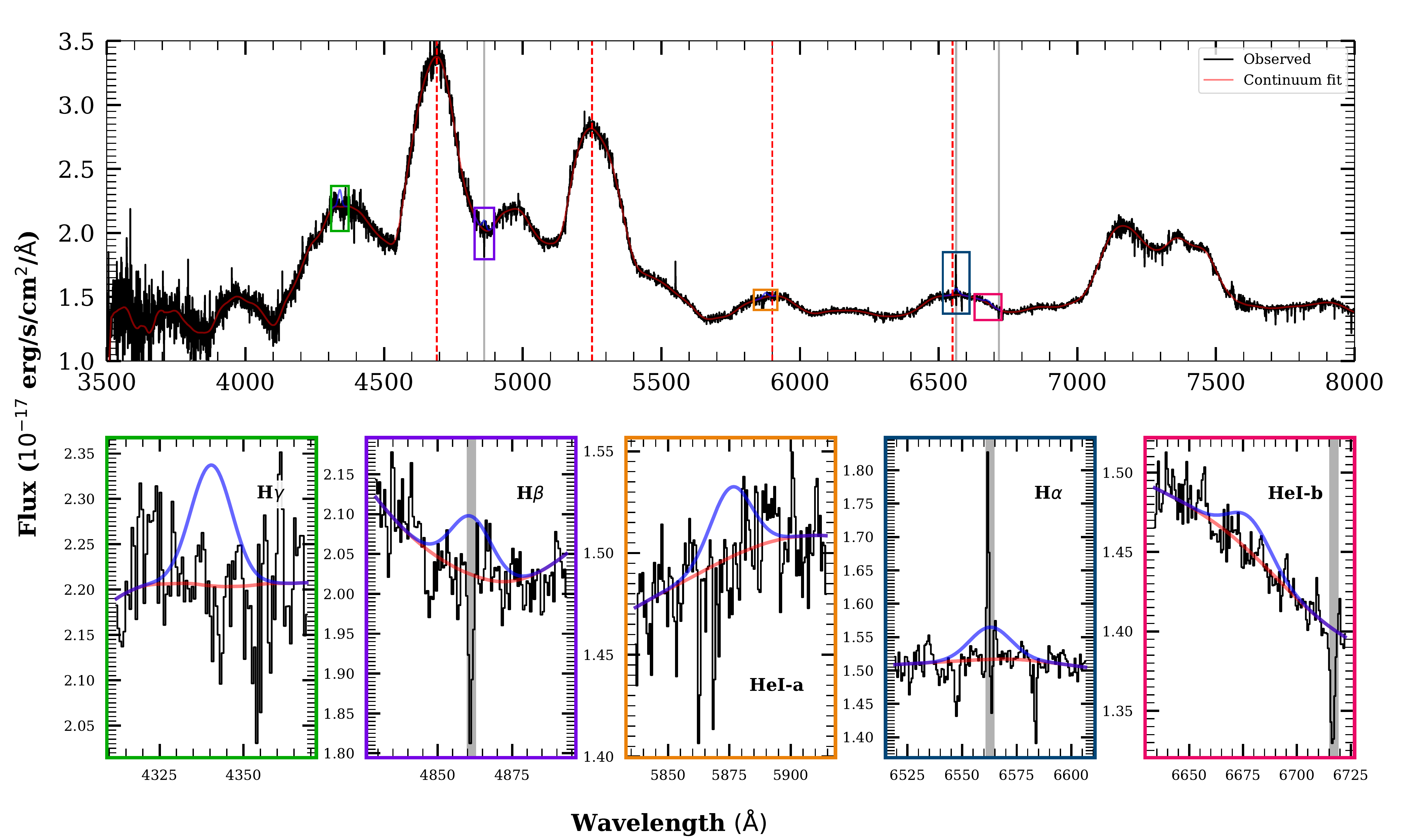}
    \caption{Nebular phase spectrum (black), continuum fit (red), and derived $10\sigma$ flux limits (blue) for the Baade/MagE +295~d spectrum of SN~2015F. The bottom panels show the regions near each possible emission line from Table \ref{tab:ModelInfo} and correspond to the coloured boxes in the top panel. Gray shaded areas indicate masked spectral regions due to host galaxy contamination or instrumental effects, and vertical dashed lines indicate \sne emission lines used to estimate the smoothing width for each spectrum (see \S\ref{sec:Hsearch}). Similar spectral cutouts for all \sne are included as supplementary figures (see Appendix \ref{app:tables}).
    } 
    \label{fig:example}
\end{figure*}

\citet{botyanszki18} estimated the line luminosity at 200 days as a function of the amount of stripped mass ($M_{st}$). Table \ref{tab:ModelInfo} provides the expected luminosity $L_{200}(M_{st})$ of various lines for $M_{st} = 0.25~M_\odot$. The dependence on the amount of stripped mass is well approximated by $\log_{10} L_{200}(M) \simeq \log_{10} L_{200}(0.25M_\odot) + 0.17 M - 0.2 M^2$, where  $M = \log_{10}(M_{st}/M_\odot)$. \citet{botyanszki18} do not provide the time dependence of the line emission specifically, but note that the H$\alpha$ emission is proportional to the Fe\texttt{III} emission over the $200 < (t-t_0) < 500$~day period they consider. Utilising the synthetic spectra models from \citet{botyanszki17}, we find the Fe\texttt{III} emission is well fit by an exponential (Fig. \ref{fig:FeIIIevolution}), which leads to an estimate for the line luminosity of 

\begin{multline}\label{eq:masslim}
    \log_{10} L(M,t) = \log_{10} L_{200} + 0.17 M\\ -0.20 M^2 - \Big(\frac{ t -t_0}{200~\hbox{days} } -1 \Big)
\end{multline}

\noindent provided $M_{st} < 2 M_\odot$. This should hold well for the Balmer lines, and is at least a better approximation for the Paschen and HeI lines than assuming their luminosities are temporally constant.

The models from \citet{botyanszki17} used to derive the luminosity decay in Fig. \ref{fig:FeIIIevolution} are truncated at 400 days after explosion. However, the ratio between Fe\texttt{III} and \Ha remains roughly constant out to 500 days after explosion \citep{botyanszki18}. To incorporate spectra taken between $400-500~\rm{days}$ after explosion, we extrapolate the exponential fit to the later epoch. Since the models used to derive the Fe\texttt{III} emission decay are generalized to non-peculiar \sne and independent of any possible H or He emission, we consider this assumption valid. However, we do not extrapolate to earlier epochs for two reasons. The onset of the nebular phase is not clearly defined in the literature \citep[e.g., ][]{black16}, leading to ambiguities in when stripped material might become visible. Additionally, the luminosities for stripped material given in Table \ref{tab:ModelInfo} are taken from \citet{botyanszki18} who explicitly state that their models are valid in the regime of $200 \leq (t-t_0)  \leq 500$~days. If these concerns are mitigated by future work our results can be expanded to include objects with earlier spectra.

\subsection{Placing Statistical Limits on Stripped Mass}

Once each spectrum is flux calibrated and corrected for the reddening, we place statistical limits on the presence of emission lines listed in Table \ref{tab:ModelInfo}, roughly following the methods of \citet{leonard07}. Each spectrum is rebinned to the approximate spectral resolution, and the spectral continuum is fit with a 2$^{\rm{nd}}$-order Savitsky-Golay polynomial \citep{press92}, excluding pixels within $2\times\rm{FWHM}$ of line centers to prevent biasing our continuum fit, as done in previous studies \citep[e.g., ][]{maguire16}. However, since we are inspecting multiple lines for emission signatures, we apply our continuum model in velocity space instead of wavelength space to incorporate this modification. 

No single continuum width adequately fits the continuum for all \sne in our sample, especially considering the spectroscopic and temporal diversity. We tailored the continuum fit width for each spectrum based on the observed \sn expansion velocity, measured from the prominent emission lines in the spectrum. Since most of the major emission lines in nebular \sne are blended to some extent \citep[e.g., ][Fig. 5]{mazzali15}, we compute the weighted average from the fitted line profiles assuming a Gaussian emission profile + linear continuum. The lines considered for deriving the expansion velocity are the major Fe\texttt{II}, Fe\texttt{III}, and Co\texttt{III} lines indicated by the vertical dashed lines in Fig. \ref{fig:example}. If the SNR of the spectrum is too low for the widths of at least 2 lines to be measured confidently, we assume a typical width of $3\,000~\rm{km}~\rm s^{-1}$. For velocities lower than this value, we risk biasing our continuum fit to include possible weak emission, and implement $3\,000~\rm{km}~\rm s^{-1}$ as a strict lower bound. Additionally, since SNe Iax are known to have narrow line profiles in the nebular phase compared to typical \sne \citep{foley16}, we adopt this lower bound for SNe-Iax as well. Because these velocities are simply a proxy for the width of the continuum fit, this method neglects the intricacies of \sne emission profiles, especially since spectroscopically bi-modal \sne are not uncommon \citep{dong15, vallely}. However, these complications are unimportant for our analysis, and we consider these simple velocity approximations adequate.

When applying the continuum fit to each spectrum, we minimize biasing our continuum by using $3\sigma$ clipping to exclude narrow host galaxy lines, telluric absorption, or instrumental artefacts. After fitting the continuum model to the data, we subtract off this continuum and inspect the residuals for emission-line signatures from unbound companion material. For each line in Table \ref{tab:ModelInfo}, we compute $10\sigma$ bounds on the integrated line flux in each region similar to Eq. 4 from \citet{leonard01}. However, for flux-calibrated spectra,

\begin{equation}\label{eq:original}
    F(10\sigma) \equiv EW(10\sigma) \times C_\lambda = 10 C_\lambda \Delta I \sqrt{W_{\rm{line}}\Delta X}
\end{equation}

\noindent where $F(10\sigma)$ is the $10\sigma$ upper limit on the integrated flux, $EW(10\sigma)$ is the corresponding upper limit on the equivalent width, $C_\lambda$ is the continuum flux at wavelength $\lambda$, $\Delta I$ is the RMS scatter around a normalised continuum, $W_{\rm{line}}$ is the width of the line profile, and $\Delta X$ is the bin size of the spectrum. We assume $W_{\rm{line}}$ is equal to the FWHM of a $\sim 1\,000~\rm{km}~\rm s^{-1}$ emission line and provide these values in Table \ref{tab:ModelInfo}. Eq. \ref{eq:original} can be re-written as

\begin{equation}\label{eq:oldlimit}
    F(10\sigma) = 10\,\Delta f_\lambda \, \mathcal{F}^{-1}\,\sqrt{ W_{\rm{line}}\Delta X}
\end{equation}

\noindent where $\Delta f_\lambda$ is the $1\sigma$ RMS scatter of the spectrum around the continuum in flux units (erg s$^{-1}$ cm$^{-2}$ \AA$^{-1}$) and $\mathcal{F}^{-1}$ is the correction term for masked pixels (see \S\ref{subsec:masking}). Our $10\sigma$ statistical limit may seem overly conservative but it does correspond to a line profile that would be visibly obvious (e.g., Fig. \ref{fig:randomcuts}). Additionally, other studies have run injection-recovery tests to determine the true detection threshold for $\sim 1\,000~\rm{km}~\rm s^{-1}$ emission lines in \sne nebula spectra and a purely statistical $F(3\sigma)$ is difficult to recover \citep[e.g., ][]{sand18}. 

$F(10\sigma)$ is then converted into a luminosity via the distance moduli listed in Table \ref{tab:snparams}. 
Distance moduli computed from the \sn light curves are used except where more reliable methods are available, such as Cepheid or Tip of the Red Giant Branch (TRGB) distances. Eq. \ref{eq:masslim} is inverted to numerically calculate a limit on $M_{st}$, which we consider a conservative upper bound on the amount of mass removed from a non-degenerate companion undergoing RLOF. This is done for each H/He line, retaining the best mass limit for both the H-rich and He-rich models. Note that the strictest mass limit for each model can come from different spectra, as each spectrum will have varying amounts of contamination from host galaxy and telluric lines.

\subsection{Mitigating Host Galaxy Emission and Other Contaminants}\label{subsec:masking}

Due to the comprehensive nature of our sample, some spectra have poor quality, significant host-galaxy emission and/or other contaminants. Pixels affected by host-galaxy emission, telluric absorption, or instrumental artefacts are masked in the ensuing flux limit calculation, ensuring only informative pixels are used in placing our flux upper limit. Masking these pixels also reduces the effective number of pixels used in the non-detection limit calculation and weakens our statistical limit. In Eq. \ref{eq:oldlimit} we include the masked pixel correction term $\mathcal{F}^{-1}$ from \citet{tucker18} to correct our limit to a more robust estimate. Concisely, the correction term $\mathcal{F}$ is the fraction of unmasked line flux to total line flux ($\mathcal{F} \in [0,1]$). Thus, masked pixels decrease $\mathcal{F}$ and increase $F(10\sigma)$, but the effect is weighted by the location of the masked pixels relative to the line centre. For example, the masked narrow host galaxy \Ha and \Hb in the bottom panels of Fig. \ref{fig:example} have larger effects on $\mathcal{F}$ than the masked [S\texttt{II}] line at $6713$\AA{} since [S\texttt{II}] is on the outskirts of the He\texttt{I}-a line profile. Masking is only implemented when the derived $F(10\sigma)$ is not representative of the true flux limit due to contaminated pixels, we leave weak or minor contamination unmasked as it only solidifies our conservative flux limit and does not introduce extra steps in our analysis.

Another difficulty occurs when the expected emission line is blended with the edge of a steep SN spectral feature.   This is especially problematic for 91bg-like and Iax SNe which have intrinsically narrow emission line profiles.    If the continuum near H/He varies by more than the amplitude of our flux limit over its FWHM, we increase our flux limit to match the continuum level variation. This results in an unambiguous line profile that would be definitively detected and prevents questionable limits from being included in our statistical analysis.


Some spectra in our study have resolutions of order $\sim 500~\rm{km}~\rm s^{-1}$, which approaches the lower end of the expected stripped mass velocity distribution \citep[e.g., ][]{boehner17}. If broad, unresolved H emission was present in a spectrum, we confirm the host galaxy source with other typical galaxy emission lines such as [O\texttt{II}] ($3727$\AA), [O\texttt{III}] ($4959$, $5007$\AA), [N\texttt{II}] (6548, 6583\AA), and [S\texttt{II}] (6713, 6731\AA). Any unresolved \Ha emission with velocity widths $\gtrsim 300~\rm{km}~\rm s^{-1}$ had at least one other unresolved galaxy emission line in the spectrum, indicating the observed H emission was not from stripped material. Additionally, the recent discovery of broad \Ha emission in ATLAS18qtd \citep{prieto19} affirms our treatment of galaxy emission lines, as none of the galaxy emission lines discussed previously were present in the discovery spectrum (see \S\ref{sec:discuss}).

\section{Results}\label{sec:results}

\begin{figure*}
    \centering
    \includegraphics[width=0.95\linewidth]{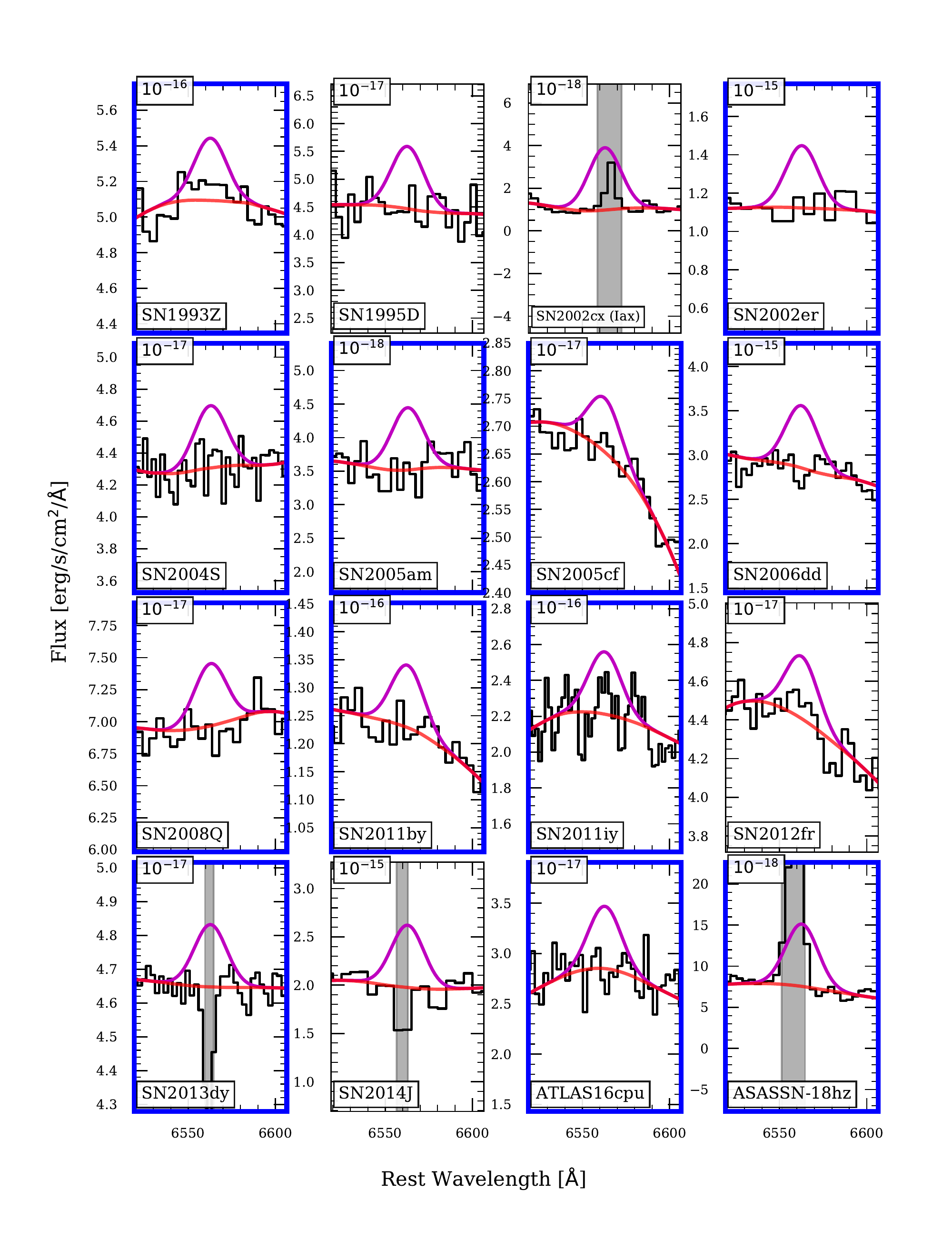}
    \caption{Randomly selected cutouts around \Ha for a portion of the \sne in this work, including the observed spectrum (black), the continuum fit (red), and the empirical $10\sigma$ line limit (purple). The scale for each spectrum is denoted in the top-left of each panel. Light grey areas mark masked regions (see text) and completely grey boxes signify \sne with no spectra covering the wavelength range. Thick blue axes indicate this spectrum was used for the best $M_{st}$ limit provided in Table \ref{tab:results}. Cutouts for all \sne and all H/He lines are provided as supplementary material (see Appendix \ref{app:tables}).}
    \label{fig:randomcuts}
\end{figure*}

\begin{figure}
    \centering
    \includegraphics[width=\linewidth]{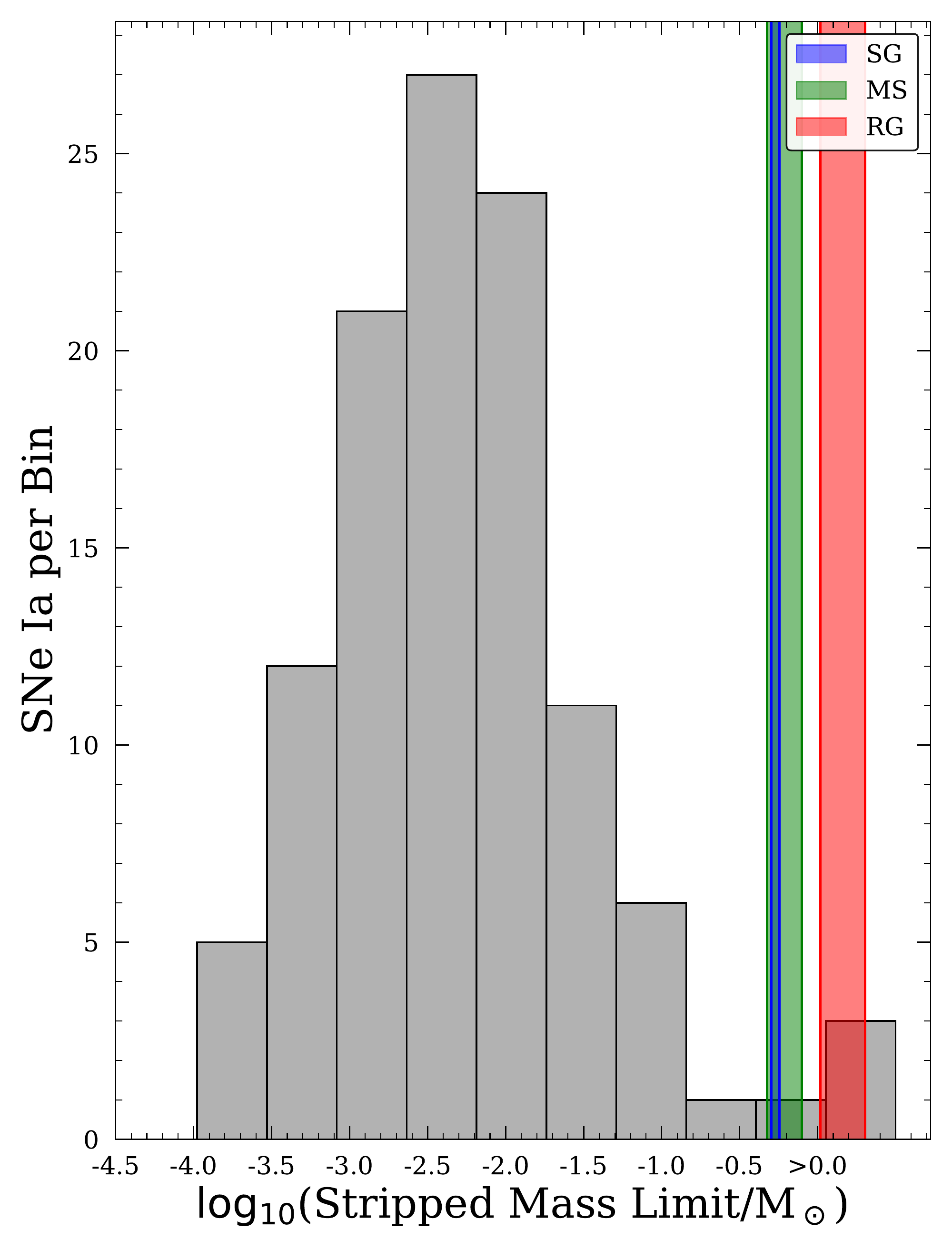}
    \caption{Distribution of mass limits on stripped H-rich material for all \sne in our sample. Colour-shaded areas indicate expected amounts of unbound mass for sub-giant (SG, blue), main-sequence (MS, green), and red-giant (RG, red) companions, taken from \citet{marietta00}, \citet{pan12}, and \citet{boehner17}.
    }
    \label{fig:NewMassHist}
\end{figure}

\begin{figure}
    \centering
    \includegraphics[width=\linewidth]{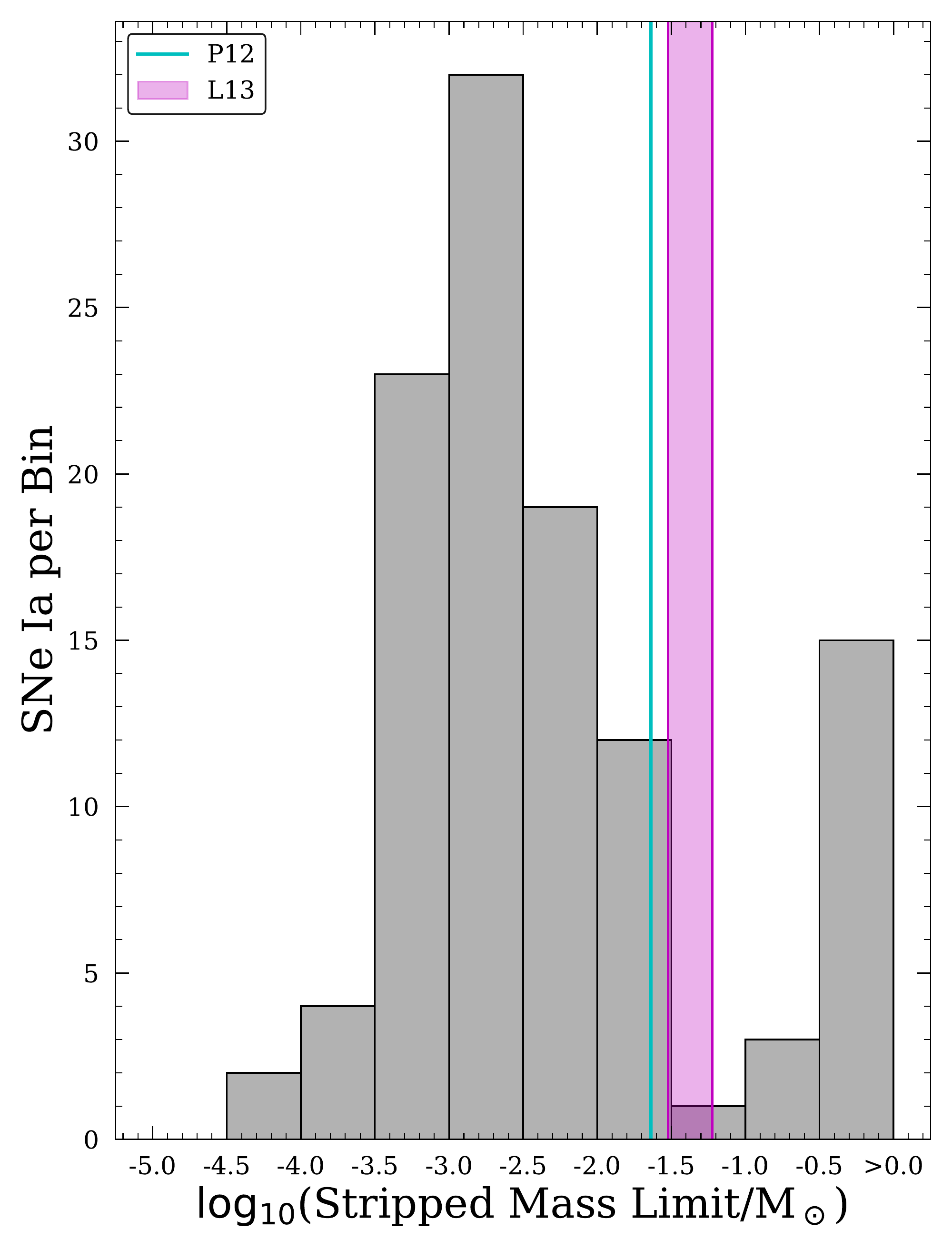}
    \caption{Similar to Fig. \ref{fig:NewMassHist}, except for the He-rich model. The magenta shaded region corresponds to stripped mass estimates from \citet{liu13} and the cyan line marks the estimate from \citet{pan12}.
    }
    \label{fig:HeMassHist}
\end{figure}

\begin{figure}
    \centering
    \includegraphics[width=\linewidth]{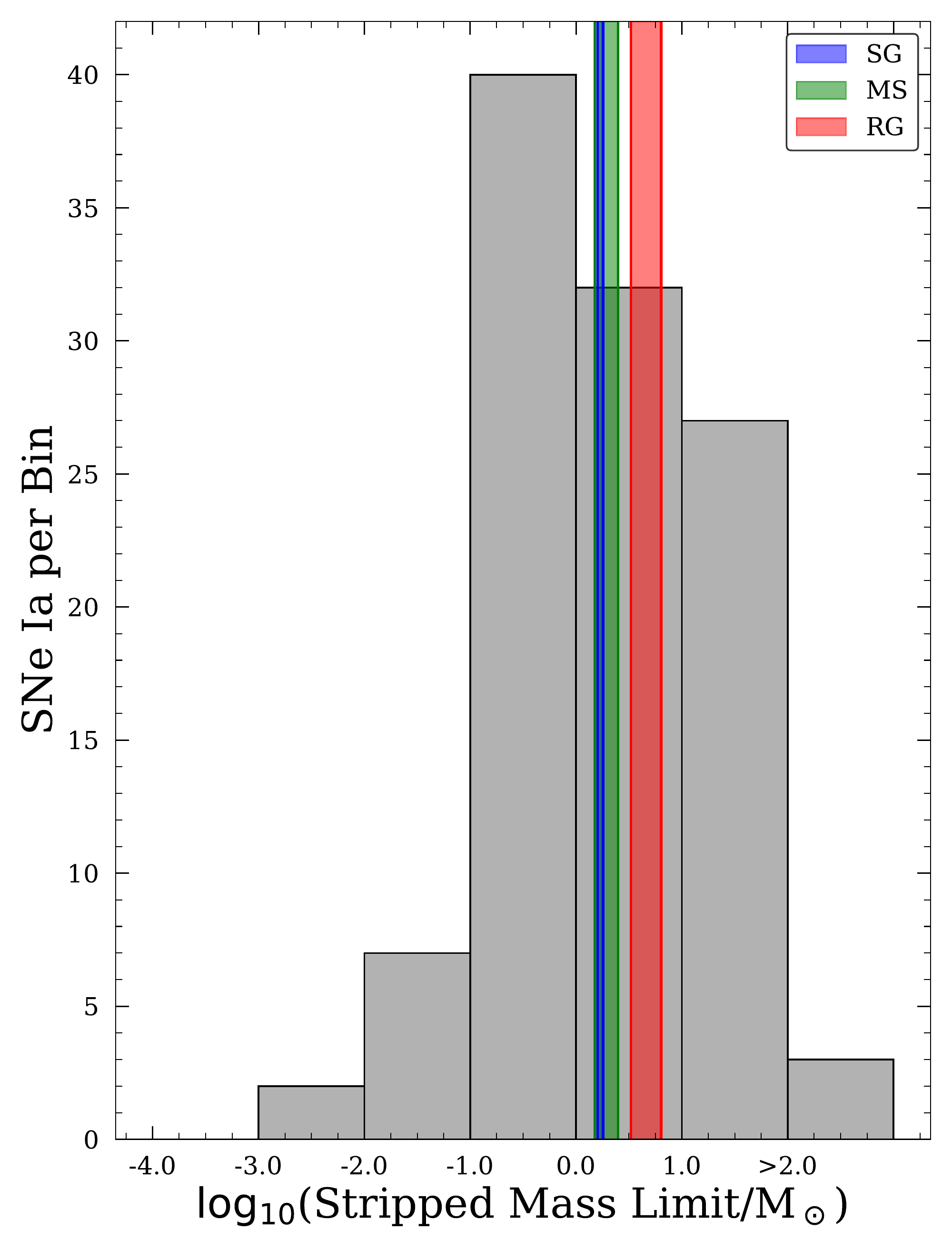}
    \caption{Similar to Fig. \ref{fig:NewMassHist}, except using the models of \citet{mattila05}. The assumed linear scaling between luminosity and stripped mass leads to higher derived $M_{st}$.
    }
    \label{fig:OldMassHist}
\end{figure}

We find no evidence of emission from stripped/ablated companion material in any of our nebular phase spectra. Fig. \ref{fig:example} provides an example Baade/MagE spectrum of SN~2015F at +295 days after maximum light, including the observed spectrum, the continuum fit, and the $10\sigma$ flux limits for each line. We provide a random selection of \Ha flux limit cutouts in Fig. \ref{fig:randomcuts} and the spectral cutouts for all H and He lines are provided as supplementary material.

The distribution of stripped mass limits are shown for the H-rich and the He-rich cases in Figs. \ref{fig:NewMassHist} and \ref{fig:HeMassHist}, respectively, with colour-shaded regions indicating the expected amounts of stripped mass from various studies in the literature. Fig. \ref{fig:OldMassHist} shows the H-rich results using the methods and models of \citet{mattila05} and \citet{leonard07} for comparison with previous estimates. Table \ref{tab:results} gives the phases, flux limits, and derived H-rich and He-rich mass limits for each \sn in our study.

We include the range of mass estimates from an H-rich RLOF companion in Figs. \ref{fig:NewMassHist} and \ref{fig:OldMassHist} as shaded regions for main-sequence (MS, blue), sub-giant (SG, green), and red giant (RG, red) companions taken from \citet{marietta00}, \citet{pan12}, and \citet{boehner17}. We take $0.15~M_\odot$ as the minimum amount of mass stripped from a companion in the \sd scenario, \sne with $M_{st,H} < 0.15~M_\odot$ are considered unlikely to have an H-rich \sd progenitor system.  

For the He-rich \sd channel, only \citet{pan12} and \citet{liu13} have published models. We include their expected values for mass stripped from a RLOF helium-star companion in Fig. \ref{fig:HeMassHist} as the magenta shaded area \citep{liu12} and the cyan line \citep{pan12}. However, there are several caveats when considering the He-rich model. The expected line luminosities given in Table \ref{tab:ModelInfo} are for $0.25~M_\odot$ of stripped He-rich material, more mass than expected for a true He-donor star. We compare our mass limits to the dedicated He-rich models from \citet{pan12}, \citet{liu13} and take a limit of $M_{st, He} < 0.023~M_\odot$ as our upper limit for He-rich \sd systems.  

\input{StatsTable.tex}

If we assume \sne with $M_{st,H} < 0.15~M_\odot$ and $M_{st,He} < 0.023~M_\odot$ exclude H-rich and He-rich \sd progenitor systems, respectively, we can constrain the observed fraction of \sd systems. Based on the non-detections in our sample, we can place observed upper limits on the fraction of \sd \sne. For a binomial distribution with $N$ trials and no successes, the upper limit $f$ at a confidence level $P$ can be expressed as 

\begin{equation}\label{eq:uplim}
    f < 1 - (1 - P)^{(N+1)^{-1}}
\end{equation}

\noindent with the results for our sample provided in Table \ref{tab:compstats}. $f_{1\sigma}$ and $f_{3\sigma}$ correspond to the $1\sigma$ and $3\sigma$ fractional upper limits on \sd \sne. For our null detections of unbound mass emission, we place statistical constraints on the fraction of \sne that can form through the classical \sd scenario for H-rich and He-rich companions. We do not consider \sne with inadequate limits on $M_{st}$ ``successes'', as the spectra do not show any evidence of the expected emission signatures, so these objects are simply omitted from our statistical analysis. 

\section{Discussion}\label{sec:discuss}

\subsection{Statistical Implications}

With our updated modelling and comprehensive sample, we place strict constraints on the fraction of \sne that can form through the classical SD scenario. At most, $\sim 6\%$ of \sne (at $3\sigma$ confidence) can stem from the H-rich formation channel, placing the majority of the production of \sne on the \dd channel, unless a modification on the \sd scenario can prevent nearly all \sne from exhibiting these expected H and He emission signatures such as the spin-up/spin-down scenario \citep{justham11, distefano11, meng13}. Considering the simplest case of only spectroscopically normal \sne, we place a $1\sigma$ ($3\sigma$) upper limit on \sd progenitors of $< \NCnormpercOne\%$ ($< \NCnormpercThree\%$). The full statistical results are provided in Table \ref{tab:compstats}, and we use the \textit{Normal+91T+91bg+SC} sample as the most representative sample from our survey. Unfortunately, our sample prevents an analysis of under- versus over-luminous \sne, as we are biased towards brighter \sne (Fig. \ref{fig:dmhist}). This highlights the importance of volume-limited surveys such as 100IAs \citep{dong15}. Still, these stringent constraints on the observed rate of \sd \sne provide strong evidence for the \dd channel producing the majority of \sne. 

We separately consider spectroscopic sub-classes at the extreme edges of the Phillips Relation. Because these \sne are thought to be on the edges of typical \sn formation, we compare the derived stripped mass limits to the same expected stripped mass values as normal \sne. Our sample has \countbg 91bg-like and \countT 91T-like \sne, for which we place $1\sigma$ ($3\sigma$) upper limits on H-rich \sd progenitors at $< \NCbgpercOne\%$ ($< \NCbgpercThree\%$)  and $< \NCTpercOne\%$ ($< \NCTpercThree\%$), respectively. It is worth mentioning that the stripped mass models assume a normal \sn explosion and the effects of under- and over-luminous \sne on the amount of stripped material has yet to be investigated.

For SNe Iax and ``Super Chandrasekhar'' (SC) \sne, it is worth discussing their characteristics and applicability to our study. The Iax sub-type \citep{foley13} is thought to stem from an entirely different formation mechanism and are not observed to enter a nebular phase but instead have photospheric properties \citep{foley16}. Our study includes \countIax such systems: SNe 2002cx, 2005hk, 2008A, and 2012Z.  \citet{liu13Iax} investigated the expected values of unbound mass for these systems if in a \sd system, finding significantly lower values of $M_{st,H} \approx 0.013-0.016~M_\odot$ compared to the typical $\sim 0.1-0.5~M_\odot$ range. All Iax \sne in our sample have $M_{st,H} < 0.013~M_\odot$, so the statistics are unchanged if the more stringent mass limit is employed. However, even if material is unbound from non-degenerate donor stars in these \sne, it is unclear if this material would be visible at late times. For these reasons, our main statistical analysis excludes these objects. 

Our sample also includes \countSC ``Super Chandrasekhar" (SC) \sne explosions (SNe~2006gz, 2007if, 2009dc, and SNF~20080723-012), where the inferred ejecta mass, $M_{ej}$, is higher than the Chandrasekhar mass of $\approx 1.4~M_\odot$ \citep[e.g., ][]{howell06, scalzo18} although the nomenclature is currently debated \citep{chen19}. The preferred formation theory of SC \sne involves a DD merger of two WDs with a combined mass above the Chandrasekhar mass \citep{tutukov94, howell01}, although SD progenitors have also been proposed \citep{yoon05}. Because these objects do enter a nebular phase and have possible SD progenitors, we include these \sne in our preferred sample.

For completeness and comparison to the literature, we also derive mass limits using the prior models of \citet{marietta00} and \citet{mattila05} which are shown in Fig. \ref{fig:OldMassHist}. Considering the same preferred \textit{Normal+91T+91bg+SC} sample, we still rule out H-rich non-degenerate companions ($M_{st,H} < 0.15~M_\odot$) for \OldnormTbgSCN \sne, corresponding to a $1\sigma$ ($3\sigma$) fractional upper limit of $< \OldnormTbgSCpercOne\%$ ($<\OldnormTbgSCpercThree\%$). This result differs slightly from the upper limit provided in Table \ref{tab:compstats} due to the assumed linear scaling between stripped mass and emitted luminosity \citep[e.g., ][]{leonard07}.     
 
In addition to the observational limitations discussed in \S\ref{sec:Hsearch}, the models used in this work are developed for normal \sne. Over- and under-luminous explosions will likely differ in the amount of stripped material from a companion star due to the differing expansion velocities \citep[e.g., ][]{benetti05, blondin12,folatelli13} and amount of ejecta mass \citep[e.g., ][]{cappellaro97, scalzo14, scalzo18}. Additionally, the SN luminosity depends on the amount of Ni synthesized in the explosion \citep[e.g., ][]{arnett82, cappellaro97, stritzinger06-ni}, indicating \sne with lower Ni mass will have less gamma-ray production to power the \Ha emission (i.e., a reduced $L_{\rm H \alpha}$). These effects likely superimpose, as under-luminous \sne will strip less mass and synthesize less Ni, but the magnitude of these effects is currently unexplored in the literature. We encourage the modeling of other \sne sub-types in future works.

\subsection{The Exclusion of SNe Ia-CSM}\label{sec:CSM}
 
SNe Ia-CSM, which show interaction with a nearby circumstellar environment, are a rare class of thermonuclear explosions for which SN~2002ic is the prototype \citep{wang04}. These events preferentially occur in star-forming host galaxies and generally have broad, over-luminous light curves \citep[$M_R\sim -20$~mag, ][]{silverman13}. The observed \Ha emission in SNe Ia-CSM usually appears near maximum light, has luminosities of $L_{\rm H \alpha}\sim  10^{40-41}~\rm{ergs}~\rm{s}^{-1}$, and have velocity widths on the order of $\sim 2\,000~\rm{km}~\rm s^{-1}$. SNe Ia-CSM are broadly thought to stem from SD progenitor systems \citep[e.g., ][]{han06}, although DD progenitors have also been proposed \citep{livio03}.

Even among this rare class of \sne there are peculiar events that do not conform to the ``standard'' properties. In particular, ASASSN-18tb \citep{kollmeier19, vallely18tb} was an under-luminous explosion, occurred in an elliptical host galaxy with little star formation, and had (comparatively) weak \Ha emission ($L_{\rm{H}\alpha}\sim 10^{38}~\rm{ergs}~\rm{s}^{-1}$), inconsistent with typical SNe Ia-CSM. Additionally, there are cases where the \Ha emission does not appear until later in the SN's evolution, referred to as ``delayed-onset'' SNe Ia-CSM \citep[e.g., ][]{graham19}. 

\Ha emission is also expected for material stripped from a companion, therefore differentiating between SNe Ia-CSM and SNe Ia with stripped material emission is an important distinction. Stripped companion material will have significantly lower velocities than the expanding SN ejecta ($v_{\rm{strip}} \sim 10^3~\rm{km}~\rm s^{-1}$ versus $v_{\rm{ej}}\sim 10^4 ~\rm{km}~\rm{s}^{-1}$), shrouding the H-emitting material with the optically-thick photosphere until the SN enters the nebular phase ($\sim 150-180~\rm{days}$ after maximum light). Thus, we exclude all objects with broad \Ha detected $<100~\rm{days}$ after maximum light. The only SN Ia-CSM that passes this criterion is SN~2015cp, a delayed-onset \sn-CSM \citep{graham19}.

While it is possible that the \Ha emission observed in SN~2015cp is from material stripped from a companion, we find this scenario unlikely. The classification spectrum taken by PESSTO \citep{PESSTO} at $\sim 45~\rm{days}$ after maximum excludes the presence of PTF11kx-like \Ha emission at $10\sigma$ \citep{graham19}. The next spectrum was acquired at $\sim 700$~days after maximum light and exhibited broad \Ha emission with $v_{\rm{FWHM}}\approx 2\,400~\rm{km}~\rm{s}^{-1}$ and $L_{\rm H\alpha}\approx 10^{38}~\rm{ergs}~\rm{s}^{-1}$. This measured $L_{\rm{H}\alpha}$ is an order of magnitude higher than the \Ha luminosity extrapolated from Eq. \ref{eq:masslim} at $\sim 710$~days after explosion, although the models have not been tested at these epochs. Additionally, the \Ha flux decreases sharply at $\sim 730~\rm{days}$ after peak by a factor of $\approx 3$ over the span of $\sim 90~\rm{days}$, inconsistent with \Ha emission powered by radioactively-decaying SN ejecta which would roughly follow the SN bolometric luminosity. These properties are consistent with CSM interaction, attributing the abrupt flux decrement to the shock passing through the CSM material. For these reasons we consider SN~2015cp a likely SN Ia-CSM and exclude it from our analysis, although further modeling is encouraged to definitively determine mass estimates.

\subsection{Time-Variable and Blue-Shifted Sodium Absorption}

Another subset of \sne with interesting properties are objects with time-varying and blue-shifted \NaID absorption, for which SN~2006X is the prototype \citep{patat07}. The \NaID absorption is thought to stem from \NaID material near the explosion that photo-ionizes during the early phases of the explosion and produces the absorption lines as the \NaID material cools and recombines \citep[e.g., ][]{simon09}. However, the origin of the \NaID material is unclear and proposed sources include wind from a SD progenitor system \citep[e.g., ][]{patat07}, circumstellar debris from a DD merger \citep[e.g., ][]{raskin13}, or even nearby gas clouds within the host galaxy \citep{chugai08}.

\citet{sternberg11} analyzed a set of 35 SNe Ia and found 22 exhibit some form of \NaID absorption profiles, with 12 having blue-shifted (relative the host-galaxy velocity) \NaID profiles. However, comparing their SNe Ia sample to a sample of core collapse (CC) SNe could not statistically confirm the two sets come from different parent populations (i.e., the source for SNe Ia and CC SNe \NaID absorption could be the same). Additionally, the blue-shifted \NaID profiles are preferentially observed in spiral galaxies, indicating age or host-galaxy environment may play a role in the \NaID interpretation. 

Another prediction for \NaID absorption associated with the \sn progenitor system is a time-variable \NaID equivalent width, as the \NaID material will recombine at different times depending on the distance from the explosion. \citet{sternberg13} searched for time-varying \NaID absorption in a sample of 14 objects and found 3 \sne that meet this criterion (PTF11kx, SNe 2006X and 2007le). With these detections, \citet{sternberg13} found $18\pm11$\% of SNe Ia have variable \NaID profiles and thus could be produced by a SD progenitor system, in conflict with the results presented here. One of these objects is a known SNe Ia-CSM that we exclude from the sample \citep[PTF11kx, ][]{dilday12}, but SNe~2006X and 2007le are both in our nebular sample and have strict constraints on stripped companion material ($M_{st} < 4\times10^{-5} ~M_\odot$ for SN~2006X and $M_{st}< 3\times10^{-3} ~M_\odot$ for SN~2007le). We discuss these discrepancies further in \S\ref{subsec:discrep}.

\subsection{The Lack of a Consistent Theory for SNe Ia Progenitors}\label{subsec:discrep}

Besides this work, there are several other studies which place quantitative or qualitative limits on the fraction of \sd progenitor systems using a range of wavelengths and techniques \citep[e.g., ][]{gilfanov10, hayden10, bianco11, brown12, chomiuk16}. Most of these studies focus on WD+RG systems, as these are the easiest to observationally detect. Each study individually does not definitively rule out \sd \sne progenitors, however, when considered as a whole it is clear that most \sne cannot form through the classical \sd scenario. Thus, the \dd scenario likely account for the majority of normal \sne. However, detecting and characterising double WD binaries is exceptionally difficult \citep[e.g., ][]{rebassa18}. 

Reconciling limits on SD progenitors with SD-favored SNe (i.e., SNe Ia-CSM and \sne with variable \NaID) has long been a difficulty for the community. Especially interesting are the systems with conflicting interpretations, such as SN~2007le and ASASSN-18bt. SN~2007le exhibits time-variable, blue-shifted \NaID absorption \citep{sternberg13} but has stringent limits on stripped material emission (this work, Table \ref{tab:results}). Similarly, ASASSN-18bt showed a two-component rising light curve \citep{shappee18bt, dimitriadis18}, a potential signature for SN ejecta impacting a nearby companion \citep{kasen10}. However, nebular spectra rule out any stripped material emission with strict upper limits \citep[][ this work]{tucker18, dimitriadis19}. While the discrepancies for both \sne can also be explained with alternative theories (e.g., Ni$^{56}$ mixing for the two-component rise in ASASSN-18bt and DD merger debris for the \NaID absorption in SN~2007le), these objects highlight the uncertainties that still surround \sn progenitors. 

Recently, ATLAS18qtd (SN~2018cqj) was discovered to exhibit time-variable \Ha emission in the nebular phase \citep{prieto19}. The spectra were posted after the submission of this manuscript, however, it warrants a brief discussion here. The classification spectrum taken at $19$~days after maximum has no evidence for H emission, and the next spectrum was not acquired until $\sim 190$~days after maximum. The measured \Ha luminosity declines contemporaneously with the \sn luminosity, a key expectation for material stripped from a \sd progenitor. However, there are only two measurements of the \Ha emission and the inferred mass of the stripped material ($M_{st} \approx 10^{-3}~M_\odot$) is far lower than expected ($\sim 0.1-0.5M_\odot$). ATLAS18qtd is an under-luminous explosion, so it is possible the stripped material will be lower than expected from simulations in the literature, but the extent of these effects is still unexplored. Late-time observations of this unique object may yet further constrain the evolution of the \Ha emission and help elucidate its origin. If we assume ATLAS18qtd is indeed a SD progenitor system, this discovery highlights the inherent rarity of such events compared to the typical \sn population.

There is also the possibility that stripped material does exist in our sample but is invisible due to observational factors. The radioactively-decaying \sn ejecta provides a power source for the stripped material, namely gamma-ray deposition. Since we only analyze spectra that are in the nebular phase we are probing the innermost regions of the explosion. \sne at these epochs essentially have no photosphere, and thus have no way of shrouding the \Ha-emitting material. No such mechanism or process has been proposed to suppress the expected \Ha emission, so we consider this possibility unlikely. We conclude most \sne cannot have formed from a classical SD progenitor system based on this work.

\section{Conclusion}\label{sec:conclusion}

We present a large, comprehensive search for emission expected from stripped companion material in the SD formation scenario of \sne. Using \GoodSpeccount spectra of \Nsn \sne from a variety of telescopes and instruments, we find no evidence for any stripped material emission in our sample. Using these null detections, we place statistical constraints on the fraction of \sne that can form through the classical RLOF SD scenario, finding $< \NCnormTbgSCpercThree\%$ and $<\HenormTbgSCpercThree\%$ of \sne can form through the H-rich and He-rich channels, respectively, at $3\sigma$ confidence. The lack of emission is difficult to reconcile with the classical SD formation scenario for \sne, and provides unique constraints on the production mechanism of these phenomena.

Thus far, there has not been a proposed formation mechanism that adequately reproduces all aspects of \sn properties. There seems to be contributions from both SD and DD progenitors to the total \sn rate \citep[e.g., Sec. 4.1 from ][]{maoz14}, yet the exact distribution is widely debated. Reconciling seemingly conflicting results (e.g., this work versus \citealp{sternberg13}) has long been a difficulty. Any unifying theory for \sne formation must account for all the observed characteristics of these phenomena and the seemingly conflicting results across various methodologies. If the SD channel does produce a significant fraction of \sne, there must be an unincorporated physical process in previous modeling efforts to explain our non-detections of stripped material. Alternatively, most \sne must form from DD systems to match the results presented here.

\textit{Facilities}: duPont, Magellan, Very Large Telescope

\textit{Software}: Python2.7, astropy \citep{astropyref}, astroquery \citep{astroqueryref} numpy, scipy, PyMUSE \citep{pymuseref}, SpectRes \citep{spectresref}, extinction\footnote{\url{https://github.com/kbarbary/extinction}}, SExtractor \citep{sextractorref}, Montage\footnote{\url{http://montage.ipac.caltech.edu/}}, Lpipe, IDL8.6, LowRedux, \texttt{IRAF}, SNooPy \citep{burns11}

\section*{Acknowledgements}

We thank K. Maguire, K. Graham, K. Motohara, K. Maeda, S. Taubenberger, T. Diamond, G. Dimitriadis, C. Ashall, and D. Sand for supplying nebular spectra. We thank A. Laity for assisting with the KOA data search and retrieval. Additionally, we thank C. Auge, G. Anand, A. Payne, and O. Graur for useful conversations about the project. We thank P. Chen from providing several \sne light curves prior to publication. Also, many thanks to Bev Wills for her help tracking down the information regarding the nebular spectrum of SN~1981B. 

M.A.T. acknowledges support from the United States Department of Energy through the Computational Sciences Graduate Fellowship (DOE CSGF) through grant DE-SC0019323. B.J.S is supported by NSF grants AST-1908952, AST-1920392, and AST-1911074. M.D.S. is supported  in part by a generous grant (13261) from VILLUM FONDEN and a project grant from the Independent Research Fund Denmark. Support for J.L.P. is provided in part by FONDECYT through the grant 1191038 and by the Ministry of Economy, Development, and Tourism's Millennium Science Initiative through grant IC120009, awarded to The Millennium Institute of Astrophysics, MAS. P.J.V. is supported by the National Science Foundation Graduate Research Fellowship Program Under Grant No. DGE-1343012. C.S.K. is supported by NSF grants AST-1515876, AST-1515927 and AST-181440. C.S.K. is also supported by a fellowship from the Radcliffe Institute for Advanced Studies at Harvard University.

This work is Based on (in part) observations collected at the European Organisation for Astronomical Research in the Southern Hemisphere under ESO programmes 0100.D-0191(A), 0101.D-0173(A), 0102.D-0287(A), and 096.D-0296(A). This paper includes data gathered with the 6.5 meter Magellan Telescopes located at Las Campanas Observatory, Chile. The CSP has been supported by the National Science Foundation under grants AST0306969, AST0607438, AST1008343, AST1613426, and AST1613472.

This paper made use of the modsIDL spectral data reduction reduction pipeline developed in part with funds provided by NSF Grant AST-1108693.

This research made use of Montage. It is funded by the National Science Foundation under Grant Number ACI-1440620, and was previously funded by the National Aeronautics and Space Administration's Earth Science Technology Office, Computation Technologies Project, under Cooperative Agreement Number NCC5-626 between NASA and the California Institute of Technology.

This work has made use of data from the European Space Agency (ESA) mission
{\it Gaia} (\url{https://www.cosmos.esa.int/gaia}), processed by the {\it Gaia}
Data Processing and Analysis Consortium (DPAC,
\url{https://www.cosmos.esa.int/web/gaia/dpac/consortium}). Funding for the DPAC
has been provided by national institutions, in particular the institutions
participating in the {\it Gaia} Multilateral Agreement.

The Pan-STARRS1 Surveys (PS1) and the PS1 public science archive have been made possible through contributions by the Institute for Astronomy, the University of Hawaii, the Pan-STARRS Project Office, the Max-Planck Society and its participating institutes, the Max Planck Institute for Astronomy, Heidelberg and the Max Planck Institute for Extraterrestrial Physics, Garching, The Johns Hopkins University, Durham University, the University of Edinburgh, the Queen's University Belfast, the Harvard-Smithsonian Center for Astrophysics, the Las Cumbres Observatory Global Telescope Network Incorporated, the National Central University of Taiwan, the Space Telescope Science Institute, the National Aeronautics and Space Administration under Grant No. NNX08AR22G issued through the Planetary Science Division of the NASA Science Mission Directorate, the National Science Foundation Grant No. AST-1238877, the University of Maryland, Eotvos Lorand University (ELTE), the Los Alamos National Laboratory, and the Gordon and Betty Moore Foundation.

Based on observations obtained at the Gemini Observatory acquired through the Gemini Observatory Archive and processed with the Gemini PyRAF package, which is operated by the Association of Universities for Research in Astronomy, Inc., under a cooperative agreement with the NSF on behalf of the Gemini partnership: the National Science Foundation (United States), the National Research Council (Canada), CONICYT (Chile), Ministerio de Ciencia, Tecnolog\'{i}a e Innovaci\'{o}n Productiva (Argentina), and Minist\'{e}rio da Ci\^{e}ncia, Tecnologia e Inova\c{c}\~{a}o (Brazil).

This paper makes use of data obtained from the Isaac Newton Group of Telescopes Archive which is maintained as part of the CASU Astronomical Data Centre at the Institute of Astronomy, Cambridge.

The LBT is an international collaboration among institutions in the United States, Italy and Germany. LBT Corporation partners are: The University of Arizona on behalf of the Arizona Board of Regents; Istituto Nazionale di Astrofisica, Italy; LBT Beteiligungsgesellschaft, Germany, representing the Max-Planck Society, The Leibniz Institute for Astrophysics Potsdam, and Heidelberg University; The Ohio State University, and The Research Corporation, on behalf of The University of Notre Dame, University of Minnesota and University of Virginia.



\bibliographystyle{mnras}
\bibliography{references}

\newpage
\appendix

\section{A Nebular Phase Phillip's Relation}\label{app:LTPR}

\begin{figure*}
    \centering
    \includegraphics[width=\linewidth]{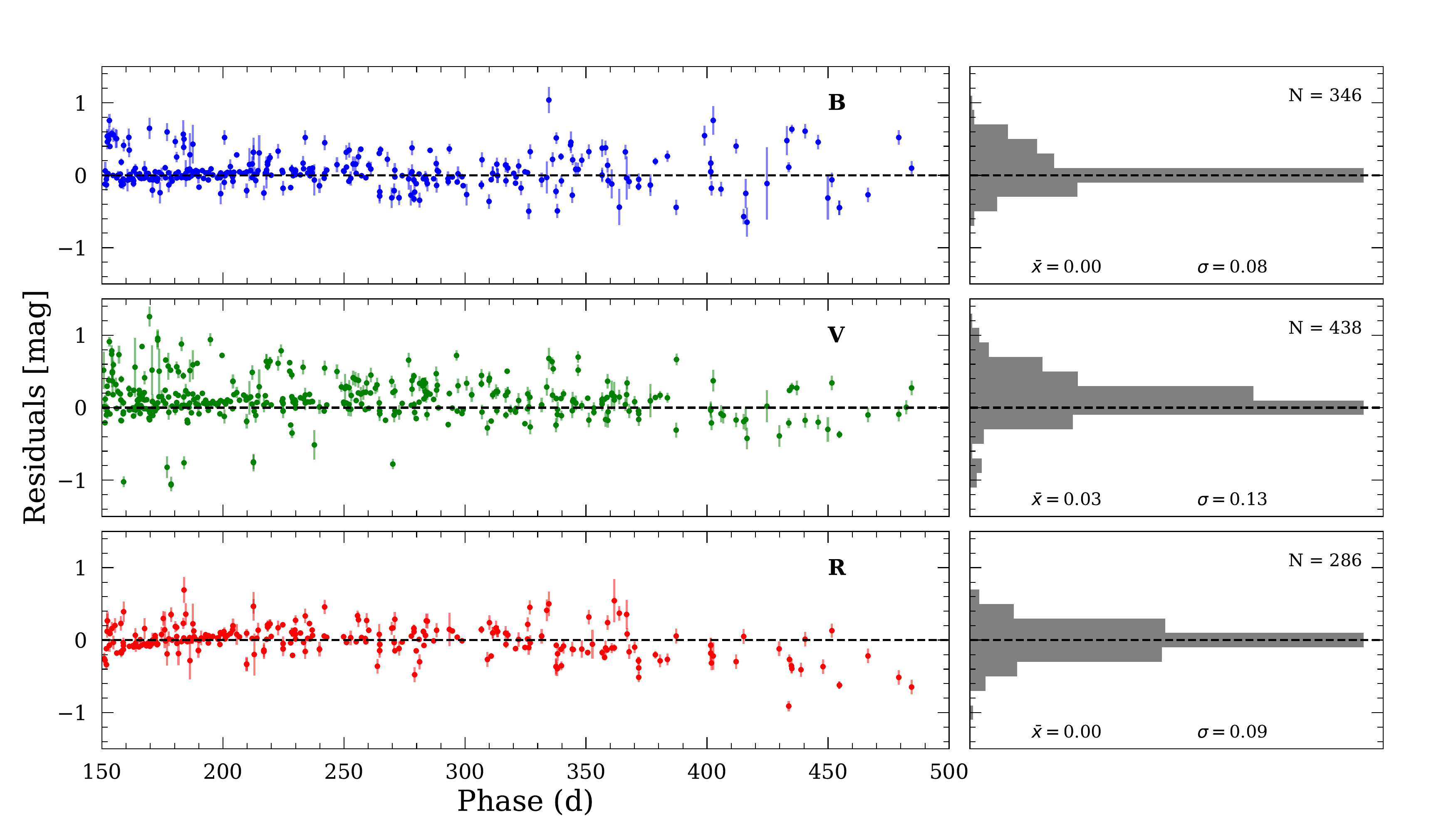}
    \caption{\textit{Left}: Residuals of the late-time relation bootstrap fit from Eqs. \ref{eq:NPPR}-\ref{eq:NPPR2} using the values in Table \ref{tab:SNPYfitValues}. \textit{Right}: Collapsed residual distribution of the best-fit solution.}
    \label{fig:late-time-residuals}
\end{figure*}

\begin{table*}
\caption{Values of the coefficients for Eqs. \ref{eq:NPPR}-\ref{eq:NPPR2} and fit statistics. $N_{tot}$ is the total number of photometric points used in each filter from $N_{SN}$ \sne.}
\label{tab:SNPYfitValues}
\begin{tabular}{c|rccccccc}
Filter & $a_\lambda$ & $b_\lambda$ & $c_\lambda$ & $d_\lambda$ & $N_{SN} $ & $N_{\rm{tot}}$ & $\bar{x}$ & $\sigma$ \\
 & $[10^{-3}~\rm{day}^{-1}]$ & [$10^{-2}$~mag day$^{-1}$] & & [mag]\\
\hline
\input{InputFiles/late-time-data.txt}
\hline
\end{tabular}
\end{table*}

In most \sne, the peak luminosity and photospheric phase decline rate (e.g., \dm) are correlated with the amount of $^{56}$Ni produced in the explosion \citep[e.g., ][]{stritzinger06, scalzo18}. Therefore, these same observables should also correlate with the magnitude of \sne as they enter the nebular phase. For \sne with nebular spectra but no usable nebular photometry, this relation provides a method for estimating the nebular magnitude using near-peak photometry.

The photometric sample used in deriving the Nebular Phase Phillips Relation (NPPR) excludes Iax, CSM, and SC \sne. Although 91T- and 91bg-like do not strictly follow the relation between luminosity and decline rate of normal \sne, they are powered by the radioactive decay of $^{56}$Ni to stable $^{56}$Fe. As mentioned before, \dm is indicative of the amount of $^{56}$Ni synthesised in the explosion, and therefore our parameterization described below still accurately models 91T- and 91bg-like \sne. However, SC and Iax \sne have unique ionisation properties, which is further exemplified by their nebular spectra which lack the prominent [Fe\texttt{II}/\texttt{III}] and [Co\texttt{II}/\texttt{III}] emission features of their normal, 91T-, and 91bg-like counterparts \citep[e.g., ][]{taubenberger13, foley16}. It is possible that the photometric intricacies of 91T- and 91bg-like \sne are washed out by our heterogeneous sample, and more precise results can be attained with distinct samples of \sne spectral types.

Taking all available nebular phase photometry of viable \sne from this work and the literature, we derive an approximate functional form for calculating the apparent magnitude of a \sn with a measured $m_{\rm{max}}$ and \dm. Since \sne have nearly linear decays in magnitude space at nebular epochs, we use the functional form

\begin{equation}
m_{\lambda, \rm{neb}}(t_p) =m_{\lambda, \rm{max}} + \Delta m_{\lambda} (t_p),
\end{equation}

\noindent where $m_{\lambda,\rm{neb}}(t_p)$ is the nebular magnitude in filter $\lambda$ at phase $t_p = t-t_{\rm{max}}$, $m_{\lambda, \rm{max}}$ the magnitude at peak in filter $\lambda$, and $\Delta m_\lambda$ is the change in brightness between maximum light and $t_p$ for that filter. By formulating our relation for individual filters, we can neglect extinction from the Milky Way and the host galaxy since the maximum light and nebular magnitude of a \sn will be affected equally. Thus, we parameterize $\Delta m_\lambda$ as a function a \sn's \dm,

\begin{equation}\label{eq:NPPR}
    \Delta m_{\lambda} = A_\lambda (\Delta m_{15}) \times (t_p-250~\rm d) + B_\lambda (\Delta m_{15}),
\end{equation}
\noindent where
\begin{equation}
    A_\lambda(\dm) = a_\lambda(\dm - 1.1~\rm{mag}) + b_\lambda,
\end{equation}
\noindent and
\begin{equation}\label{eq:NPPR2}
    B_\lambda(\dm) = c_\lambda(\dm - 1.1~\rm{mag}) + d_\lambda.
\end{equation}

Here, $m_{\lambda, \rm{max}}$ is the apparent magnitude at maximum light in filter $\lambda$, $m_\lambda(t)$ is the nebular magnitude, $t_p$ is the phase of the observations, \dm the decline rate, and measured coefficients $\{a,b,c,d\}$ which are provided in Table \ref{tab:SNPYfitValues}. $t_p$ and $\Delta m_{15}$ are offset by typical values to reduce their covariance in the fitting process.

The coefficients in Table \ref{tab:SNPYfitValues} were computed using all available nebular photometry between $150-500~\rm d$ after maximum light. The coefficients were first approximated using a sample of well-studied \sne with $\geq 5$ measurements in a given filter in the temporal bounds listed above, such as SNe 2011fe, 2012fr, 2013gy, and 2015F, then expanded to include all photometric points. The \sne used in deriving the NPPR have decline rates that span $\Delta m_{15} \sim 0.8-1.8~\rm{mag}$ and are denoted with a $^\star$ in Table \ref{tab:photdata}. For publicly available photometry for which there are no reported uncertainties, we assign a nominal uncertainty of $0.1~\rm{mag}$. In fitting the data, we implement non-linear least squares fitting coupled with a bootstrap-resampling technique to derive reasonable estimates for the uncertainties. The residuals of the best-fit solution are shown in the left panel of Fig. \ref{fig:late-time-residuals}, and the collapsed distribution is provided in the right panel.

For \sne with a measured peak magnitude and \dm, we show the nebular $BVR$ magnitude can be approximated to $\sim 20\%$. These results were derived using a heterogeneous data set and likely can be improved with a consistent photometric system and targeted observations across a reasonable span of \dm. This technique can also be used in identifying peculiar or strange \sne that deviate from their expected brightness at a given epoch, such as ``late-onset'' CSM interaction \citep[e.g., ][]{graham19}. Additionally, we attempted to expand this methodology to other photometric filters (e.g., $g$, $r$), but there were too few observations to build a quality model.

\section{Supplementary Tables and Figures}\label{app:tables}

In Table \ref{tab:targets}, we provide the name of the \sn, redshift, and references for discovery and classification. Table \ref{tab:snparams} provides the parameters from the light curve fits, including time of maximum light, \dm, and the distance modulus. We also include the total number of nebular phase spectra for that \sn and the corresponding phases. For information on each spectrum, including the date, telescope, instrument, and reference, see Table \ref{tab:spectra-params}. Flux limits and derived mass limits are given in Table \ref{tab:results}. New photometry presented in this work is provided in Table \ref{tab:newphot} and all photometry references are given in Table \ref{tab:photdata}.

\subsection{Data Tables}

For \sne with redshifts measured from the supernova lines near maximum light, we tweak the redshift using host galaxy emission lines when necessary. Major host galaxy lines such as \Ha, \Hb, [N\texttt{II}], and [O\texttt{III}] are fit with Gaussian line profiles to estimate the line centre and then used to measure the host redshift. 


For \sne with insufficient photometry for a reliable light curve fit in \texttt{SNooPy}, we consider two approaches. If there are $\geq 3$ photometric points near maximum light, we use linear least-squares coupled with bootstrap-resampling to fit a quadratic curve to the data and estimate $t_{max}$ and the associated uncertainty. Otherwise, the value for $t_{max}$ is taken from the spectroscopic classification reference given in Table \ref{tab:targets} and assigned a nominal uncertainty of $\pm 5~\rm d$.

SNe-Iax do not conform to the standard \sne templates utilised by \texttt{SNooPy} and other \sn light curve fitters. Thus, we compute \dm and $t_{\rm{max}}$ using spline fits in the \texttt{SNooPy} environment. This prevents us from deriving the host reddening $E(B-V)_{\rm{host}}$, however, the Iax SNe in our sample have negligible reddening \citep{li03, phillips07, mccully14, stritzinger15, foley15}.

The ``Quality'' column in Table \ref{tab:spectra-params} provides a rough estimate of the quality of the spectrum. This is mostly qualitative, and intended to provide readers with an estimate of the spectral quality for each \sn in our sample. The rankings are as follows:
\begin{itemize}
    \item[] \textit{High}: The spectrum clearly shows the major Fe and Co emission lines between $\sim 4\,000-7\,000$\AA. The spectrum exhibits little to no host contamination or instrumental artefacts. 
    \item[] \textit{Med(ium)}: The major Fe lines are visible, while the Co lines are noisy or absent. The spectrum may also suffer from minor to moderate host galaxy contamination and/or instrumental artefacts.
    \item[] \textit{Low}: The major Fe lines are barely detectable above the spectral noise, and the Co lines mostly below the detection threshold. This category also includes overall medium-quality spectra with significant host galaxy contamination and/or instrumental artefacts.
\end{itemize}

\subsection{Special Cases}\label{app:specialSNe}

\begin{table}
    \centering
    \caption{$R_V$ values and references for \sne with $\geq3\sigma$ deviations from the assumed $R_V=3.1$.}
    \label{tab:reddenrefs}
    \begin{tabular}{ccp{3cm}}
        Name & $R_V$ & Ref. \\\hline
        SN2002bo & 1.2 & \citet{phillips13} \\
        SN2004eo & 0.8 & \citet{burns14} \\
        SN2006X & 1.5 & \citet{wang08, phillips13,burns14} \\
        SN2007le & 1.6 & \citet{phillips13, burns14} \\
        SN2014J & 1.5 & \citet{amanullah14, foley14, gao15, brown15} \\
         \hline               
    \end{tabular}
\end{table}

We discuss any extenuating circumstances or any other relevant details about specific \sne that differ from the general methodology described in \S\ref{sec:data}. Examples include alternative flux calibration methods, spectroscopic oddities noticed in our analysis, and spectrum reference discrepancies. \sne with $R_V$ values known to deviate from the standard $R_V = 3.1$ are listed in Table \ref{tab:reddenrefs}. For ensemble studies \citep[e.g., ][]{phillips13, burns14}, we require a $\geq 3\sigma$ deviation from $R_V = 3.1$ to include the value in our calculation. When drawing $R_V$ values from \citet{burns14}, we implement the F99+uniform prior results.


\textit{SN1998bu}: The two nebular spectra from \citet{cappellaro01} do not have any specific mention in that manuscript, however, the reference on WiseRep points to this paper. Thus, we include the reference, but acknowledge we could not verify this paper was the true source for these spectra. 

\textit{SN2002bo}: The OSC and WiseRep also report several nebular phase NIR spectra for this SN. However, cross-referencing the reported spectra with the observational parameters given in \citet{benetti04}, we believe the dates provided for the NIR spectra are off by a year, and these spectra are closer to a few months after maximum light instead of several hundred days after maximum light. We exclude these spectra from our sample.

\subsection{Supplementary Figures}
We provide cutouts around each spectral line inspected for H/He emission (Table \ref{tab:ModelInfo}) for the spectrum used in calculating the limits provided in Table \ref{tab:results} for each \sn as supplementary figures. An example of the format of these figures is provided in Fig. \ref{fig:randomcuts}. The black line is the observed spectrum, with the continuum fit and flux upper limit in red and purple, respectively. Gray shaded areas indicate masked spectral regions and completely gray boxes indicate that particular \sn had no spectra covering that spectral region. When multiple nebular spectra of a \sn cover the same expected H/He line, we provide the spectrum corresponding to the best mass limit \textit{for that line}. Therefore, the panel for \Ha may show a different spectrum than the panel for \Hb for the same \sn. This ensures all adequate spectra are presented, even when some spectra do not cover all the optical and NIR lines considered in this study. The border colour of a given panel indicates whether it is used in the final stripped mass determination, the results of which are provided in Table \ref{tab:results}. Blue borders indicate the panels used in the H-rich mass limit, red borders indicate He-rich limits, and purple borders indicate He lines used for both H- and He-rich limits.

\label{lastpage}

\begin{table*}
\caption{All \sne studied in this work. }
\label{tab:targets}
\input{InputFiles/basictable-0.tex}
\end{table*}

\begin{table*}
\contcaption{All \sne studied in this work.}
\begin{threeparttable}
\input{InputFiles/basictable-1.tex}
\end{threeparttable}
\begin{tablenotes}
\item[]$^a$PSN~J1149 = PSN~J11492548-0507138 
\item[]$^b$SNF-012 = SNF20080723-012 
\end{tablenotes}
\end{table*}


\begin{table*}
\caption{\sne light curve parameters, number of late-time spectra and the corresponding phases, ordered by $t_{\rm{max}}$. See \S\ref{sec:Hsearch} for fitting methods.}
\label{tab:snparams}
\begin{threeparttable}
\input{InputFiles/targettable-0.tex}
\end{threeparttable}
\end{table*}

\begin{table*}
\begin{threeparttable}
\contcaption{Basic \sne light curve parameters, number of late-time spectra and the corresponding phases. See \S\ref{sec:Hsearch} for fitting methods.}
\centering
\input{InputFiles/targettable-1.tex}
\end{threeparttable}
\end{table*}

\begin{table*}
\caption{Spectra observations.}
\label{tab:spectra-params}
 \input{InputFiles/spectable-0.tex}
\end{table*}

\begin{table*}
\contcaption{Spectra observations.}
 \input{InputFiles/spectable-1.tex}
\end{table*}

\begin{table*}
\contcaption{Spectra observations.}
 \input{InputFiles/spectable-2.tex}
\end{table*}

\begin{table*}
\begin{threeparttable}
\contcaption{Spectra observations.}
 \input{InputFiles/spectable-3.tex}
\end{threeparttable}
\end{table*}


\begin{table*}
\caption{Flux limits for each line in Table \ref{tab:ModelInfo} and the corresponding H/He mass limit.}
\label{tab:results}
\centering
\begin{threeparttable}

 \input{InputFiles/resultstable-0.tex}
\end{threeparttable}
\end{table*}

\begin{table*}
\contcaption{Flux limits for each line in Table \ref{tab:ModelInfo} and the corresponding H/He mass limit if computed. Flux limits are given in [erg/s/cm$^2$] and stripped mass limits are given in [$M_{\odot}$].}
\centering
\begin{threeparttable}

 \input{InputFiles/resultstable-1.tex}
\end{threeparttable}
\end{table*}

\input{texfiles/newphot.tex}

\begin{table}
    \centering
    \caption{Photometry data for each \sn studied in this work. $N_{tot}$ refers to the total number of photometric points for a given SN. Phases are given relative to maximum light. Objects denoted with a $^\star$ are used in deriving the NPPR (Appendix \ref{app:LTPR}). }
    \label{tab:photdata}
    \input{texfiles/phottable-0.tex}

\end{table}

\begin{table}
    \centering
    \contcaption{Continuation of Table \ref{tab:photdata}.}
    \input{texfiles/phottable-1.tex}

\end{table}

\begin{table}
    \centering
    \contcaption{Continuation of Table \ref{tab:photdata}.}
    \input{texfiles/phottable-2.tex}

\end{table}

\begin{table}
    \centering
    \contcaption{Continuation of Table \ref{tab:photdata}.}
    \input{texfiles/phottable-3.tex}

\end{table}

\begin{table}
    \centering
    \contcaption{Continuation of Table \ref{tab:photdata}.}
    \input{texfiles/phottable-4.tex}

\end{table}

\begin{table}
    \centering
    \contcaption{Continuation of Table \ref{tab:photdata}.}
    \input{texfiles/phottable-5.tex}

\end{table}

\begin{table}
    \centering
    \contcaption{Continuation of Table \ref{tab:photdata}.}
    \input{texfiles/phottable-6.tex}

\end{table}

\begin{table}
    \centering
    \contcaption{Continuation of Table \ref{tab:photdata}.}
    \input{texfiles/phottable-7.tex}

\end{table}

\begin{table}
    \centering
    \contcaption{Continuation of Table \ref{tab:photdata}.}
    \input{texfiles/phottable-8.tex}

\end{table}

\begin{table}
    \centering
    \contcaption{Continuation of Table \ref{tab:photdata}.}
    \input{texfiles/phottable-9.tex}

\end{table}

\begin{table}
    \centering
    \contcaption{Continuation of Table \ref{tab:photdata}.}
    \input{texfiles/phottable-10.tex}

\end{table}


\end{document}

%% file: InputFiles/FinalNumbers.tex
\newcommand{\HeallpercOne}{\ensuremath{1.2}\xspace}
\newcommand{\HeallpercThree}{\ensuremath{6.2}\xspace}
\newcommand{\HeallN}{\ensuremath{92}\xspace}
\newcommand{\NCallpercOne}{\ensuremath{1.1}\xspace}
\newcommand{\NCallpercThree}{\ensuremath{5.3}\xspace}
\newcommand{\NCallN}{\ensuremath{107}\xspace}

\newcommand{\numall}{\ensuremath{111}\xspace}
\newcommand{\HenormpercOne}{\ensuremath{1.5}\xspace}
\newcommand{\HenormpercThree}{\ensuremath{7.7}\xspace}
\newcommand{\HenormN}{\ensuremath{73}\xspace}
\newcommand{\NCnormpercOne}{\ensuremath{1.3}\xspace}
\newcommand{\NCnormpercThree}{\ensuremath{6.6}\xspace}
\newcommand{\NCnormN}{\ensuremath{86}\xspace}

\newcommand{\numnorm}{\ensuremath{90}\xspace}
\newcommand{\HeTpercOne}{\ensuremath{20.5}\xspace}
\newcommand{\HeTpercThree}{\ensuremath{69.4}\xspace}
\newcommand{\HeTN}{\ensuremath{4}\xspace}
\newcommand{\NCTpercOne}{\ensuremath{17.4}\xspace}
\newcommand{\NCTpercThree}{\ensuremath{62.7}\xspace}
\newcommand{\NCTN}{\ensuremath{5}\xspace}

\newcommand{\numT}{\ensuremath{5}\xspace}
\newcommand{\HebgpercOne}{\ensuremath{13.4}\xspace}
\newcommand{\HebgpercThree}{\ensuremath{52.3}\xspace}
\newcommand{\HebgN}{\ensuremath{7}\xspace}
\newcommand{\NCbgpercOne}{\ensuremath{12.0}\xspace}
\newcommand{\NCbgpercThree}{\ensuremath{48.2}\xspace}
\newcommand{\NCbgN}{\ensuremath{8}\xspace}

\newcommand{\numbg}{\ensuremath{8}\xspace}
\newcommand{\HeIaxpercOne}{\ensuremath{20.5}\xspace}
\newcommand{\HeIaxpercThree}{\ensuremath{69.4}\xspace}
\newcommand{\HeIaxN}{\ensuremath{4}\xspace}
\newcommand{\NCIaxpercOne}{\ensuremath{20.5}\xspace}
\newcommand{\NCIaxpercThree}{\ensuremath{69.4}\xspace}
\newcommand{\NCIaxN}{\ensuremath{4}\xspace}

\newcommand{\numIax}{\ensuremath{4}\xspace}
\newcommand{\HeSCpercOne}{\ensuremath{20.5}\xspace}
\newcommand{\HeSCpercThree}{\ensuremath{69.4}\xspace}
\newcommand{\HeSCN}{\ensuremath{4}\xspace}
\newcommand{\NCSCpercOne}{\ensuremath{20.5}\xspace}
\newcommand{\NCSCpercThree}{\ensuremath{69.4}\xspace}
\newcommand{\NCSCN}{\ensuremath{4}\xspace}

\newcommand{\numSC}{\ensuremath{4}\xspace}
\newcommand{\HenormTpercOne}{\ensuremath{1.5}\xspace}
\newcommand{\HenormTpercThree}{\ensuremath{7.3}\xspace}
\newcommand{\HenormTN}{\ensuremath{77}\xspace}
\newcommand{\NCnormTpercOne}{\ensuremath{1.2}\xspace}
\newcommand{\NCnormTpercThree}{\ensuremath{6.2}\xspace}
\newcommand{\NCnormTN}{\ensuremath{91}\xspace}

\newcommand{\numnormT}{\ensuremath{95}\xspace}
\newcommand{\HenormbgpercOne}{\ensuremath{1.4}\xspace}
\newcommand{\HenormbgpercThree}{\ensuremath{7.0}\xspace}
\newcommand{\HenormbgN}{\ensuremath{80}\xspace}
\newcommand{\NCnormbgpercOne}{\ensuremath{1.2}\xspace}
\newcommand{\NCnormbgpercThree}{\ensuremath{6.0}\xspace}
\newcommand{\NCnormbgN}{\ensuremath{94}\xspace}

\newcommand{\numnormbg}{\ensuremath{98}\xspace}
\newcommand{\HenormTbgpercOne}{\ensuremath{1.3}\xspace}
\newcommand{\HenormTbgpercThree}{\ensuremath{6.7}\xspace}
\newcommand{\HenormTbgN}{\ensuremath{84}\xspace}
\newcommand{\NCnormTbgpercOne}{\ensuremath{1.1}\xspace}
\newcommand{\NCnormTbgpercThree}{\ensuremath{5.7}\xspace}
\newcommand{\NCnormTbgN}{\ensuremath{99}\xspace}

\newcommand{\numnormTbg}{\ensuremath{103}\xspace}
\newcommand{\HenormTbgSCpercOne}{\ensuremath{1.3}\xspace}
\newcommand{\HenormTbgSCpercThree}{\ensuremath{6.4}\xspace}
\newcommand{\HenormTbgSCN}{\ensuremath{88}\xspace}
\newcommand{\NCnormTbgSCpercOne}{\ensuremath{1.1}\xspace}
\newcommand{\NCnormTbgSCpercThree}{\ensuremath{5.5}\xspace}
\newcommand{\NCnormTbgSCN}{\ensuremath{103}\xspace}
\newcommand{\OldnormTbgSCpercOne}{\ensuremath{1.9}\xspace}
\newcommand{\OldnormTbgSCpercThree}{\ensuremath{9.2}\xspace}
\newcommand{\OldnormTbgSCN}{\ensuremath{60}\xspace}
\newcommand{\numnormTbgSC}{\ensuremath{107}\xspace}
\newcommand{\countbg}{\ensuremath{8}\xspace}

\newcommand{\countIax}{\ensuremath{4}\xspace}
\newcommand{\countSC}{\ensuremath{4}\xspace}
\newcommand{\countT}{\ensuremath{5}\xspace}

\newcommand{\Nsn}{\ensuremath{111}\xspace}

%% file: InputFiles/InstrCounts.tex
\newcommand{\HFOSCcount}{2\xspace}
\newcommand{\GMOScount}{13\xspace}
\newcommand{\MOSCAcount}{2\xspace}
\newcommand{\WFCCDcount}{4\xspace}
\newcommand{\BCcount}{2\xspace}
\newcommand{\IMACScount}{3\xspace}
\newcommand{\KASTcount}{11\xspace}

\newcommand{\EFOSCcount}{11\xspace}
\newcommand{\DPBCcount}{1\xspace}
\newcommand{\DOLOREScount}{2\xspace}
\newcommand{\LDSScount}{5\xspace}

\newcommand{\WHTFOScount}{6\xspace}
\newcommand{\NewSpeccount}{14\xspace}
\newcommand{\MUSEcount}{8\xspace}
\newcommand{\LRIScount}{27\xspace}
\newcommand{\FORScount}{43\xspace}
\newcommand{\HSTFOScount}{1\xspace}
\newcommand{\XSHcount}{16\xspace}
\newcommand{\INTFOScount}{2\xspace}
\newcommand{\ESIcount}{3\xspace}
\newcommand{\BCScount}{6\xspace}
\newcommand{\EMMIcount}{1\xspace}
\newcommand{\SOFIcount}{1\xspace}
\newcommand{\MODScount}{10\xspace}
\newcommand{\ISIScount}{5\xspace}
\newcommand{\DFOSCcount}{1\xspace}
\newcommand{\DBSPcount}{4\xspace}
\newcommand{\CAFOScount}{1\xspace}
\newcommand{\GoodSpeccount}{227\xspace}
\newcommand{\OSIRIScount}{1\xspace}
\newcommand{\FASTcount}{7\xspace}
\newcommand{\DEIMOScount}{9\xspace}
\newcommand{\CISCOcount}{3\xspace}
\newcommand{\MAGEcount}{2\xspace}
\newcommand{\FOCAScount}{2\xspace}

\newcommand{\RSScount}{3\xspace}
\newcommand{\ACAMcount}{1\xspace}
\newcommand{\WIFEScount}{7\xspace}

\newcommand{\UniqueSNe}{10\xspace}
\newcommand{\NewSNe}{13\xspace}

%% file: InstrumentTable.tex
\begin{tabular}{p{4cm}lp{5cm}llclc}
Telescope & Abbrev.$^a$ & Instrument & Abbrev.$^a$  & Ref. & $N_{\rm{spec}}$ \\
\hline
\hline
Australian National University 2.3m & ANU2.3m & Wide-Field Spectrograph & WiFeS & \citet{WiFeSref1, WiFeSref2} & \WIFEScount \\
Calar Alto 2.2m & CA2.2m & Calar Alto Faint Object Spectrograph & CAFOS & $\ldots^b$ & \CAFOScount \\

Calar Alto 3.5m & CA3.5m & Multi-Object Spectrograph at Calar Alto & MOSCA & $\ldots^c$ & \MOSCAcount \\

Danish 1.54m & D1.54m & Danish Faint Object Spectrograph and Camera & DFOSC & \citet{DFOSCref} & \DFOSCcount \\

du Pont Telescope & duPont & Wide Field Reimaging CCD Camera & WFCCD & $\ldots^d$ & \WFCCDcount \\
 & & Boller and Chivens Spectrograph & BC & $\ldots^e$ & \DPBCcount \\

ESO 1.5m & ESO1.5m & Boller and Chivens Spectrograph & BC & $\ldots^f$ & \BCcount \\

ESO 3.6m & ESO3.6m & ESO Faint Object Spectrograph and Camera & EFOSC1/2 & \citet{EFOSCref1} & \EFOSCcount \\

Harlan J Smith Telescope & HJST & UltraViolet Image Tube Spectrograph & UVITS & \citet{UVITSref} & 1 \\

Himalayan Chandra Telescope & HCT & Himalayan Faint Object Spectrograph & HFOSC & $\ldots^g$ & \HFOSCcount \\

Hubble Space Telescope & HST & Faint Object Spectrograph & FOS & $\ldots^h$ & \HSTFOScount \\ 

Isaac Newton Telescope & INT & Faint Object Spectrograph (1st Gen.) & FOS1 & \citet{INGFOSref} & \INTFOScount \\

Gemini North/South & GN/S & Gemini Multi-Object Spectrograph & GMOS & \citet{GMOSref} & \GMOScount \\


Gran Telescopio Canarias & GTC & Optical System for Imaging and low-Intermediate-Resolution Integrated Spectroscopy & OSIRIS & \citet{OSIRISref} & \OSIRIScount \\

Keck I & KeckI & Low Resolution Imaging Spectrograph & LRIS & \citet{LRISref} & \LRIScount \\

Keck II & KeckII & DEep Imaging Multi-Object Spectrograph & DEIMOS & \citet{DEIMOSref} & \DEIMOScount \\
    & & Echelette Imager and Spectrograph & ESI & \citet{ESIref} & \ESIcount \\

Large Binocular Telescope & LBT & Multi-Object Double Spectrograph & MODS1/2 & \citet{MODSref} & \MODScount \\

Magellan Baade Telescope & Baade & Inamori-Magellan Areal Camera and Spectrograph & IMACS & \citet{IMACSref} & \IMACScount \\
    & & Magellan Echellette Spectrograph & MagE & \citet{MagEref} & \MAGEcount \\

Magellan Clay Telescope & Clay & Low Dispersion Survey Spectrograph & LDSS & $\ldots^i$ & \LDSScount \\

Multiple Mirror Telescope & MMT & Blue Channel Spectrograph & BCS & \citet{BCSref} & \BCScount \\

New Technology Telescope & NTT & ESO Multi-Mode Instrument & EMMI & \citet{EMMIref} & \EMMIcount \\
 & & SOFI & $\ldots$ & \citet{SOFIref} & \SOFIcount \\

Palomar 200-inch & P200 & Double Spectrograph & DBSP & \citet{DBSPref} & \DBSPcount \\

Shane 3m Telescope & Shane3m & Kast Spectrograph & KAST & \citet{silverman13} & \KASTcount \\

Southern African Large Telescope & SALT & Robert Stobie Spectrograph & RSS & \citet{SALTref} & \RSScount\\

Subaru & Sub & OH-Airglow Suppressor/Cooled Infrared Spectrograph and Camera for OHS & CISCO & \citet{CISCOref} & \CISCOcount\\
                & & Faint Object Spectrograph and Camera & FOCAS & \citet{FOCASref} & \FOCAScount \\
    
Tillinghast 1.5m & Till & FAst Spectrograph for the Tillinghast telescope & FAST & \citet{FASTref} & \FASTcount \\

Telescopio Nazionale Galileo & TNG & Device Optimized for LOw RESolution & DOLORES & \citet{DOLORESref} & \DOLOREScount \\

Very Large Telescope & VLT & FOcal Reducer and low dispersion Spectrograph & FORS1/2 & \citet{FORS1ref} & \FORScount \\
    & & Multi-Unit Spectroscopic Explorer & MUSE & \citet{MUSEref} & \MUSEcount \\
    & & XSHOOTER & XSH & \citet{XSHOOTERref} & \XSHcount \\

William Herschel Telescope & WHT & Intermediate dispersion Spectrograph and Imaging System & ISIS & \citet{ISISref} & \ISIScount \\
        & & ACAM & $\ldots$ & \citet{ACAMref} & \ACAMcount \\
        & & Faint Object Spectrograph (2nd Gen.) & FOS2 & \citet{INGFOSref} & \WHTFOScount\\
        
\\
Total & 29 & & 40 & & \GoodSpeccount \\

\hline
\end{tabular}
\begin{tablenotes}
\scriptsize
\item[a]Abbreviations used in Table \ref{tab:spectra-params}.
\item[b]\url{http://w3.caha.es/CAHA/Instruments/CAFOS/cafos_overview.html} 
\item[c]\url{http://www.caha.es/CAHA/Instruments/MOSCA/index.html} 
\item[d]\url{http://www.lco.cl/telescopes-information/lco/telescopes-information/irenee-du-pont/instruments/website/wfccd/wfccd-manuals}
\item[e]\url{http://www.lco.cl/telescopes-information/irenee-du-pont/instruments/website/boller-chivens-spectrograph-manuals/user-manual/the-boller-and-chivens-spectrograph}
\item[f]\url{http://www.ls.eso.org/lasilla/Telescopes/2p2/E1p5M/BC/BC.html} 
\item[g]\url{https://www.iiap.res.in/iao/hfosc.html}
\item[h]\url{http://stecf-poa.stsci.edu/poa/FOS/fos_doc_access.html} 
\item[i]\url{http://www.lco.cl/telescopes-information/magellan/instruments/ldss-3} 
\end{tablenotes}

%% file: ModelInfoTable.tex
\begin{table}
\caption{Line luminosities for both the hydrogen-rich (H-rich) model and the helium-rich (He-rich) model corresponding to the MS38 and helium models from \citet{botyanszki18}. Helium lines are given letter designations to ease identification in Table \ref{tab:results}. FWHM refers to the expected FWHM of a line profile broadened by $\sim 10^3 ~\rm{km}\,\rm s^{-1}$.}
\label{tab:ModelInfo}
\begin{tabular}{cccc}
Line & $\lambda$ & $L_{200}$[$10^{38}$ erg/s] & FWHM [\AA] \\

\hline
H-rich Model \\
\hline
H$\gamma$ & 4341\AA & 0.271 & 14.5\\
H$\beta$ & 4831\AA & 4.38 & 16.1\\
He\texttt{I}-a & 5875\AA & 4.27 & 19.6\\
H$\alpha$ & 6563\AA & 68.0 & 21.9\\
He\texttt{I}-b & 6678\AA& 2.24 & 22.3\\
He\texttt{I}-c & 1.08$\mu\rm m$& 10.5 & 36.0\\
Pa$\beta$ & 1.281$\mu\rm m$ & 14.6 & 42.7\\
Pa$\alpha$ & 1.875$\mu \rm m$ & 14.6 & 62.5\\
He\texttt{I}-d & 2.06$\mu\rm m$& 8.48 & 68.7\\
\hline
He-rich Model \\
\hline
He\texttt{I}-a & 5875\AA& 8.26 & 19.6\\
He\texttt{I}-b & 6678\AA& 6.90 & 22.3\\
He\texttt{I}-c & 1.08$\mu\rm m$& 18.2 & 36.0\\
He\texttt{I}-d & 2.06$\mu\rm m$ & 12.9 & 68.7\\
\hline
\end{tabular}
\end{table}

%% file: StatsTable.tex
\begin{table*}
\caption{
Statistics for each sample considered in our study (see \S\ref{sec:results}). $N$ is the number of \sne with $M_{st} < M_{\rm{cut}}$ and $f_{n\sigma}$ is the $n\sigma$ fractional upper limit on their occurrence. $N_{tot}$ refers to the total number of \sne in that sample.
}
\label{tab:compstats}
\begin{tabular}{lrrrrrrr}
& & \multicolumn{3}{c}{H-rich} & \multicolumn{3}{c}{He-rich} \\ 

 & & \multicolumn{3}{c}{\scriptsize{($M_{st,H} < 0.15~M_\odot$)}} & \multicolumn{3}{c}{\scriptsize{($M_{st,He} < 0.023~M_\odot$)}} \\ \cmidrule(lr){3-5} \cmidrule(lr){6-8} 
Sample & $N_{tot}$ & $N$ & $f_{1\sigma}$ &$f_{3\sigma}$ & $N$ & $f_{1\sigma}$ & $f_{3\sigma}$ \\

\hline
\textit{Normal} & \numnorm & \NCnormN & $< \NCnormpercOne\%$ & $< \NCnormpercThree\%$ & \HenormN & $<\HenormpercOne\%$ & $< \HenormpercThree\%$\\
\textit{91T-like} & \numT & \NCTN & $<\NCTpercOne\%$ & $< \NCTpercThree\%$ & \HeTN & $<\HeTpercOne\%$ & $<\HeTpercThree\%$\\
\textit{91bg-like} & \numbg & \NCbgN & $<\NCbgpercOne\%$ & $<\NCbgpercThree\%$ & \HebgN & $<\HebgpercOne\%$ & $<\HebgpercThree\%$ \\
\textit{SC} & \numSC & \NCSCN & $<\NCSCpercOne\%$ & $< \NCSCpercThree\%$ & \HeSCN & $< \HeSCpercOne\%$ & $<\HeSCpercThree\%$ \\
\textit{Iax} & \numIax & \NCIaxN & $< \NCIaxpercOne\%$ & $< \NCIaxpercThree\%$  & \HeIaxN & $<\HeIaxpercOne\%$ & $< \HeIaxpercThree\%$  \\
\textit{Normal+91T} & \numnormT & \NCnormTN & $< \NCnormTpercOne\%$ & $< \NCnormTpercThree\%$ & \HenormTN & $< \HenormTpercOne\%$ & $<\HenormTpercThree\%$ \\
\textit{Normal+91bg} & \numnormbg & \NCnormbgN & $< \NCnormbgpercOne\%$ & $< \NCnormbgpercThree\%$ & \HenormbgN & $< \HenormbgpercOne\%$ & $< \HenormbgpercThree\%$ \\ 
\textit{Normal+91T+91bg} & \numnormTbg & \NCnormTbgN & $<\NCnormTbgpercOne\%$ & $< \NCnormTbgpercThree\%$ & \HenormTbgN & $< \HenormTbgpercOne\%$ & $<\HenormTbgpercThree\%$ \\
\textbf{\textit{Normal+91T+91bg+SC}} & \numnormTbgSC & \NCnormTbgSCN & $<\NCnormTbgSCpercOne\%$ & $< \NCnormTbgSCpercThree\%$ & \HenormTbgSCN & $<\HenormTbgSCpercOne\%$ & $< \HenormTbgSCpercThree\%$ \\
\textit{All} & \numall & \NCallN & $<\NCallpercOne\%$ & $< \NCallpercThree\%$& \HeallN & $<\HeallpercOne\%$ & $<\HeallpercThree\%$ \\

\hline
\end{tabular}
\end{table*}

%% file: InputFiles/late-time-data.txt
$B$ & $-7.66_{-0.62}^{+0.53}$ & $1.443_{-0.005}^{+0.005}$ & $0.48_{-0.04}^{+0.04}$ & $6.168_{-0.004}^{+0.003}$ & 42 & 346 & 0.00 & 0.08 \\[5pt]
$V$ & $-13.68_{-0.37}^{+0.33}$ & $1.620_{-0.003}^{+0.003}$ & $2.58_{-0.02}^{+0.02}$ & $6.050_{-0.002}^{+0.002}$ & 67 & 438 & 0.03 & 0.13 \\[5pt]
$R$ & $4.22_{-0.33}^{+0.41}$ & $1.650_{-0.003}^{+0.004}$ & $1.00_{-0.03}^{+0.03}$ & $6.776_{-0.002}^{+0.002}$ & 34 & 286 & 0.00 & 0.09 \\[5pt]

%% file: InputFiles/basictable-0.tex
\begin{tabular}{lcrcccc}
Disc. Name & IAU Name & Pec? & z & Redshift Ref. & Discovery & Classification \\\hline
$\ldots$ & ASASSN-14hr & N & $0.03362$ & \citet{jones09} & \citet{ATel6500}& \citet{ATel6508} \\
$\ldots$ & ASASSN-14jc & N & $0.01132$ & \citet{jones09} & \citet{ATel6594}& \citet{ATel6618} \\
$\ldots$ & ASASSN-14jg & N & $0.01483$ & \citet{jones09} & \citet{ATel6637}& \citet{ATel6661} \\
$\ldots$ & ASASSN-14jz & N & $0.01550$ & This Work & \citet{ATel6683}& \citet{ATel6749} \\
$\ldots$ & ASASSN-14kq & N & $0.03360$ & \citet{jones09} & \citet{ATel6737}& \citet{ATel6765} \\
$\ldots$ & ASASSN-14lt & N & $0.03205$ & \citet{springob14} & \citet{ATel6802}& \citet{ATel6813} \\
$\ldots$ & ASASSN-14lu & N & $0.02700$ & \citet{collobert06} & \citet{ATel6803}& \citet{ATel6812} \\
$\ldots$ & ASASSN-14lv & N & $0.04919$ & \citet{jones06} & \citet{ATel6809}& \citet{ATel6882} \\
$\ldots$ & ASASSN-14me & N & $0.01800$ & \citet{ATel6882} & \citet{ATel6830}& \citet{ATel6882} \\
$\ldots$ & ASASSN-15be & N & $0.02190$ & \citet{colless03} & \citet{ATel6950}& \citet{ATel6988} \\
$\ldots$ & ASASSN-15hx & N & $0.00810$ & This Work & \citet{ATel7447}& \citet{ATel7452} \\
$\ldots$ & PSN~J1149$^a$ & N & $0.00560$ & \citet{meyer04} & \citet{ATel7732}& \citet{ATel7825} \\
$\ldots$ & SN1972E & N & $0.00136$ & \citet{koribalski04} & \citet{IAUC2405}& \citet{IAUC2407} \\
$\ldots$ & SN1981B & N & $0.00603$ & \citet{grogin98} & \citet{IAUC3580}& \citet{IAUC3584} \\
$\ldots$ & SN1986G & 91bg-like & $0.00182$ & \citet{graham78} & \citet{IAUC4208}& \citet{IAUC4210} \\
$\ldots$ & SN1990N & N & $0.00340$ & \citet{meyer04} & \citet{IAUC5039}& \citet{IAUC5039} \\
$\ldots$ & SN1991T & 91T-like & $0.00579$ & \citet{strauss92} & \citet{IAUC5239}& \citet{IAUC5239} \\
$\ldots$ & SN1991bg & 91bg-like & $0.00339$ & \citet{cappellari11} & \citet{IAUC5400}& \citet{IAUC5403} \\
$\ldots$ & SN1992A & N & $0.00626$ & \citet{donofrio95} & \citet{IAUC5428}& \citet{IAUC5428} \\
$\ldots$ & SN1993Z & N & $0.00450$ & \citet{epinat08} & \citet{IAUC5870}& \citet{IAUC5870} \\
$\ldots$ & SN1994ae & N & $0.00427$ & \citet{krumm80} & \citet{IAUC6105}& \citet{IAUC6108} \\
$\ldots$ & SN1995D & N & $0.00656$ & \citet{cappellari11} & \citet{IAUC6134}& \citet{IAUC6135} \\
$\ldots$ & SN1996X & N & $0.00694$ & \citet{ogando08} & \citet{IAUC6380}& \citet{IAUC6381} \\
$\ldots$ & SN1998aq & N & $0.00370$ & \citet{devaucouleurs91} & \citet{IAUC6875}& \citet{IAUC6878} \\
$\ldots$ & SN1998bu & N & $0.00299$ & \citet{devaucouleurs91} & \citet{IAUC6899}& \citet{IAUC6905} \\
$\ldots$ & SN1999aa & 91T-like & $0.01444$ & \citet{devaucouleurs91} & \citet{IAUC7108}& \citet{IAUC7108-1} \\
$\ldots$ & SN1999by & 91bg-like & $0.00213$ & \citet{devaucouleurs91} & \citet{IAUC7156}& \citet{IAUC7158} \\
$\ldots$ & SN2000cx & 91T-like & $0.00802$ & \citet{cappellari11} & \citet{IAUC7458}& \citet{IAUC7463} \\
$\ldots$ & SN2001el & N & $0.00389$ & \citet{koribalski04} & \citet{IAUC7720}& \citet{IAUC7723} \\
$\ldots$ & SN2002bo & N & $0.00424$ & \citet{theureau98} & \citet{IAUC7847}& \citet{IAUC7848} \\
$\ldots$ & SN2002cx & Iax & $0.02396$ & \citet{falco99} & \citet{IAUC7902}& \citet{IAUC7903} \\
$\ldots$ & SN2002dj & N & $0.00939$ & \citet{rothberg06} & \citet{IAUC7918}& \citet{IAUC7919} \\
$\ldots$ & SN2002er & N & $0.00857$ & \citet{devaucouleurs91} & \citet{IAUC7959}& \citet{IAUC7961} \\
$\ldots$ & SN2003cg & N & $0.00413$ & \citet{vandriel01} & \citet{IAUC8097}& \citet{IAUC8099} \\
$\ldots$ & SN2003du & N & $0.00638$ & \citet{schneider92} & \citet{IAUC8121}& \citet{IAUC8122} \\
$\ldots$ & SN2003gs & 91bg-like & $0.00477$ & \citet{smith00} & \citet{IAUC8171}& \citet{IAUC8171} \\
$\ldots$ & SN2003hv & N & $0.00562$ & \citet{ogando08} & \citet{IAUC8197}& \citet{IAUC8198} \\
$\ldots$ & SN2003kf & N & $0.00739$ & \citet{theureau98} & \citet{IAUC8245}& \citet{IAUC8245} \\
$\ldots$ & SN2004S & N & $0.00936$ & \citet{theureau07} & \citet{IAUC8282}& \citet{IAUC8283} \\
$\ldots$ & SN2004eo & N & $0.01570$ & \citet{theureau98} & \citet{IAUC8406}& \citet{IAUC8409} \\
$\ldots$ & SN2005W & N & $0.00889$ & \citet{devaucouleurs91} & \citet{IAUC8475}& \citet{IAUC8479} \\
$\ldots$ & SN2005am & N & $0.00790$ & \citet{theureau98} & \citet{IAUC8490}& \citet{IAUC8491} \\
$\ldots$ & SN2005cf & N & $0.00646$ & \citet{devaucouleurs91} & \citet{CBET158}& \citet{CBET160} \\
$\ldots$ & SN2005hk & Iax & $0.01299$ & \citet{SDSS-DR14} & \citet{IAUC8625}& \citet{CBET269} \\
$\ldots$ & SN2006X & N & $0.00524$ & \citet{rand95} & \citet{IAUC8667}& \citet{CBET393} \\
$\ldots$ & SN2006dd & N & $0.00587$ & \citet{longhetti98} & \citet{IAUC8723}& \citet{CBET557} \\
$\ldots$ & SN2006gz & SC & $0.02800$ & \citet{falco99} & \citet{IAUC8754}& \citet{CBET651} \\
$\ldots$ & SN2007af & N & $0.00546$ & \citet{koribalski04} & \citet{CBET863}& \citet{CBET865} \\
$\ldots$ & SN2007gi & N & $0.00462$ & \citet{cappellari11} & \citet{CBET1017}& \citet{CBET1021} \\
$\ldots$ & SN2007if & SC & $0.07416$ & \citet{scalzo10} & \citet{CBET1059}& \citet{CBET1059} \\
$\ldots$ & SN2007le & N & $0.00672$ & \citet{koribalski04} & \citet{CBET1100}& \citet{CBET1101} \\
$\ldots$ & SN2007on & N & $0.00649$ & \citet{graham98} & \citet{CBET1121}& \citet{CBET1131} \\
$\ldots$ & SN2008A & Iax & $0.01643$ & \citet{theureau98} & \citet{CBET1193}& \citet{CBET1198} \\
$\ldots$ & SN2008Q & N & $0.00802$ & \citet{cappellari11} & \citet{CBET1228}& \citet{CBET1232} \\
$\ldots$ & SN2009dc & SC & $0.02139$ & \citet{falco99} & \citet{CBET1762}& \citet{CBET1768} \\
$\ldots$ & SN2009ig & N & $0.00877$ & \citet{meyer04} & \citet{CBET1918}& \citet{CBET1918} \\
$\ldots$ & SN2009le & N & $0.01779$ & \citet{theureau98} & \citet{CBET2022}& \citet{CBET2025} \\
$\ldots$ & SN2010ev & N & $0.00921$ & \citet{meyer04} & \citet{CBET2344}& \citet{CBET2346} \\
$\ldots$ & SN2010gp & N & $0.02450$ & \citet{downes93} & \citet{CBET2388}& \citet{CBET2390} \\
$\ldots$ & SN2010hg & N & $0.00822$ & \citet{meyer04} & \citet{CBET2434}& \citet{CBET2453} \\
$\ldots$ & SN2010lp & 91bg-like & $0.01015$ & \citet{huchra99} & \citet{CBET2612}& \citet{CBET2613} \\
\hline
\end{tabular}

%% file: InputFiles/basictable-1.tex
\begin{tabular}{lcrcccc}
Disc. Name & IAU Name & Pec? & z & Redshift Ref. & Discovery & Classification \\\hline
$\ldots$ & SN2011K & N & $0.01450$ & \citet{CBET2635} & \citet{CBET2636}& \citet{CBET2636} \\
SNhunt37 & SN2011ae & N & $0.00605$ & \citet{meyer04} & \citet{CBET2658}& \citet{CBET2658} \\
$\ldots$ & SN2011at & N & $0.00676$ & \citet{theureau98} & \citet{CBET2676}& \citet{CBET2676} \\
$\ldots$ & SN2011by & N & $0.00284$ & \citet{verheijen01} & \citet{CBET2703}& \citet{CBET2708} \\
$\ldots$ & SN2011ek & N & $0.00503$ & \citet{rhee96} & \citet{CBET2783}& \citet{CBET2783} \\
PTF11kly & SN2011fe & N & $0.00080$ & \citet{maguire14} & \citet{ATel3581}& \citet{ATel3583} \\
$\ldots$ & SN2011im & N & $0.01623$ & \citet{catinella05} & \citet{CBET2928}& \citet{CBET2928} \\
$\ldots$ & SN2011iv & N & $0.00649$ & \citet{graham98} & \citet{CBET2940}& \citet{CBET2940} \\
$\ldots$ & SN2011iy & N & $0.00427$ & \citet{corsini03} & \citet{CBET2943}& \citet{CBET2943} \\
$\ldots$ & SN2012Z & Iax & $0.00712$ & \citet{koribalski04} & \citet{CBET3014}& \citet{CBET3014} \\
$\ldots$ & SN2012cg & N & $0.00146$ & \citet{kent08} & \citet{CBET3111}& \citet{CBET3111} \\
SNhunt136 & SN2012cu & N & $0.00347$ & \citet{devaucouleurs91} & \citet{CBET3146}& \citet{CBET3146} \\
$\ldots$ & SN2012ei & N & $0.00672$ & \citet{galbany14} & \citet{CBET3209}& \citet{CBET3209} \\
$\ldots$ & SN2012fr & N & $0.00546$ & \citet{bureau96} & \citet{CBET3277}& \citet{CBET3277} \\
$\ldots$ & SN2012hr & N & $0.00756$ & \citet{tully08} & \citet{CBET3346}& \citet{CBET3346} \\
$\ldots$ & SN2012ht & N & $0.00356$ & \citet{guthrie96} & \citet{CBET3349}& \citet{CBET3349} \\
$\ldots$ & SN2013aa & N & $0.00400$ & \citet{huchra99} & \citet{CBET3416}& \citet{CBET3416} \\
SNhunt196 & SN2013cs & N & $0.00924$ & \citet{pisano11} & \citet{CBET3533}& \citet{CBET3533} \\
$\ldots$ & SN2013ct & N & $0.00384$ & \citet{smoker00} & \citet{CBET3539}& \citet{CBET3539} \\
$\ldots$ & SN2013dy & N & $0.00389$ & \citet{pan15} & \citet{CBET3588}& \citet{CBET3588} \\
$\ldots$ & SN2013gy & N & $0.01402$ & \citet{catinella05} & \citet{CBET3743}& \citet{CBET3743} \\
$\ldots$ & SN2014J & N & $0.00068$ & \citet{devaucouleurs91} & \citet{CBET3792}& \citet{CBET3792-1} \\
$\ldots$ & SN2014bv & N & $0.00559$ & \citet{devaucouleurs91} & \citet{CBET3911}& \citet{CBET3911} \\
$\ldots$ & SN2015F & N & $0.00489$ & \citet{meyer04} & \citet{CBET4081}& \citet{CBET4081} \\
$\ldots$ & SN2015I & N & $0.00759$ & \citet{giovanelli97} & \citet{CBET4106}& \citet{ATel7476} \\
$\ldots$ & SN2016brx & 91bg-like & $0.01017$ & \citet{jones09} & \citet{2016TNSTR304}& \citet{ATel9170} \\
$\ldots$ & SN2016bry & N & $0.01602$ & \citet{rhee96} & \citet{2016TNSTR305}& \citet{ATel9018} \\
ASASSN-16eq & SN2016bsa & N & $0.01431$ & \citet{paturel03} & \citet{ATel8979}& \citet{ATel8992} \\
Gaia16avm & SN2016ehy & N & $0.04500$ & \citet{ATel9309} & \citet{TNSTR485}& \citet{ATel9309} \\
ATLAS16cpu & SN2016ffh & N & $0.01820$ & \citet{SDSS-DR3} & \citet{TNSTR583}& \citet{ATel9403} \\
$\ldots$ & SN2016gxp & 91T-like & $0.01785$ & \citet{huchra99} & \citet{2016TNSTR761}& \citet{2016TNSCR793} \\
ASASSN-17cs & SN2017azw & N & $0.02000$ & \citet{ATel10131} & \citet{ATel10108}& \citet{ATel10131} \\
DLT17u & SN2017cbv & N & $0.00399$ & \citet{koribalski04} & \citet{ATel10158}& \citet{ATel10164} \\
ATLAS17dfo & SN2017ckq & N & $0.00989$ & \citet{mathewson92} & \citet{2017TNSTR361}& \citet{ATel10225} \\
DLT17ar & SN2017cyy & N & $0.00978$ & \citet{meyer04} & \citet{ATel10260}& \citet{ATel10261} \\
DLT17bk & SN2017ejb & 91bg-like & $0.00987$ & \citet{devaucouleurs91} & \citet{ATel10439}& \citet{TCRN1020} \\
ASASSN-18hz & SN2017evn & N & $0.01716$ & \citet{adelman08} & \citet{ATel10521}& \citet{ATel10518} \\
$\ldots$ & SN2017ezd & N & $0.01808$ & \citet{jones09} & \citet{ATRN12243}& \citet{ATel10605} \\
DLT17bx & SN2017fgc & N & $0.00772$ & \citet{cappellari11} & \citet{ATel10569}& \citet{ATel10569} \\
DLT17cd & SN2017fzw & 91bg-like & $0.00540$ & \citet{devaucouleurs91} & \citet{ATel10629}& \citet{ATel10639} \\
ATLAS17jiv & SN2017gah & N & $0.00891$ & \citet{lauberts89} & \citet{TNSTR860}& \citet{ATel10639} \\
$\ldots$ & SN2017glq & N & $0.01176$ & \citet{woods06} & \citet{2017TNSTR961}& \citet{2017TNSCR978} \\
ATLAS17nmh & SN2017isq & N & $0.00939$ & \citet{ATel11036} & \citet{TNSTR1371}& \citet{ATel11036} \\
ATLAS17nse & SN2017iyb & N & $0.01011$ & \citet{meyer04} & \citet{2017TNSTR15357}& \citet{ATel11092} \\
ASASSN-18hb & SN2018aqi & N & $0.01251$ & \citet{theureau98} & \citet{ATel11516}& \citet{ATel11521} \\
ASASSN-18bt & SN2018oh & N & $0.01098$ & \citet{schneider90} & \citet{ATel11253}& \citet{TNSCR159} \\
ASASSN-18da & SN2018vw & 91T-like & $0.02000$ & \citet{ATel11346} & \citet{2018TNSTR16974}& \citet{ATel11343} \\
DLT18h & SN2018xx & N & $0.00999$ & \citet{smith00} & \citet{ATel11328}& \citet{ATel11330} \\
DLT18i & SN2018yu & N & $0.00811$ & \citet{devaucouleurs91} & \citet{ATel11371}& \citet{2018TNSCR1755} \\
$\ldots$ & SNF-012$^b$ & SC & $0.07454$ & \citet{taubenberger13} & $\ldots$& $\ldots$ \\
\hline
\end{tabular}

%% file: InputFiles/targettable-0.tex
\begin{tabular}{lcccrrcc}
SN &$t_{\rm{max}}^{a}$ & $\Delta m_{15}(B)$ &$\mu$ & E(B-V)& $N_{\rm{spec}}$& Phase \\
& (MJD)& (mag)&(mag) & &  & (days) \\\hline
SN1972E & $41445.9\pm0.4$ & $0.93\pm0.06$ & $27.75\pm0.06^{1}$ & $-0.03\pm0.06$ & 4 & $205-418$ \\ 
SN1981B & $44672.7\pm0.4$ & $1.12\pm0.06$ & $30.91\pm0.05^{2}$ & $0.06\pm0.06$ & 1 & 267 \\ 
SN1986G & $46561.0\pm0.4$ & $1.57\pm0.07$ & $27.82\pm0.06^{1}$ & $0.91\pm0.06$ & 4 & $256-325$ \\ 
SN1990N & $48082.5\pm0.3$ & $1.09\pm0.06$ & $31.53\pm0.07^{2}$ & $0.02\pm0.06$ & 5 & $186-333$ \\ 
SN1991T & $48374.5\pm0.3$ & $0.97\pm0.06$ & $30.67\pm0.09^{1}$ & $0.17\pm0.06$ & 6 & $258-552$ \\ 
SN1991bg & $48604.1\pm0.4$ & $1.76\pm0.06$ & $31.07\pm0.06^{3}$ & $0.07\pm0.07$ & 2 & $198-202$ \\ 
SN1992A & $48639.3\pm0.3$ & $1.27\pm0.06$ & $31.22\pm0.06^{3}$ & $0.04\pm0.06$ & 1 & 292 \\ 
SN1993Z & $49247.1\pm0.8$ & $0.87\pm0.06$ & $33.10\pm0.17$ & $0.00\pm0.06$ & 2 & $181-213$ \\ 
SN1994ae & $49685.3\pm0.3$ & $1.04\pm0.06$ & $32.07\pm0.05^{2}$ & $0.05\pm0.06$ & 2 & $219-369$ \\ 
SN1995D & $49767.4\pm0.4$ & $0.90\pm0.06$ & $32.67\pm0.18$ & $0.01\pm0.06$ & 2 & $278-286$ \\ 
SN1996X & $50189.8\pm0.3$ & $1.20\pm0.06$ & $32.29\pm0.25$ & $-0.02\pm0.06$ & 2 & $247-299$ \\ 
SN1998aq & $50930.8\pm0.3$ & $1.11\pm0.07$ & $31.74\pm0.07^{2}$ & $0.01\pm0.06$ & 3 & $210-240$ \\ 
SN1998bu & $50951.9\pm0.3$ & $1.05\pm0.06$ & $30.11\pm0.06^{4}$ & $0.41\pm0.06$ & 11 & $191-341$ \\ 
SN1999aa & $51232.5\pm0.3$ & $0.90\pm0.06$ & $34.19\pm0.23$ & $-0.01\pm0.06$ & 2 & $258-284$ \\ 
SN1999by & $51308.9\pm0.3$ & $1.76\pm0.06$ & $30.75\pm0.06^{5}$ & $0.05\pm0.06$ & 1 & 185 \\ 
SN2000cx & $51752.8\pm0.3$ & $1.27\pm0.06$ & $32.14\pm0.09^{6}$ & $0.02\pm0.06$ & 2 & $182-451$ \\ 
SN2001el & $52182.3\pm0.4$ & $1.12\pm0.06$ & $31.31\pm0.05^{2}$ & $0.22\pm0.06$ & 3 & $310-398$ \\ 
SN2002bo & $52356.8\pm0.3$ & $1.10\pm0.06$ & $31.73\pm0.09^{1}$ & $0.36\pm0.06$ & 2 & $227-311$ \\ 
SN2002cx & $52415.1\pm0.3$ & $1.13\pm0.06$ & $35.31\pm0.15^{7}$ & $0.07\pm0.06$ & 4 & $232-317$ \\ 
SN2002dj & $52451.3\pm0.4$ & $1.02\pm0.08$ & $33.23\pm0.16$ & $0.03\pm0.07$ & 2 & $220-273$ \\ 
SN2002er & $52525.3\pm0.3$ & $1.23\pm0.06$ & $33.15\pm0.17$ & $0.16\pm0.06$ & 1 & 214 \\ 
SN2003cg & $52729.2\pm0.3$ & $1.14\pm0.06$ & $31.83\pm0.10^{1}$ & $1.32\pm0.06$ & 1 & 385 \\ 
SN2003du & $52766.6\pm0.3$ & $1.02\pm0.06$ & $32.92\pm0.06^{2}$ & $0.00\pm0.06$ & 6 & $194-375$ \\ 
SN2003gs & $52842.2\pm0.4$ & $1.59\pm0.06$ & $32.13\pm0.20$ & $0.00\pm0.06$ & 1 & 207 \\ 
SN2003hv & $52891.5\pm0.3$ & $1.55\pm0.06$ & $31.51\pm0.20$ & $0.00\pm0.06$ & 2 & $319-393$ \\ 
SN2003kf & $52980.3\pm0.3$ & $1.03\pm0.06$ & $32.43\pm0.10^{1}$ & $-0.03\pm0.06$ & 1 & 401 \\ 
SN2004S & $53039.7\pm0.4$ & $1.06\pm0.06$ & $33.41\pm0.18$ & $0.00\pm0.06$ & 1 & 315 \\ 
SN2004eo & $53278.7\pm0.3$ & $1.31\pm0.06$ & $34.03\pm0.15$ & $0.01\pm0.06$ & 1 & 227 \\ 
SN2005W & $53412.6\pm0.3$ & $1.02\pm0.06$ & $33.01\pm0.20$ & $0.23\pm0.06$ & 1 & 213 \\ 
SN2005am & $53435.0\pm0.3$ & $1.30\pm0.06$ & $32.66\pm0.17$ & $0.03\pm0.06$ & 2 & $300-383$ \\ 
SN2005cf & $53534.3\pm0.3$ & $1.11\pm0.06$ & $32.26\pm0.10^{2}$ & $-0.02\pm0.06$ & 4 & $266-383$ \\ 
SN2005hk & $53684.8\pm0.3$ & $1.58\pm0.10$ & $33.91\pm0.15^{7}$ & $\ldots$ & 2 & $378-408$ \\ 
SN2006X & $53786.3\pm0.3$ & $1.08\pm0.06$ & $30.72\pm0.06^{1}$ & $1.38\pm0.06$ & 3 & $277-360$ \\ 
SN2006dd & $53918.6\pm0.3$ & $1.05\pm0.06$ & $31.52\pm0.13$ & $0.00\pm0.06$ & 2 & $188-195$ \\ 
SN2006gz & $54021.7\pm0.1$ & $0.88\pm0.06$ & $35.22\pm0.15^{7}$ & $0.20\pm0.06$ & 1 & 339 \\ 
SN2007af & $54174.0\pm0.3$ & $1.08\pm0.06$ & $31.79\pm0.05^{2}$ & $0.12\pm0.06$ & 1 & 302 \\ 
SN2007gi & $54327.9\pm0.3$ & $1.21\pm0.06$ & $32.17\pm0.12$ & $0.12\pm0.06$ & 1 & 223 \\ 
SN2007if & $54340.3\pm0.4$ & $0.88\pm0.06$ & $37.55\pm0.15^{7}$ & $-0.03\pm0.06$ & 2 & $393-421$ \\ 
SN2007le & $54399.0\pm0.3$ & $1.05\pm0.06$ & $32.44\pm0.10$ & $0.30\pm0.06$ & 1 & 307 \\ 
SN2007on & $54419.8\pm0.3$ & $1.81\pm0.06$ & $31.34\pm0.07^{3}$ & $\ldots$ & 3 & $286-381$ \\ 
SN2008A & $54478.0\pm0.4$ & $1.55\pm0.09$ & $34.26\pm0.15^{7}$ & $\ldots$ & 3 & $204-288$ \\ 
SN2008Q & $54505.6\pm0.3$ & $1.09\pm0.06$ & $32.66\pm0.18$ & $0.06\pm0.06$ & 1 & 200 \\ 
SNF-012 & $54682.1\pm1.1$ & $1.29\pm0.16$ & $37.66\pm0.15^{7}$ & $\ldots$ & 3 & $265-319$ \\ 
SN2009dc & $54946.4\pm0.4$ & $0.80\pm0.06$ & $35.07\pm0.15^{7}$ & $\ldots$ & 2 & $287-380$ \\ 
SN2009ig & $55080.1\pm0.3$ & $0.88\pm0.06$ & $32.50\pm0.08^{2}$ & $0.12\pm0.06$ & 1 & 405 \\ 
SN2009le & $55165.6\pm0.4$ & $0.91\pm0.06$ & $34.47\pm0.17$ & $0.20\pm0.06$ & 1 & 317 \\ 
SN2010ev & $55384.8\pm0.3$ & $1.12\pm0.06$ & $33.47\pm0.20$ & $0.19\pm0.06$ & 1 & 270 \\ 
SN2010gp & $55406.0\pm0.4$ & $1.10\pm0.06$ & $34.61\pm0.36$ & $0.44\pm0.08$ & 1 & 276 \\ 
SN2010hg & $55451.0\pm5.0$ & $\ldots$ & $32.82\pm0.50^{8}$ & $\ldots$ & 1 & 203 \\ 
SN2010lp & $55568.0\pm5.0$ & $\ldots$ & $33.04\pm0.40^{8}$ & $\ldots$ & 1 & 264 \\ 
SN2011K & $55577.9\pm0.3$ & $1.30\pm0.06$ & $33.95\pm0.09$ & $0.01\pm0.06$ & 1 & 344 \\ 
SNhunt37 & $55620.1\pm0.6$ & $1.03\pm0.07$ & $32.16\pm0.08$ & $0.05\pm0.13$ & 1 & 306 \\ 
SN2011at & $55625.0\pm0.4$ & $0.92\pm0.06$ & $32.63\pm0.16$ & $0.17\pm0.06$ & 1 & 359 \\ 
SN2011by & $55690.5\pm0.5$ & $1.11\pm0.08$ & $31.59\pm0.07^{2}$ & $0.09\pm0.07$ & 1 & 207 \\ 
SN2011ek & $55788.8\pm0.4$ & $1.00\pm0.07$ & $32.47\pm0.44$ & $0.59\pm0.06$ & 1 & 422 \\ 
PTF11kly & $55815.1\pm0.3$ & $1.18\pm0.06$ & $29.14\pm0.05^{2}$ & $0.04\pm0.06$ & 9 & $204-346$ \\ 
SN2011iy & $55893.2\pm0.4$ & $1.03\pm0.08$ & $31.33\pm0.17$ & $0.25\pm0.06$ & 1 & 205 \\ 
SN2011im & $55902.3\pm0.4$ & $1.06\pm0.07$ & $34.83\pm0.15$ & $0.11\pm0.06$ & 1 & 313 \\ 
SN2011iv & $55906.0\pm0.3$ & $1.63\pm0.05$ & $31.53\pm0.07^{9}$ & $0.01\pm0.06$ & 5 & $244-304$ \\ 
SN2012Z & $55967.7\pm0.1$ & $1.43\pm0.01$ & $32.52\pm0.06^{2}$ & $\ldots$ & 4 & $193-254$ \\ 
\hline
\end{tabular}

%% file: InputFiles/targettable-1.tex
\begin{tabular}{lcccrrcc}
SN &$t_{\rm{max}}^{a}$ & $\Delta m_{15}(B)$ &$\mu$ & E(B-V)& $N_{\rm{spec}}$& Phase \\
& (MJD)& (mag)&(mag) & &  & (days) \\\hline
SN2012cg & $56082.1\pm0.3$ & $0.98\pm0.06$ & $31.03\pm0.15$ & $0.20\pm0.06$ & 3 & $286-342$ \\ 
SNhunt136 & $56105.1\pm0.1$ & $\ldots$ & $31.11\pm0.15^{10}$ & $\ldots$ & 1 & 319 \\ 
SN2012ei & $56160.0\pm5.0$ & $\ldots$ & $32.01\pm0.46^{11}$ & $\ldots$ & 1 & 254 \\ 
SN2012fr & $56244.0\pm0.3$ & $0.90\pm0.06$ & $31.31\pm0.06^{2}$ & $-0.02\pm0.06$ & 8 & $222-415$ \\ 
SN2012hr & $56287.0\pm0.3$ & $1.07\pm0.06$ & $33.19\pm0.16$ & $0.00\pm0.06$ & 3 & $284-458$ \\ 
SN2012ht & $56296.0\pm0.3$ & $1.56\pm0.06$ & $31.91\pm0.04^{2}$ & $0.00\pm0.06$ & 1 & 422 \\ 
SN2013aa & $56343.2\pm0.4$ & $0.90\pm0.06$ & $30.60\pm0.22$ & $0.02\pm0.06$ & 6 & $189-426$ \\ 
SN2013ct & $56416.1\pm5.0$ & $\ldots$ & $30.27\pm0.20^{1}$ & $\ldots$ & 1 & 229 \\ 
SNhunt196 & $56437.2\pm0.3$ & $1.07\pm0.06$ & $32.78\pm0.17$ & $0.14\pm0.06$ & 3 & $263-304$ \\ 
SN2013dy & $56501.0\pm0.3$ & $0.94\pm0.06$ & $31.50\pm0.08^{2}$ & $0.10\pm0.06$ & 4 & $334-480$ \\ 
SN2013gy & $56648.9\pm0.3$ & $1.10\pm0.06$ & $33.75\pm0.15^{12}$ & $0.20\pm0.06$ & 4 & $235-424$ \\ 
SN2014J & $56690.3\pm0.3$ & $1.01\pm0.06$ & $27.74\pm0.08^{1}$ & $1.22\pm0.06$ & 4 & $212-350$ \\ 
SN2014bv & $56840.3\pm0.4$ & $1.79\pm0.06$ & $31.66\pm0.36$ & $0.46\pm0.06$ & 1 & 294 \\ 
ASASSN-14hr & $56937.4\pm1.7$ & $\ldots$ & $35.70\pm0.36^{7}$ & $\ldots$ & 1 & 468 \\ 
ASASSN-14jg & $56959.7\pm0.4$ & $0.89\pm0.06$ & $34.00\pm0.16$ & $0.03\pm0.06$ & 3 & $223-325$ \\ 
ASASSN-14jc & $56960.3\pm4.0$ & $\ldots$ & $33.44\pm0.15^{7}$ & $\ldots$ & 1 & 390 \\ 
ASASSN-14jz & $56979.0\pm2.6$ & $\ldots$ & $33.50\pm0.50^{7}$ & $\ldots$ & 1 & 204 \\ 
ASASSN-14kq & $56991.1\pm1.5$ & $\ldots$ & $35.69\pm0.15^{7}$ & $\ldots$ & 1 & 413 \\ 
ASASSN-14lv & $56995.1\pm1.7$ & $\ldots$ & $36.57\pm0.15^{7}$ & $\ldots$ & 1 & 398 \\ 
ASASSN-14lu & $57000.7\pm4.8$ & $\ldots$ & $34.46\pm0.46^{13}$ & $\ldots$ & 1 & 477 \\ 
ASASSN-14lt & $57011.1\pm2.1$ & $\ldots$ & $35.76\pm0.50^{14}$ & $\ldots$ & 1 & 389 \\ 
ASASSN-14me & $57023.1\pm0.5$ & $\ldots$ & $34.42\pm0.15^{7}$ & $\ldots$ & 2 & $304-362$ \\ 
ASASSN-15be & $57052.5\pm0.6$ & $\ldots$ & $34.97\pm0.15^{7}$ & $\ldots$ & 1 & 265 \\ 
SN2015F & $57106.7\pm0.3$ & $\ldots^a$ & $\ldots^a$ & $\ldots^a$ & 7 & $193-295$ \\ 
ASASSN-15hx & $57152.2\pm0.3$ & $\ldots^a$ & $\ldots^a$ & $\ldots^a$ & 3 & $250-445$ \\ 
SN2015I & $57157.4\pm0.4$ & $\ldots^a$ & $\ldots^a$ & $\ldots^a$ & 1 & 269 \\ 
PSN~J1149 & $57216.6\pm0.4$ & $\ldots^a$ & $\ldots^a$ & $\ldots^a$ & 1 & 205 \\ 
SN2016brx & $57497.0\pm5.0$ & $\ldots$ & $33.01\pm0.40^{8}$ & $\ldots$ & 1 & 184 \\ 
SN2016bry & $57506.4\pm0.0$ & $\ldots^a$ & $\ldots^a$ & $\ldots^a$ & 1 & 206 \\ 
ASASSN-16eq & $57507.1\pm0.5$ & $\ldots^a$ & $\ldots^a$ & $\ldots^a$ & 1 & 202 \\ 
Gaia16avm & $57584.8\pm0.4$ & $\ldots^a$ & $\ldots^a$ & $\ldots^a$ & 1 & 229 \\ 
ATLAS16cpu & $57632.0\pm0.4$ & $\ldots^a$ & $\ldots^a$ & $\ldots^a$ & 1 & 182 \\ 
SN2016gxp & $57682.1\pm0.4$ & $\ldots^a$ & $\ldots^a$ & $\ldots^a$ & 1 & 218 \\ 
ASASSN-17cs & $57817.2\pm0.3$ & $\ldots^a$ & $\ldots^a$ & $\ldots^a$ & 1 & 220 \\ 
DLT17u & $57842.3\pm0.3$ & $\ldots^a$ & $\ldots^a$ & $\ldots^a$ & 1 & 316 \\ 
ATLAS17dfo & $57850.3\pm0.3$ & $\ldots^a$ & $\ldots^a$ & $\ldots^a$ & 1 & 288 \\ 
DLT17ar & $57870.7\pm0.4$ & $\ldots^a$ & $\ldots^a$ & $\ldots^a$ & 1 & 228 \\ 
DLT17bk & $57911.1\pm0.4$ & $\ldots^a$ & $\ldots^a$ & $\ldots^a$ & 1 & 284 \\ 
ASASSN-18hz & $57935.7\pm0.4$ & $\ldots^a$ & $\ldots^a$ & $\ldots^a$ & 1 & 229 \\ 
SN2017ezd & $57941.3\pm0.4$ & $\ldots^a$ & $\ldots^a$ & $\ldots^a$ & 1 & 255 \\ 
DLT17bx & $57959.9\pm0.3$ & $\ldots^a$ & $\ldots^a$ & $\ldots^a$ & 1 & 379 \\ 
ATLAS17jiv & $57985.4\pm0.3$ & $\ldots^a$ & $\ldots^a$ & $\ldots^a$ & 1 & 295 \\ 
DLT17cd & $57989.5\pm0.3$ & $\ldots^a$ & $\ldots^a$ & $\ldots^a$ & 1 & 229 \\ 
SN2017glq & $58016.0\pm0.4$ & $\ldots^a$ & $\ldots^a$ & $\ldots^a$ & 1 & 331 \\ 
ATLAS17nmh & $58058.4\pm1.4$ & $\ldots^a$ & $\ldots^a$ & $\ldots^a$ & 1 & 195 \\ 
ATLAS17nse & $58118.4\pm0.3$ & $\ldots^a$ & $\ldots^a$ & $\ldots^a$ & 2 & $253-307$ \\ 
ASASSN-18bt & $58163.3\pm0.3$ & $\ldots^a$ & $\ldots^a$ & $\ldots^a$ & 3 & $236-267$ \\ 
ASASSN-18da & $58178.8\pm0.4$ & $\ldots^a$ & $\ldots^a$ & $\ldots^a$ & 1 & 220 \\ 
DLT18h & $58185.3\pm0.4$ & $\ldots^a$ & $\ldots^a$ & $\ldots^a$ & 1 & 336 \\ 
DLT18i & $58195.2\pm0.3$ & $\ldots^a$ & $\ldots^a$ & $\ldots^a$ & 1 & 237 \\ 
ASASSN-18hb & $58222.9\pm0.4$ & $\ldots^a$ & $\ldots^a$ & $\ldots^a$ & 1 & 237 \\ 
\hline
\end{tabular}
\begin{tablenotes}
\item[] References: (1)\citet{tully13}; (2)\citet{riess16}; (3)\citet{villegas10}; (4)\citet{freedman01}; (5)\citet{saha06}; (6)\citet{larsen01}; (7)Derived from redshift; (8)\citet{theureau07}; (9)\citet{blakeslee10}; (10)\citet{huang17}; (11)\citet{tonry01}; (12)\citet{holmbo18}; (13)\citet{saulder16}; (14)\citet{springob14}; (15)\citet{li18}.
\item[a] To be presented in P. Chen et al. (in prep).
\end{tablenotes}

%% file: InputFiles/spectable-0.tex
\begin{tabular}{lcrccccccc}
SN & Obs. Date & Phase$^{a}$ & Telescope$^{b}$ & Instrument$^{b}$ & Range & Expt.  & $v_{exp}$ & Qual. & Ref. \\
& (MJD) & (days) & & & (\AA) & (s) & [km~s$^{-1}$] & & \\
\hline
SN1972E & 41651.00 & 205.1 & P200 & DBSP & $4000-10000$ & $\ldots$ & $5100\pm800$ & Med & \citet{kirshner75} \\
 & 41682.00 & 236.1 & P200 & DBSP & $3300-10300$ & $\ldots$ & $5900\pm400$ & Med & \citet{kirshner75} \\
 & 41795.00 & 349.1 & P200 & DBSP & $3400-10200$ & $\ldots$ & $6300\pm100$ & Med & \citet{kirshner75} \\
 & 41864.00 & 418.1 & P200 & DBSP & $3400-9200$ & $\ldots$ & $6600\pm200$ & Med & \citet{kirshner75} \\
SN1981B & 44940.00 & 267.3 & HJST & UVITS & $3400-7100$ & $3600$ & $6200\pm300$ & Med & \citet{branch93} \\
SN1986G & 46816.50 & 255.5 & ESO2.2m & BC & $4500-6800$ & $\ldots$ & $5500\pm100$ & Med & \citet{cristiani92} \\
\hline
\end{tabular}

%% file: InputFiles/spectable-1.tex
\begin{tabular}{lcrccccccc}
SN & Obs. Date & Phase$^{a}$ & Telescope$^{b}$ & Instrument$^{b}$ & Range & Expt.  & $v_{exp}$ & Qual. & Ref. \\
& (MJD) & (days) & & & (\AA) & (s) & [km~s$^{-1}$] & & \\
\hline
SN2002er & 52739.00 & 213.7 & TNG & DOLORES & $3500-8000$ & $\ldots$ & $5800\pm100$ & Med & \citet{kotak05} \\
SN2003cg & 53114.00 & 384.8 & VLT & FORS1 & $3900-8000$ & $3800$ & $6000\pm500$ & Med & \citet{eliasrosa06} \\
SN2003du & 52961.00 & 194.4 & WHT & ISIS & $4400-7000$ & $\ldots$ & $5800\pm100$ & Med & \citet{stanishev07} \\
 & 52974.00 & 207.4 & CA3.5m & MOSCA & $3200-9200$ & $\ldots$ & $5500\pm200$ & Med & \citet{stanishev07} \\
 & 52986.00 & 219.4 & CA2.2m & CAFOS & $3200-8700$ & $\ldots$ & $6300\pm900$ & Med & \citet{stanishev07} \\
 & 53037.00 & 270.4 & CA2.5m & MOSCA & $3200-9200$ & $\ldots$ & $6600\pm100$ & Med & \citet{stanishev07} \\
\hline
\end{tabular}

%% file: InputFiles/spectable-2.tex
\begin{tabular}{lcrccccccc}
SN & Obs. Date & Phase$^{a}$ & Telescope$^{b}$ & Instrument$^{b}$ & Range & Expt.  & $v_{exp}$ & Qual. & Ref. \\
& (MJD) & (days) & & & (\AA) & (s) & [km~s$^{-1}$] & & \\
\hline
 & 56044.00 & 228.9 & LBT & MODS & $3200-10000$ & $\ldots$ & $5800\pm300$ & High & \citet{shappee13} \\
 & 56046.25 & 231.1 & KeckI & LRIS & $3100-7300$ & $300$ & $5900\pm100$ & Med & PI-Filippenko \\
 & 56073.00 & 257.9 & WHT & ISIS & $3500-9500$ & $\ldots$ & $5500\pm100$ & High & \citet{mazzali15} \\
 & 56103.00 & 287.9 & WHT & ISIS & $3400-10000$ & $\ldots$ & $5700\pm200$ & High & \citet{mazzali15} \\
 & 56125.00 & 309.9 & Shane3m & KAST & $3500-10000$ & $\ldots$ & $5800\pm200$ & High & \citet{mazzali15} \\
 & 56128.00 & 312.9 & GTC & OSIRIS & $3600-10400$ & $\ldots$ & $5800\pm100$ & High & \citet{taubenberger15} \\
\hline
\end{tabular}

%% file: InputFiles/spectable-3.tex
\begin{tabular}{lcrccccccc}
SN & Obs. Date & Phase$^{a}$ & Telescope$^{b}$ & Instrument$^{b}$ & Range & Expt.  & $v_{exp}$ & Qual. & Ref. \\
& (MJD) & (days) & & & (\AA) & (s) & [km~s$^{-1}$] & & \\
\hline
 & 57228.15 & 268.4 & GS & GMOS & $4000-9400$ & $14400$ & $5900\pm1000$ & Med & \citet{graham17} \\
 & 57285.20 & 325.5 & VLT & XSH & $3600-10100$ & $2900$ & $6300\pm1100$ & Med & \citet{maguire18} \\
ASASSN-14jc & 57350.26 & 390.0 & VLT & MUSE & $4700-9200$ & $2465$ & $(3\,000)$ & Low & This work \\
ASASSN-14jz & 57182.56 & 203.5 & Baade & IMACS & $4000-8100$ & $1800$ & $6000\pm900$ & High & This work \\
ASASSN-14kq & 57404.02 & 412.9 & VLT & MUSE & $4600-9000$ & $2450$ & $(3\,000)$ & Low & This work \\
ASASSN-14lv & 57393.04 & 398.0 & VLT & MUSE & $4500-8900$ & $2460$ & $(3\,000)$ & Low & This work \\
\hline
\end{tabular}
\begin{tablenotes}
\item[a] Relative to $t_{\rm{max}}$
\item[b] See \S\ref{sec:data} for definitions and references.
\end{tablenotes}

%% file: InputFiles/resultstable-0.tex
\begin{tabular}{l|rrcc|rrcc}
& \multicolumn{4}{c}{H-rich Model} & \multicolumn{4}{c}{He-rich Model} \\ \cmidrule(lr){2-5} \cmidrule(lr){6-9}
SN & Phase$^{a}$ & Line$^{b}$ & Flux Limit & Mass Limit & Phase$^a$ & Line$^b$ & Flux Limit & Mass Limit\\
& (days) & & (erg/s/cm$^2$) & ($M_\odot$) & (days) & & (erg/s/cm$^2$) & ($M_\odot$) \\
\hline
SN1972E &418.1 &H$\alpha$ &$2.52\times 10^{-16}$ &$4.95\times 10^{-5}$ &349.1 &HeI-a &$3.07\times 10^{-17}$ &$9.84\times 10^{-5}$ \\
SN1981B &267.3 &H$\alpha$ &$5.54\times 10^{-16}$ &$1.49\times 10^{-3}$ &267.3 &HeI-a &$5.86\times 10^{-16}$ &$2.25\times 10^{-3}$ \\
SN1986G &324.5 &H$\alpha$ &$1.68\times 10^{-15}$ &$2.84\times 10^{-4}$ &256.5 &HeI-b &$6.57\times 10^{-16}$ &$5.29\times 10^{-4}$ \\
SN1990N &280.0 &H$\alpha$ &$9.52\times 10^{-17}$ &$5.68\times 10^{-4}$ &280.0 &HeI-a &$2.35\times 10^{-16}$ &$1.53\times 10^{-3}$ \\
SN1991T &258.0 &H$\alpha$ &$1.99\times 10^{-14}$ &$4.78\times 10^{-2}$ &349.3 &HeI-a &$4.84\times 10^{-14}$ &$>2.0$ \\
\hline
\end{tabular}

%% file: InputFiles/resultstable-1.tex
\begin{tabular}{l|rrcc|rrcc}
& \multicolumn{4}{c}{H-rich Model} & \multicolumn{4}{c}{He-rich Model} \\ \cmidrule(lr){2-5} \cmidrule(lr){6-9}
SN & Phase$^{a}$ & Line$^{b}$ & Flux Limit & Mass Limit & Phase$^a$ & Line$^b$ & Flux Limit & Mass Limit\\
& (days) & & (erg/s/cm$^2$) & ($M_\odot$) & (days) & & (erg/s/cm$^2$) & ($M_\odot$) \\
\hline
SN2011iv &304.4 &H$\alpha$ &$1.16\times 10^{-16}$ &$5.36\times 10^{-4}$ &304.4 &HeI-a &$1.25\times 10^{-16}$ &$7.79\times 10^{-4}$ \\
SN2012Z &254.3 &H$\beta$ &$6.99\times 10^{-16}$ &$1.25\times 10^{-2}$ &247.3 &HeI-a &$5.71\times 10^{-16}$ &$1.19\times 10^{-2}$ \\
SN2012cg &338.1 &Pa$\alpha$ &$1.33\times 10^{-16}$ &$2.08\times 10^{-4}$ &286.3 &HeI-a &$2.30\times 10^{-16}$ &$1.13\times 10^{-3}$ \\
SNhunt136 &319.1 &H$\alpha$ &$3.21\times 10^{-16}$ &$7.37\times 10^{-4}$ &319.1 &HeI-a &$3.48\times 10^{-16}$ &$1.09\times 10^{-3}$ \\
SN2012ei &254.4 &H$\alpha$ &$1.71\times 10^{-17}$ &$2.99\times 10^{-4}$ &254.4 &HeI-a &$2.43\times 10^{-17}$ &$5.08\times 10^{-4}$ \\
\hline
\end{tabular}
\begin{tablenotes}
\item[a] Relative to $t_{\rm{max}}$. \\
\item[b] Corresponds to the lines listed in Table \ref{tab:ModelInfo}.\\
\end{tablenotes}

%% file: texfiles/newphot.tex
\begin{table}
\caption{New photometry presented in this study. Some data stem from archival images processed by this study for flux calibration purposes.}
\label{tab:newphot}
\scriptsize
\begin{tabular}{lrcrr}
SN & MJD & Filter & AB Mag \\\hline
ASASSN-14jg & 57183.78 & V & $20.49\pm0.08$ \\
ASASSN-15hx & 57396.33 & V & $21.33\pm0.04$ \\
SN2001el & 52492.37 & I & $19.82\pm0.07$ \\
SN2001el & 52549.24 & I & $20.24\pm0.07$ \\
SN2001el & 52552.37 & I & $20.23\pm0.08$ \\
SN2001el & 52618.04 & I & $22.08\pm0.15$ \\
SN2001el & 52492.36 & R & $20.68\pm0.10$ \\
\hline
\end{tabular}
\end{table}

\begin{table}
\contcaption{New photometry presented in this study. Some data stem from archival images processed by this study for flux calibration purposes.}
\scriptsize
\begin{tabular}{lrcrr}
SN2015F & 57273.30 & V & $18.90\pm0.02$ \\
SN2015F & 57306.35 & V & $19.31\pm0.02$ \\
SN2015F & 57334.29 & V & $19.66\pm0.02$ \\
SN2015F & 57403.20 & V & $20.87\pm0.07$ \\
SN2015F & 57424.11 & V & $20.99\pm0.03$ \\
SN2015F & 57443.15 & V & $21.33\pm0.07$ \\
SN2015F & 57494.06 & V & $22.28\pm0.08$ \\
SN2015F & 57273.30 & g & $18.74\pm0.02$ \\
\hline
\end{tabular}
\end{table}

%% file: texfiles/phottable-0.tex
\begin{tabular}{lrcp{0.6cm}p{2.5cm}}
\hline
SN & $N_{tot}$ & Phases & Filters & Refs. \\\hline\hline
ASASSN-14jg$^{\star}$ & 113 & $5-268$ & B, V, g, r, i & ASAS-SN; \citet{graham17}; This work\\\hline
ASASSN-15hx & 202 & $-16-325$ & V, g, i, SwiftU, SwiftB, SwiftV & ASAS-SN; This work\\\hline
PSN~J1149 & 16 & $-12-141$ & B, V, I & ASAS-SN\\\hline
SN1972E$^{\star}$ & 468 & $-23-359$ & U, B, V, R, I & \citet{IAUC2434}; \citet{cousins72}; \citet{IAUC2411}; \citet{ardeburg73}; \citet{vangenderen75}; \citet{osmer72}; \citet{lee72}; \citet{IAUC2398}; \citet{IAUC2434}\\\hline
SN1980N$^{\star}$ & 238 & $-11-368$ & U, B, V, R, I, H & \citet{hamuy91}; \citet{IAUC3556}; \citet{IAUC3556}; \citet{olsz92}; \citet{elias83}\\\hline
SN1981B & 281 & $-9-136$ & U, B, V, R, J, H, K & \citet{IAUC3589}; \citet{tsvetkov82}; \citet{buta83}; \citet{tsvetkov82}; \citet{barbon82}; \citet{IAUC3584}; \citet{elias83}\\\hline
SN1981D$^{\star}$ & 30 & $-14-274$ & B, V, J, H & \citet{IAUC3583}; \citet{hamuy91}; \citet{elias83}\\\hline
SN1983W$^{\star}$ & 57 & $-7-161$ & U, B, V & \\\hline
SN1986G & 93 & $-7-403$ & U, B, V, R, I & \citet{IAUC4183}; \citet{phillips87}; \citet{IAUC4215}; \citet{IAUC4224}; \citet{turatto90}; \citet{IAUC4421}\\\hline
SN1989B$^{\star}$ & 350 & $-5-263$ & U, B, V, R, I, J, H, K & \citet{barbon90}; \citet{IAUC4728}; \citet{IAUC4743}; \citet{tsvetkov90}; \citet{wells94}; \citet{IAUC4742}; \citet{volkov91}; \citet{IAUC4761}\\\hline
SN1990N$^{\star}$ & 289 & $-12-376$ & U, B, V, R, I & \citet{lira98}\\\hline
\hline
\end{tabular}

%% file: texfiles/phottable-1.tex
\begin{tabular}{lrcp{0.6cm}p{2.5cm}}
\hline
SN & $N_{tot}$ & Phases & Filters & Refs. \\\hline\hline
SN1990O$^{\star}$ & 41 & $2-309$ & B, V, R, I & \citet{hamuy96}\\\hline
SN1991T$^{\star}$ & 669 & $-13-454$ & U, B, V, R, I, J, H, K & \citet{tsvetkov86}; \citet{altavilla04}; \citet{IAUC5239}; \citet{lira98}; \citet{IAUC5246}; \citet{krisciunas04}; \citet{IAUC5270}; \citet{ford93}; \citet{IAUC5309}; \citet{schmidt94}\\\hline
SN1991bg$^{\star}$ & 171 & $-10-528$ & B, V, R, I, J, H, K & \citet{leibendgut93}; \citet{turatto96}; \citet{krisciunas04}\\\hline
SN1992A$^{\star}$ & 62 & $-7-536$ & B, V, R, I & \citet{IAUC5431}; \citet{altavilla04}; \citet{IAUC5432}\\\hline
SN1992bc$^{\star}$ & 145 & $-11-326$ & B, V, R, I & \citet{hamuy96}\\\hline
SN1993Z & 78 & $6-239$ & B, V, R, I & \citet{ho01}\\\hline
SN1994D$^{\star}$ & 371 & $-14-432$ & U, B, V, R, I, J, H, K & \citet{IAUC5946}; \citet{richmond95}; \citet{IAUC5951}; \citet{IAUC5950}; \citet{IAUC5952}; \citet{IAUC5976}; \citet{IAUC5958}; \citet{altavilla04}\\\hline
SN1994ae$^{\star}$ & 301 & $-15-530$ & U, B, V, R, I & \citet{IAUC6115}; \citet{riess05}; \citet{riess99}; \citet{IAUC6115}; \citet{ho01}; \citet{tsvetkov97}; \citet{altavilla04}\\\hline
SN1995D$^{\star}$ & 225 & $-8-424$ & B, V, R, I & \citet{IAUC6166}; \citet{riess99}; \citet{altavilla04}; \citet{ho01}\\\hline
SN1995ac$^{\star}$ & 100 & $-5-250$ & B, V, R, I & \citet{riess99}; \citet{altavilla04}\\\hline
SN1995al$^{\star}$ & 115 & $-13-182$ & U, B, V, R, I & \citet{riess99}\\\hline
SN1996X$^{\star}$ & 241 & $-4-502$ & U, B, V, R, I, J, H, K & \citet{IAUC6484}; \citet{salvo01}; \citet{riess99}\\\hline
SN1997bp$^{\star}$ & 101 & $0-416$ & U, B, V, R, I & CfA; \citet{altavilla04}\\\hline
\hline
\end{tabular}

%% file: texfiles/phottable-2.tex
\begin{tabular}{lrcp{0.6cm}p{2.5cm}}
\hline
SN & $N_{tot}$ & Phases & Filters & Refs. \\\hline\hline
SN1997br$^{\star}$ & 223 & $-9-402$ & U, B, V, R, I & \citet{li99}; CfA; \citet{altavilla04}\\\hline
SN1997cw$^{\star}$ & 75 & $13-181$ & U, B, V, R, I & CfA\\\hline
SN1998V$^{\star}$ & 60 & $2-187$ & U, B, V, R, I & CfA\\\hline
SN1998aq$^{\star}$ & 257 & $-10-344$ & U, B, V, R, I & \citet{riess05}\\\hline
SN1998bu & 914 & $-7-56$ & U, B, V, R, I, J, H, K & \citet{jha99}; \citet{hernandez00}; \citet{suntzeff99}\\\hline
SN1998ec$^{\star}$ & 59 & $5-186$ & U, B, V, R, I & BSNIP; CfA\\\hline
SN1999aa & 289 & $-11-79$ & U, B, V, R, I, J, K & \citet{kowalski08}; BSNIP; CfA; \citet{kriciunas00}; \citet{altavilla04}\\\hline
SN1999ac$^{\star}$ & 356 & $-13-183$ & U, B, V, R, I, H, Ks & \citet{phillips06}; BSNIP; CfA\\\hline
SN1999by$^{\star}$ & 377 & $-12-212$ & U, B, V, R, I, J, H, K & \citet{garnavich04}; BSNIP; \citet{toth00}\\\hline
SN1999dq$^{\star}$ & 247 & $-11-154$ & U, B, V, R, I & BSNIP; CfA\\\hline
SN1999gd$^{\star}$ & 35 & $-3-150$ & U, B, V, R, I & CfA\\\hline
SN2000E$^{\star}$ & 287 & $-16-484$ & U, B, V, R, I, J, H, K & \citet{valentini03}; \citet{lair06}\\\hline
SN2000cx$^{\star}$ & 1067 & $-9-500$ & U, B, V, R, I, J, H, K & BSNIP; \citet{li01}; \citet{candia03}; CfA; \citet{altavilla04}; \citet{lair06}; \citet{sollerman04}\\\hline
SN2001C$^{\star}$ & 80 & $20-415$ & U, B, V, R, I & \citet{lair06}; BSNIP; \citet{hicken09}\\\hline
SN2001V$^{\star}$ & 370 & $-14-351$ & U, B, V, R, I & \citet{hicken09}; \citet{vinko03}; \citet{lair06}; BSNIP\\\hline
\hline
\end{tabular}

%% file: texfiles/phottable-3.tex
\begin{tabular}{lrcp{0.6cm}p{2.5cm}}
\hline
SN & $N_{tot}$ & Phases & Filters & Refs. \\\hline\hline
SN2001bg$^{\star}$ & 91 & $-2-358$ & B, V, R, I & BSNIP; \citet{lair06}\\\hline
SN2001el$^{\star}$ & 566 & $-13-435$ & U, B, V, R, I, J, H, K & \citet{krisciunas03}; This work\\\hline
SN2002bo & 656 & $-14-100$ & U, B, V, R, I, J, H, K & BSNIP; \citet{benetti04}; \citet{hicken09}; \citet{szabo03}; \citet{krisciunas04}\\\hline
SN2002cx & 110 & $-8-50$ & B, V, R, I & BSNIP\\\hline
SN2002dj & 166 & $-14-65$ & U, B, V, R, I, J, H, K & \citet{pignata08}; BSNIP; \citet{hicken09}\\\hline
SN2002dp$^{\star}$ & 190 & $-6-183$ & U, B, V, R, I & BSNIP; \citet{hicken09}\\\hline
SN2002er$^{\star}$ & 339 & $-16-323$ & U, B, V, R, I & \citet{pignata04}; BSNIP; \citet{christensen03}\\\hline
SN2002ha$^{\star}$ & 162 & $-12-237$ & U, B, V, R, I & BSNIP; \citet{hicken09}\\\hline
SN2003K$^{\star}$ & 38 & $1-169$ & U, B, V, R, I & \citet{hicken09}\\\hline
SN2003cg & 439 & $-18-413$ & U, B, V, R, I, J, H, K & \citet{eliasrosa06}; \citet{ganeshalingam10}; \citet{hicken09}\\\hline
SN2003du & 628 & $-12-465$ & U, B, V, R, I, J, H, K & \citet{anupama05}; \citet{stanishev07}; \citet{hicken09}; BSNIP\\\hline
SN2003gs$^{\star}$ & 188 & $8-378$ & B, V, R, I, J, H, K & \citet{krisciunas09}; BSNIP\\\hline
SN2003hv & 217 & $0-538$ & B, V, R, I, J, H, K & \citet{leloudas09}; BSNIP\\\hline
SN2003kf & 180 & $-3-104$ & U, B, V, R, I & BSNIP; \citet{hicken09}\\\hline
SN2004S & 215 & $1-375$ & B, V, R, I, J, H, K & BSNIP; \citet{krisciunas07}; \citet{misra05}\\\hline
\hline
\end{tabular}

%% file: texfiles/phottable-4.tex
\begin{tabular}{lrcp{0.6cm}p{2.5cm}}
\hline
SN & $N_{tot}$ & Phases & Filters & Refs. \\\hline\hline
SN2004eo & 768 & $-12-345$ & U, B, V, R, I, J, H, K, u, g, r, i & CSP; \citet{pastorello07}; BSNIP\\\hline
SN2005W & 115 & $-8-16$ & B, V, u, g, r, i & CSP\\\hline
SN2005am & 388 & $-2-76$ & U, B, V, R, I, J, H, K, u, g, r, i & CSP; BSNIP; \citet{hicken09}\\\hline
SN2005cf & 1107 & $-13-89$ & U, B, V, R, I, J, H, K, Ks & \citet{hicken09}; BSNIP; \citet{pastorello07}; \citet{friedman15}; \citet{brown14}\\\hline
SN2005hk & 402 & $-20-381$ & U, B, V, R, I, J, H, Ks, u, g, r, i, z & \citet{sako14}; \citet{holtzman08}; \citet{hicken09}; \citet{friedman15}; \citet{brown14}; \citet{sahu08}\\\hline
SN2005ki$^{\star}$ & 344 & $-12-155$ & U, B, V, J, H, u, g, r, i & \citet{hicken09}; CSP\\\hline
SN2006X$^{\star}$ & 638 & $-11-151$ & U, B, V, R, I, J, H, K, Ks, u, g, r, i & \citet{brown14}; BSNIP; CSP; \citet{hicken09}; \citet{friedman15}\\\hline
SN2006dd$^{\star}$ & 284 & $-11-228$ & B, V, R, I, J, H, K, u, g, r, i & \citet{stritzinger10}\\\hline
SN2006gz & 107 & $-14-49$ & U, B, V, r, i & \citet{hicken07}\\\hline
SN2007af$^{\star}$ & 807 & $-12-274$ & U, B, V, R, I, J, H, K, u, g, r, i & \citet{brown14}; \citet{hicken09}; CSP; BSNIP\\\hline
SN2007bd$^{\star}$ & 120 & $-10-214$ & B, V, J, H, u, g, r, i & \citet{hicken09}; CSP\\\hline
\hline
\end{tabular}

%% file: texfiles/phottable-5.tex
\begin{tabular}{lrcp{0.6cm}p{2.5cm}}
\hline
SN & $N_{tot}$ & Phases & Filters & Refs. \\\hline\hline
SN2007gi & 109 & $-12-190$ & U, B, V, R, I & \citet{brown14}; \citet{zhang10}\\\hline
SN2007if & 224 & $14-358$ & B, V, R, I, J, H, Ks, u, g, r, i & CSP; \citet{hicken12}; \citet{friedman15}; \citet{taubenberger13}\\\hline
SN2007is$^{\star}$ & 24 & $2-163$ & B, V, r, i & \citet{hicken12}\\\hline
SN2007le & 484 & $-11-84$ & B, V, R, I, J, H, Ks, u, g, r, i & CSP; \citet{friedman15}; BSNIP; \citet{hicken12}\\\hline
SN2007on & 665 & $-9-405$ & U, B, V, Y, J, H, K, u, g, r, i & CSP; \citet{brown14}; This work; \citet{gall18}\\\hline
SN2007sr$^{\star}$ & 347 & $5-196$ & U, B, V, R, I, J, H, Ks & \citet{hicken09}; BSNIP; \citet{brown14}; \citet{friedman15}\\\hline
SN2007sw$^{\star}$ & 101 & $-2-154$ & B, V, r, i & \citet{hicken12}\\\hline
SN2008A & 201 & $-7-208$ & U, B, V, R, I, J, H, Ks, r, i & \citet{hicken12}; \citet{brown14}; \citet{friedman15}; BSNIP\\\hline
SN2008Q & 84 & $-9-208$ & U, B, V, R, I, r, i & \citet{brown14}; \citet{hicken12}; BSNIP\\\hline
SN2009dc & 672 & $-20-286$ & U, B, V, R, I, J, H, Ks, u, g, r, i & \citet{silverman11}; \citet{friedman15}; \citet{hicken12}; CSP; \citet{brown14}\\\hline
SN2009ig & 258 & $-15-404$ & U, B, V, R, I, J, H, Ks, u, r, i & \citet{brown14}; \citet{brown12}; \citet{hicken12}; \citet{friedman15}; This work\\\hline
SN2009kq$^{\star}$ & 128 & $-4-177$ & B, V, J, H, Ks, u, r, i & \citet{hicken12}; \citet{friedman15}\\\hline
SN2009le & 57 & $-12-318$ & B, V, R, I, J, H, Ks, u, r, i & \citet{hicken12}; \citet{friedman15}; This work\\\hline
\hline
\end{tabular}

%% file: texfiles/phottable-6.tex
\begin{tabular}{lrcp{0.6cm}p{2.5cm}}
\hline
SN & $N_{tot}$ & Phases & Filters & Refs. \\\hline\hline
SN2009nr$^{\star}$ & 147 & $-13-175$ & U, B, V, R, I & \citet{khan11}\\\hline
SN2010ag$^{\star}$ & 102 & $-3-179$ & B, V, J, H, Ks, u, r, i & \citet{friedman15}; \citet{hicken12}\\\hline
SN2010dd$^{\star}$ & 96 & $-23-523$ & R & \citet{maguire14}\\\hline
SN2010ev$^{\star}$ & 67 & $-7-270$ & U, B, V, R, I, u, g, r, i, z & \citet{gutierrez16}; \citet{brown14}; This work\\\hline
SN2010gp & 59 & $-4-276$ & U, B, V, R, I & \citet{brown14}; This work\\\hline
SN2011K & 40 & $-1-253$ & V, R, I, J, H, Ks & \citet{friedman15}; This work\\\hline
SNhunt37 & 90 & $1-312$ & V, R, I, J, H, Ks & \citet{friedman15}; This work\\\hline
SN2011at & 57 & $7-463$ & U, B, V, R, I, J, H, Ks & \citet{brown14}; \citet{friedman15}; This work\\\hline
SN2011by & 158 & $-11-44$ & U, B, V, J, H, Ks & \citet{brown14}; \citet{friedman15}\\\hline
SN2011ek & 68 & $-6-421$ & U, B, V, R, I & \citet{maguire14}; \citet{brown14}; This work\\\hline
PTF11kly$^{\star}$ & 1956 & $-19-571$ & U, B, V, R, I, J, H, K, g, r & \citet{firth15}; \citet{nugent11}; \citet{brown14}; \citet{OSCref}; \citet{tsvetkov13}; \citet{munari13}; \citet{richmond12}; \citet{weyant18}\\\hline
SN2011im & 63 & $-13-310$ & U, B, V, R, I & \citet{firth15}; \citet{maguire14}; \citet{brown14}; This work\\\hline
SN2011iv & 235 & $-6-301$ & B, V, R, I, Y, J, H, u, g, r, i & This work; \citet{gall18}\\\hline
\hline
\end{tabular}

%% file: texfiles/phottable-7.tex
\begin{tabular}{lrcp{0.6cm}p{2.5cm}}
\hline
SN & $N_{tot}$ & Phases & Filters & Refs. \\\hline\hline
SN2011iy & 127 & $10-150$ & B, V, Y, J, H, u, g, r, i & \citet{weyant18}; CSP\\\hline
SN2012Z & 306 & $-8-265$ & U, B, V, J, H, u, g, r, i & \citet{stritzinger15}; \citet{brown14}; \citet{foley13}\\\hline
SN2012cg & 176 & $-16-598$ & U, B, V, R, I & \citet{brown14}; \citet{munari13}; This work\\\hline
SN2012fr & 733 & $-15-415$ & U, B, V, R, I, g, r, i & \citet{brown14}; \citet{graham17}; \citet{zhang14}\\\hline
SN2012hr & 284 & $-4-457$ & U, B, V, R, I, g, r, i & \citet{brown14}; \citet{graham17}; \citet{smartt15}\\\hline
SN2012ht & 99 & $-13-102$ & U, B, V, R, I & \citet{brown14}; \citet{smartt15}\\\hline
SN2013aa & 493 & $-5-399$ & U, B, V, g, r, i & \citet{brown14}; \citet{graham17}\\\hline
SNhunt196 & 233 & $-20-261$ & U, B, V, R, I, J, H, u, g, r, i & \citet{graham17}; \citet{walker15}; \citet{brown14}; \citet{weyant18}\\\hline
SN2013ct & 7 & $80-228$ & g, r, z & \citet{inserra13}\\\hline
SN2013dy & 1210 & $-16-337$ & U, B, V, R, I, J, H, g, r, i & \citet{graham17}; \citet{pan15}; \citet{brown14}; \citet{zhai16}\\\hline
SN2013gy & 155 & $-17-234$ & U, B, V, R, g, r, i & \citet{ATel5637}; \citet{graham17}; \citet{brown14}; This work\\\hline
SN2014J & 706 & $-12-76$ & U, B, V, R, I, g, r, i, z & \citet{goobar14}; \citet{foley14}; \citet{brown14}; \citet{zhang14}; \citet{tsvetkov14}\\\hline
SN2014bv & 45 & $0-293$ & U, B, V, g & \citet{brown14}; This work\\\hline
SN2015F$^{\star}$ & 477 & $-14-594$ & U, B, V, I, g, r, i, SwiftB, SwiftV & ASAS-SN; \citet{graham17}; \citet{brown14}; \citet{nucita17}; \citet{ATel8484}; This work\\\hline
\hline
\end{tabular}

%% file: texfiles/phottable-8.tex
\begin{tabular}{lrcp{0.6cm}p{2.5cm}}
\hline
SN & $N_{tot}$ & Phases & Filters & Refs. \\\hline\hline
SN2015I & 70 & $-10-322$ & U, B, V, I, g, r, i, z, SwiftU, SwiftB, SwiftV & ASAS-SN; \citet{foley18}; \citet{brown14}; \citet{ATel7476}; This work\\\hline
SN2016brx & 11 & $10-185$ & V, R, g, r & ASAS-SN\\\hline
SN2016bry & 61 & $5-205$ & B, V, R, I, g, r & ASAS-SN; This work\\\hline
ASASSN-16eq & 55 & $-14-202$ & B, V, R, I, g, r & ASAS-SN; This work\\\hline
iPTF16auf$^{\star}$ & 198 & $-15-185$ & B, V, R, I, g, r, i, z, SwiftU, SwiftB, SwiftV & ASAS-SN; \citet{foley18}; \citet{ATel9049}\\\hline
ASASSN-16fn$^{\star}$ & 98 & $-32-189$ & B, V, R, I, g, r, i & ASAS-SN\\\hline
Gaia16avm & 5 & $-11-325$ & V, r & ASAS-SN; This work\\\hline
PS16em$^{\star}$ & 82 & $-78-195$ & B, V, r, i, SwiftU, SwiftB, SwiftV & ASAS-SN\\\hline
ATLAS16cpu & 81 & $-6-182$ & V, g, r, i, z, SwiftU, SwiftB, SwiftV & ASAS-SN; \citet{foley18}; This work\\\hline
Gaia16hj$^{\star}$ & 65 & $-11-154$ & B, V, r, i & ASAS-SN; \citet{inserra13}\\\hline
SN2016gxp$^{\star}$ & 331 & $-9-267$ & B, V, r, i, SwiftU, SwiftB, SwiftV & ASAS-SN; This work\\\hline
ASASSN-16lx$^{\star}$ & 169 & $-7-174$ & B, V, r, i, SwiftU, SwiftB, SwiftV & ASAS-SN; \citet{ATel9666}\\\hline
Gaia16bql$^{\star}$ & 90 & $-13-178$ & B, V, r, i & ASAS-SN\\\hline
\hline
\end{tabular}

%% file: texfiles/phottable-9.tex
\begin{tabular}{lrcp{0.6cm}p{2.5cm}}
\hline
SN & $N_{tot}$ & Phases & Filters & Refs. \\\hline\hline
ASASSN-17cs & 180 & $-11-219$ & B, V, r, i, SwiftU, SwiftB, SwiftV & ASAS-SN; This work\\\hline
DLT17u$^{\star}$ & 622 & $-33-315$ & U, B, V, r, i, SwiftU, SwiftB, SwiftV & ASAS-SN; \citet{brown14}; This work\\\hline
ATLAS17dfo & 166 & $-17-288$ & B, V, r, i, SwiftU, SwiftB, SwiftV & ASAS-SN; This work\\\hline
DLT17ar$^{\star}$ & 184 & $-14-227$ & B, V, r, i, SwiftU, SwiftB, SwiftV & ASAS-SN; This work\\\hline
DLT17bk$^{\star}$ & 212 & $-8-284$ & B, V, r, i, SwiftU, SwiftB, SwiftV & ASAS-SN; This work\\\hline
ASASSN-18hz$^{\star}$ & 88 & $-10-228$ & B, V, r, i, SwiftU, SwiftB, SwiftV & ASAS-SN; This work\\\hline
SN2017ezd & 139 & $-7-255$ & B, V, g, r, i, SwiftU, SwiftV & ASAS-SN; This work\\\hline
DLT17bx & 781 & $-13-379$ & B, V, g, r, i, SwiftU, SwiftB, SwiftV & ASAS-SN; This work\\\hline
DLT17cd & 427 & $-13-228$ & B, V, g, r, i, SwiftU, SwiftB, SwiftV & ASAS-SN; This work\\\hline
ATLAS17jiv & 269 & $-7-294$ & B, V, r, i, SwiftU, SwiftB, SwiftV & ASAS-SN; This work\\\hline
SN2017glq & 828 & $-17-331$ & B, V, g, r, i, SwiftU, SwiftB, SwiftV & ASAS-SN; This work\\\hline
\hline
\end{tabular}

%% file: texfiles/phottable-10.tex
\begin{tabular}{lrcp{0.6cm}p{2.5cm}}
\hline
SN & $N_{tot}$ & Phases & Filters & Refs. \\\hline\hline
ATLAS17nmh$^{\star}$ & 301 & $35-194$ & B, V, g, r, i & ASAS-SN; This work\\\hline
ATLAS17nse$^{\star}$ & 10 & $-4-306$ & V, i & \citet{inserra13}; This work\\\hline
ASASSN-17u$^{\star}$ & 129 & $-14-196$ & B, V, r, i, SwiftB, SwiftV & ASAS-SN\\\hline
ASASSN-18hb & 108 & $-7-237$ & B, V, g, r, i & ASAS-SN; This work\\\hline
ASASSN-18bt & 576 & $-9-265$ & B, V, g, r, i, SwiftU, SwiftB, SwiftV & ASAS-SN; \citet{tucker18}; \citet{dimitriadis19}\\\hline
ASASSN-18da & 31 & $-10-220$ & B, V, r, i, SwiftU, SwiftB, SwiftV & ASAS-SN; This work\\\hline
DLT18h$^{\star}$ & 200 & $-24-335$ & B, V, g, r, i, SwiftU, SwiftB, SwiftV & ASAS-SN; This work\\\hline
DLT18i & 340 & $-18-237$ & B, V, g, r, i, SwiftU, SwiftB, SwiftV & ASAS-SN; This work\\\hline
SNF-012 & 4 & $55-272$ & B, V, R, I & \citet{taubenberger13}\\\hline
\hline
\end{tabular}

%% file: main.bbl
\begin{thebibliography}{}
\makeatletter
\relax
\def\mn@urlcharsother{\let\do\@makeother \do\$\do\&\do\#\do\^\do\_\do\%\do\~}
\def\mn@doi{\begingroup\mn@urlcharsother \@ifnextchar [ {\mn@doi@}
  {\mn@doi@[]}}
\def\mn@doi@[#1]#2{\def\@tempa{#1}\ifx\@tempa\@empty \href
  {http://dx.doi.org/#2} {doi:#2}\else \href {http://dx.doi.org/#2} {#1}\fi
  \endgroup}
\def\mn@eprint#1#2{\mn@eprint@#1:#2::\@nil}
\def\mn@eprint@arXiv#1{\href {http://arxiv.org/abs/#1} {{\tt arXiv:#1}}}
\def\mn@eprint@dblp#1{\href {http://dblp.uni-trier.de/rec/bibtex/#1.xml}
  {dblp:#1}}
\def\mn@eprint@#1:#2:#3:#4\@nil{\def\@tempa {#1}\def\@tempb {#2}\def\@tempc
  {#3}\ifx \@tempc \@empty \let \@tempc \@tempb \let \@tempb \@tempa \fi \ifx
  \@tempb \@empty \def\@tempb {arXiv}\fi \@ifundefined
  {mn@eprint@\@tempb}{\@tempb:\@tempc}{\expandafter \expandafter \csname
  mn@eprint@\@tempb\endcsname \expandafter{\@tempc}}}

\bibitem[\protect\citeauthoryear{{Abazajian} et~al.,}{{Abazajian}
  et~al.}{2005}]{SDSS-DR3}
{Abazajian} K.,  et~al., 2005, \mn@doi [\aj] {10.1086/427544}, \href
  {https://ui.adsabs.harvard.edu/\#abs/2005AJ....129.1755A} {129, 1755}

\bibitem[\protect\citeauthoryear{{Abolfathi} et~al.,}{{Abolfathi}
  et~al.}{2018}]{SDSS-DR14}
{Abolfathi} B.,  et~al., 2018, \mn@doi [The Astrophysical Journal Supplement
  Series] {10.3847/1538-4365/aa9e8a}, \href
  {https://ui.adsabs.harvard.edu/\#abs/2018ApJS..235...42A} {235, 42}

\bibitem[\protect\citeauthoryear{{Adelman-McCarthy} et~al.,}{{Adelman-McCarthy}
  et~al.}{2008}]{adelman08}
{Adelman-McCarthy} J.~K.,  et~al., 2008, \mn@doi [The Astrophysical Journal
  Supplement Series] {10.1086/524984}, \href
  {https://ui.adsabs.harvard.edu/#abs/2008ApJS..175..297A} {175, 297}

\bibitem[\protect\citeauthoryear{{Akerlof} et~al.,}{{Akerlof}
  et~al.}{2007}]{CBET1059}
{Akerlof} C.,  et~al., 2007, Central Bureau Electronic Telegrams, \href
  {https://ui.adsabs.harvard.edu/#abs/2007CBET.1059....2A} {1059, 2}

\bibitem[\protect\citeauthoryear{{Aksenov}}{{Aksenov}}{1981}]{IAUC3580}
{Aksenov} E.~P.,  1981, International Astronomical Union Circular, \href
  {https://ui.adsabs.harvard.edu/#abs/1981IAUC.3580....1A} {3580, 1}

\bibitem[\protect\citeauthoryear{{Altavilla} et~al.,}{{Altavilla}
  et~al.}{2004}]{altavilla04}
{Altavilla} G.,  et~al., 2004, \mn@doi [\mnras]
  {10.1111/j.1365-2966.2004.07616.x}, \href
  {https://ui.adsabs.harvard.edu/#abs/2004MNRAS.349.1344A} {349, 1344}

\bibitem[\protect\citeauthoryear{{Amanullah} et~al.,}{{Amanullah}
  et~al.}{2014}]{amanullah14}
{Amanullah} R.,  et~al., 2014, \mn@doi [\apj] {10.1088/2041-8205/788/2/L21},
  \href {https://ui.adsabs.harvard.edu/\#abs/2014ApJ...788L..21A} {788, L21}

\bibitem[\protect\citeauthoryear{{Andersen} et~al.,}{{Andersen}
  et~al.}{1995}]{DFOSCref}
{Andersen} J.,  et~al., 1995, The Messenger, \href
  {https://ui.adsabs.harvard.edu/\#abs/1995Msngr..79...12A} {79, 12}

\bibitem[\protect\citeauthoryear{{Angel} et~al.,}{{Angel}
  et~al.}{1979}]{BCSref}
{Angel} J.~R.~P.,  et~al., 1979, in The MMT and the Future of Ground-Based
  Astronomy, Proceedings of a Symposium held to makr the dedication of the
  Multiple Mirror Telescope at the Mount Hopkins Observatory, Arizona on May 9,
  1979. Edited by Trevor C. Weekes. SAO Special Report \#385, 1979., p.87.
  p.~87

\bibitem[\protect\citeauthoryear{{Antognini} et~al.,}{{Antognini}
  et~al.}{2014}]{antogini14}
{Antognini} J.~M.,  et~al., 2014, \mn@doi [\mnras] {10.1093/mnras/stu039},
  \href {https://ui.adsabs.harvard.edu/\#abs/2014MNRAS.439.1079A} {439, 1079}

\bibitem[\protect\citeauthoryear{{Anupama} et~al.,}{{Anupama}
  et~al.}{2005}]{anupama05}
{Anupama} G.~C.,  et~al., 2005, \mn@doi [\aap] {10.1051/0004-6361:20041687},
  \href {https://ui.adsabs.harvard.edu/#abs/2005A&A...429..667A} {429, 667}

\bibitem[\protect\citeauthoryear{{Appenzeller} et~al.,}{{Appenzeller}
  et~al.}{1998}]{FORS1ref}
{Appenzeller} I.,  et~al., 1998, The Messenger, \href
  {https://ui.adsabs.harvard.edu/#abs/1998Msngr..94....1A} {94, 1}

\bibitem[\protect\citeauthoryear{{Arbour} et~al.,}{{Arbour}
  et~al.}{1999}]{IAUC7156}
{Arbour} R.,  et~al., 1999, International Astronomical Union Circular, \href
  {https://ui.adsabs.harvard.edu/#abs/1999IAUC.7156....1A} {7156, 1}

\bibitem[\protect\citeauthoryear{{Arbour} et~al.,}{{Arbour}
  et~al.}{2004}]{IAUC8406}
{Arbour} R.,  et~al., 2004, International Astronomical Union Circular, \href
  {https://ui.adsabs.harvard.edu/#abs/2004IAUC.8406....1A} {8406, 1}

\bibitem[\protect\citeauthoryear{{Arcavi} et~al.,}{{Arcavi}
  et~al.}{2014}]{ATel6661}
{Arcavi} I.,  et~al., 2014, The Astronomer's Telegram, \href
  {https://ui.adsabs.harvard.edu/#abs/2014ATel.6661....1A} {6661, 1}

\bibitem[\protect\citeauthoryear{{Ardeberg} et~al.,}{{Ardeberg} \& {de
  Groot}}{1973}]{ardeburg73}
{Ardeberg} A.,  et~al., 1973, \aap, \href
  {https://ui.adsabs.harvard.edu/\#abs/1973A&A....28..295A} {28, 295}

\bibitem[\protect\citeauthoryear{{Argyle} et~al.,}{{Argyle}
  et~al.}{1994}]{IAUC5976}
{Argyle} R.~W.,  et~al., 1994, International Astronomical Union Circular, \href
  {https://ui.adsabs.harvard.edu/\#abs/1994IAUC.5976....3A} {5976, 3}

\bibitem[\protect\citeauthoryear{{Armstrong} et~al.,}{{Armstrong} \&
  {Schwartz}}{1999}]{IAUC7108}
{Armstrong} M.,  et~al., 1999, International Astronomical Union Circular, \href
  {https://ui.adsabs.harvard.edu/#abs/1999IAUC.7108....1A} {7108, 1}

\bibitem[\protect\citeauthoryear{{Arnett}}{{Arnett}}{1982}]{arnett82}
{Arnett} W.~D.,  1982, \mn@doi [\apj] {10.1086/159681}, \href
  {https://ui.adsabs.harvard.edu/abs/1982ApJ...253..785A} {253, 785}

\bibitem[\protect\citeauthoryear{{Ayani} et~al.,}{{Ayani} \&
  {Yamaoka}}{1998}]{IAUC6878}
{Ayani} K.,  et~al., 1998, International Astronomical Union Circular, \href
  {https://ui.adsabs.harvard.edu/#abs/1998IAUC.6878....2A} {6878, 2}

\bibitem[\protect\citeauthoryear{{Ayani} et~al.,}{{Ayani}
  et~al.}{1998}]{IAUC6905}
{Ayani} K.,  et~al., 1998, International Astronomical Union Circular, \href
  {https://ui.adsabs.harvard.edu/#abs/1998IAUC.6905....1A} {6905, 1}

\bibitem[\protect\citeauthoryear{{Bacon} et~al.,}{{Bacon}
  et~al.}{2010}]{MUSEref}
{Bacon} R.,  et~al., 2010, in Proceedings of the SPIE, Volume 7735, id. 773508
  (2010).. , \mn@doi{10.1117/12.856027}

\bibitem[\protect\citeauthoryear{{Barbon} et~al.,}{{Barbon}
  et~al.}{1972}]{IAUC2411}
{Barbon} R.,  et~al., 1972, International Astronomical Union Circular, \href
  {https://ui.adsabs.harvard.edu/\#abs/1972IAUC.2411....1B} {2411, 1}

\bibitem[\protect\citeauthoryear{{Barbon} et~al.,}{{Barbon}
  et~al.}{1982}]{barbon82}
{Barbon} R.,  et~al., 1982, \aap, \href
  {https://ui.adsabs.harvard.edu/#abs/1982A&A...116...35B} {116, 35}

\bibitem[\protect\citeauthoryear{{Barbon} et~al.,}{{Barbon}
  et~al.}{1990}]{barbon90}
{Barbon} R.,  et~al., 1990, \aap, \href
  {https://ui.adsabs.harvard.edu/\#abs/1990A&A...237...79B} {237, 79}

\bibitem[\protect\citeauthoryear{{Benetti} et~al.,}{{Benetti}
  et~al.}{1995}]{IAUC6135}
{Benetti} S.,  et~al., 1995, International Astronomical Union Circular, \href
  {https://ui.adsabs.harvard.edu/#abs/1995IAUC.6135....1B} {6135, 1}

\bibitem[\protect\citeauthoryear{{Benetti} et~al.,}{{Benetti}
  et~al.}{2004}]{benetti04}
{Benetti} S.,  et~al., 2004, \mn@doi [\mnras]
  {10.1111/j.1365-2966.2004.07357.x}, \href
  {https://ui.adsabs.harvard.edu/\#abs/2004MNRAS.348..261B} {348, 261}

\bibitem[\protect\citeauthoryear{{Benetti} et~al.,}{{Benetti}
  et~al.}{2005}]{benetti05}
{Benetti} S.,  et~al., 2005, \mn@doi [\apj] {10.1086/428608}, \href
  {https://ui.adsabs.harvard.edu/abs/2005ApJ...623.1011B} {623, 1011}

\bibitem[\protect\citeauthoryear{{Benetti} et~al.,}{{Benetti}
  et~al.}{2017}]{ATel11036}
{Benetti} S.,  et~al., 2017, The Astronomer's Telegram, \href
  {https://ui.adsabs.harvard.edu/#abs/2017ATel11036....1B} {11036, 1}

\bibitem[\protect\citeauthoryear{{Benn} et~al.,}{{Benn} et~al.}{2008}]{ACAMref}
{Benn} C.,  et~al., 2008, in Ground-based and Airborne Instrumentation for
  Astronomy II. Edited by McLean, Ian S.; Casali, Mark M. Proceedings of the
  SPIE, Volume 7014, article id. 70146X, 12 pp. (2008).. ,
  \mn@doi{10.1117/12.788694}

\bibitem[\protect\citeauthoryear{{Bertin} et~al.,}{{Bertin} \&
  {Arnouts}}{1996}]{sextractorref}
{Bertin} E.,  et~al., 1996, \mn@doi [Astronomy and Astrophysics Supplement
  Series] {10.1051/aas:1996164}, \href
  {https://ui.adsabs.harvard.edu/\#abs/1996A&AS..117..393B} {117, 393}

\bibitem[\protect\citeauthoryear{{Beutler} et~al.,}{{Beutler} \&
  {Li}}{2003}]{IAUC8197}
{Beutler} B.,  et~al., 2003, International Astronomical Union Circular, \href
  {https://ui.adsabs.harvard.edu/#abs/2003IAUC.8197....1B} {8197, 1}

\bibitem[\protect\citeauthoryear{{Bianco} et~al.,}{{Bianco}
  et~al.}{2011}]{bianco11}
{Bianco} F.~B.,  et~al., 2011, \mn@doi [\apj] {10.1088/0004-637X/741/1/20},
  \href {https://ui.adsabs.harvard.edu/#abs/2011ApJ...741...20B} {741, 20}

\bibitem[\protect\citeauthoryear{{Black} et~al.,}{{Black}
  et~al.}{2016}]{black16}
{Black} C.~S.,  et~al., 2016, \mn@doi [\mnras] {10.1093/mnras/stw1680}, \href
  {https://ui.adsabs.harvard.edu/abs/2016MNRAS.462..649B} {462, 649}

\bibitem[\protect\citeauthoryear{{Blakeslee} et~al.,}{{Blakeslee}
  et~al.}{2010}]{blakeslee10}
{Blakeslee} J.~P.,  et~al., 2010, \mn@doi [\apj] {10.1088/0004-637X/724/1/657},
  \href {https://ui.adsabs.harvard.edu/\#abs/2010ApJ...724..657B} {724, 657}

\bibitem[\protect\citeauthoryear{{Blanco} et~al.,}{{Blanco}
  et~al.}{1980}]{IAUC3556}
{Blanco} V.~M.,  et~al., 1980, International Astronomical Union Circular, \href
  {https://ui.adsabs.harvard.edu/\#abs/1980IAUC.3556....2B} {3556, 2}

\bibitem[\protect\citeauthoryear{{Blanco} et~al.,}{{Blanco}
  et~al.}{1986}]{IAUC4224}
{Blanco} V.~M.,  et~al., 1986, International Astronomical Union Circular, \href
  {https://ui.adsabs.harvard.edu/#abs/1986IAUC.4224....1B} {4224, 1}

\bibitem[\protect\citeauthoryear{{Blondin} et~al.,}{{Blondin} \&
  {Berlind}}{2008}]{CBET1198}
{Blondin} S.,  et~al., 2008, Central Bureau Electronic Telegrams, \href
  {https://ui.adsabs.harvard.edu/\#abs/2008CBET.1198....1B} {1198, 1}

\bibitem[\protect\citeauthoryear{{Blondin} et~al.,}{{Blondin}
  et~al.}{2012}]{blondin12}
{Blondin} S.,  et~al., 2012, \mn@doi [\aj] {10.1088/0004-6256/143/5/126}, \href
  {https://ui.adsabs.harvard.edu/#abs/2012AJ....143..126B} {143}

\bibitem[\protect\citeauthoryear{{Boehner} et~al.,}{{Boehner}
  et~al.}{2017}]{boehner17}
{Boehner} P.,  et~al., 2017, \mn@doi [\mnras] {10.1093/mnras/stw2737}, \href
  {https://ui.adsabs.harvard.edu/#abs/2017MNRAS.465.2060B} {465, 2060}

\bibitem[\protect\citeauthoryear{{Boty{\'a}nszki} et~al.,}{{Boty{\'a}nszki} \&
  {Kasen}}{2017}]{botyanszki17}
{Boty{\'a}nszki} J.,  et~al., 2017, \mn@doi [\apj] {10.3847/1538-4357/aa81d8},
  \href {https://ui.adsabs.harvard.edu/#abs/2017ApJ...845..176B} {845}

\bibitem[\protect\citeauthoryear{{Boty{\'a}nszki} et~al.,}{{Boty{\'a}nszki}
  et~al.}{2018}]{botyanszki18}
{Boty{\'a}nszki} J.,  et~al., 2018, \mn@doi [\apj] {10.3847/2041-8213/aaa07b},
  \href {https://ui.adsabs.harvard.edu/#abs/2018ApJ...852L...6B} {852}

\bibitem[\protect\citeauthoryear{{Branch} et~al.,}{{Branch}
  et~al.}{1993}]{branch93}
{Branch} D.,  et~al., 1993, \mn@doi [\aj] {10.1086/116810}, \href
  {https://ui.adsabs.harvard.edu/\#abs/1993AJ....106.2383B} {106, 2383}

\bibitem[\protect\citeauthoryear{{Breare} et~al.,}{{Breare}
  et~al.}{1987}]{INGFOSref}
{Breare} J.~M.,  et~al., 1987, \mn@doi [\mnras] {10.1093/mnras/227.4.909},
  \href {https://ui.adsabs.harvard.edu/#abs/1987MNRAS.227..909B} {227, 909}

\bibitem[\protect\citeauthoryear{{Brimacombe} et~al.,}{{Brimacombe}
  et~al.}{2011}]{CBET2928}
{Brimacombe} J.,  et~al., 2011, Central Bureau Electronic Telegrams, \href
  {https://ui.adsabs.harvard.edu/\#abs/2011CBET.2928....1B} {2928, 1}

\bibitem[\protect\citeauthoryear{{Brimacombe} et~al.,}{{Brimacombe}
  et~al.}{2014a}]{ATel6737}
{Brimacombe} J.,  et~al., 2014a, The Astronomer's Telegram, \href
  {https://ui.adsabs.harvard.edu/\#abs/2014ATel.6737....1B} {6737, 1}

\bibitem[\protect\citeauthoryear{{Brimacombe} et~al.,}{{Brimacombe}
  et~al.}{2014b}]{ATel6803}
{Brimacombe} J.,  et~al., 2014b, The Astronomer's Telegram, \href
  {https://ui.adsabs.harvard.edu/\#abs/2014ATel.6803....1B} {6803, 1}

\bibitem[\protect\citeauthoryear{{Brimacombe} et~al.,}{{Brimacombe}
  et~al.}{2015}]{ATel6950}
{Brimacombe} J.,  et~al., 2015, The Astronomer's Telegram, \href
  {https://ui.adsabs.harvard.edu/\#abs/2015ATel.6950....1B} {6950, 1}

\bibitem[\protect\citeauthoryear{{Brimacombe} et~al.,}{{Brimacombe}
  et~al.}{2016}]{ATel8979}
{Brimacombe} J.,  et~al., 2016, The Astronomer's Telegram, \href
  {https://ui.adsabs.harvard.edu/#abs/2016ATel.8979....1B} {8979, 1}

\bibitem[\protect\citeauthoryear{{Brimacombe} et~al.,}{{Brimacombe}
  et~al.}{2017}]{ATel10108}
{Brimacombe} J.,  et~al., 2017, The Astronomer's Telegram, \href
  {https://ui.adsabs.harvard.edu/#abs/2017ATel10108....1B} {10108, 1}

\bibitem[\protect\citeauthoryear{{Brimacombe} et~al.,}{{Brimacombe}
  et~al.}{2018}]{ATel11521}
{Brimacombe} J.,  et~al., 2018, The Astronomer's Telegram, \href
  {https://ui.adsabs.harvard.edu/\#abs/2018ATel11521....1B} {11521, 1}

\bibitem[\protect\citeauthoryear{{Brown} et~al.,}{{Brown}
  et~al.}{2012}]{brown12}
{Brown} P.~J.,  et~al., 2012, \mn@doi [\apj] {10.1088/0004-637X/749/1/18},
  \href {https://ui.adsabs.harvard.edu/#abs/2012ApJ...749...18B} {749}

\bibitem[\protect\citeauthoryear{{Brown} et~al.,}{{Brown}
  et~al.}{2014}]{brown14}
{Brown} P.~J.,  et~al., 2014, \mn@doi [\apss] {10.1007/s10509-014-2059-8},
  \href {https://ui.adsabs.harvard.edu/#abs/2014Ap&SS.354...89B} {354, 89}

\bibitem[\protect\citeauthoryear{{Brown} et~al.,}{{Brown}
  et~al.}{2015}]{brown15}
{Brown} P.~J.,  et~al., 2015, \mn@doi [\apj] {10.1088/0004-637X/805/1/74},
  \href {https://ui.adsabs.harvard.edu/\#abs/2015ApJ...805...74B} {805, 74}

\bibitem[\protect\citeauthoryear{{Brown} et~al.,}{{Brown}
  et~al.}{2016}]{ATel9666}
{Brown} J.~S.,  et~al., 2016, The Astronomer's Telegram, \href
  {https://ui.adsabs.harvard.edu/\#abs/2016ATel.9666....1B} {9666, 1}

\bibitem[\protect\citeauthoryear{{Brown} et~al.,}{{Brown}
  et~al.}{2018}]{ATel11253}
{Brown} J.~S.,  et~al., 2018, The Astronomer's Telegram, \href
  {https://ui.adsabs.harvard.edu/\#abs/2018ATel11253....1B} {11253, 1}

\bibitem[\protect\citeauthoryear{{Buckley} et~al.,}{{Buckley}
  et~al.}{2006}]{SALTref}
{Buckley} D.~A.~H.,  et~al., 2006, in Society of Photo-Optical Instrumentation
  Engineers (SPIE) Conference Series. p. 62690A, \mn@doi{10.1117/12.673838}

\bibitem[\protect\citeauthoryear{{Bues} et~al.,}{{Bues}
  et~al.}{1986}]{IAUC4215}
{Bues} I.,  et~al., 1986, International Astronomical Union Circular, \href
  {https://ui.adsabs.harvard.edu/#abs/1986IAUC.4215....2B} {4215, 2}

\bibitem[\protect\citeauthoryear{{Bureau} et~al.,}{{Bureau}
  et~al.}{1996}]{bureau96}
{Bureau} M.,  et~al., 1996, \mn@doi [\apj] {10.1086/177222}, \href
  {https://ui.adsabs.harvard.edu/\#abs/1996ApJ...463...60B} {463, 60}

\bibitem[\protect\citeauthoryear{{Burns} et~al.,}{{Burns}
  et~al.}{2011}]{burns11}
{Burns} C.~R.,  et~al., 2011, \mn@doi [\aj] {10.1088/0004-6256/141/1/19}, \href
  {https://ui.adsabs.harvard.edu/#abs/2011AJ....141...19B} {141}

\bibitem[\protect\citeauthoryear{{Burns} et~al.,}{{Burns}
  et~al.}{2014}]{burns14}
{Burns} C.~R.,  et~al., 2014, \mn@doi [\apj] {10.1088/0004-637X/789/1/32},
  \href {https://ui.adsabs.harvard.edu/\#abs/2014ApJ...789...32B} {789, 32}

\bibitem[\protect\citeauthoryear{{Busko} et~al.,}{{Busko}
  et~al.}{1981}]{IAUC3589}
{Busko} I.,  et~al., 1981, International Astronomical Union Circular, \href
  {https://ui.adsabs.harvard.edu/#abs/1981IAUC.3589....2B} {3589, 2}

\bibitem[\protect\citeauthoryear{{Buta} et~al.,}{{Buta} \&
  {Turner}}{1983}]{buta83}
{Buta} R.~J.,  et~al., 1983, \mn@doi [Publications of the Astronomical Society
  of the Pacific] {10.1086/131120}, \href
  {https://ui.adsabs.harvard.edu/#abs/1983PASP...95...72B} {95, 72}

\bibitem[\protect\citeauthoryear{{Buzzoni} et~al.,}{{Buzzoni}
  et~al.}{1984}]{EFOSCref1}
{Buzzoni} B.,  et~al., 1984, The Messenger, \href
  {https://ui.adsabs.harvard.edu/#abs/1984Msngr..38....9B} {38, 9}

\bibitem[\protect\citeauthoryear{{Cacella} et~al.,}{{Cacella}
  et~al.}{2002}]{IAUC7847}
{Cacella} P.,  et~al., 2002, International Astronomical Union Circular, \href
  {https://ui.adsabs.harvard.edu/#abs/2002IAUC.7847....1C} {7847, 1}

\bibitem[\protect\citeauthoryear{{Canal} et~al.,}{{Canal}
  et~al.}{2001}]{canal01}
{Canal} R.,  et~al., 2001, \mn@doi [\apj] {10.1086/319479}, \href
  {https://ui.adsabs.harvard.edu/#abs/2001ApJ...550L..53C} {550, L53}

\bibitem[\protect\citeauthoryear{{Candia} et~al.,}{{Candia}
  et~al.}{2003}]{candia03}
{Candia} P.,  et~al., 2003, \mn@doi [Publications of the Astronomical Society
  of the Pacific] {10.1086/368229}, \href
  {https://ui.adsabs.harvard.edu/#abs/2003PASP..115..277C} {115, 277}

\bibitem[\protect\citeauthoryear{{Cappellari} et~al.,}{{Cappellari}
  et~al.}{2011}]{cappellari11}
{Cappellari} M.,  et~al., 2011, \mn@doi [\mnras]
  {10.1111/j.1365-2966.2010.18174.x}, \href
  {https://ui.adsabs.harvard.edu/#abs/2011MNRAS.413..813C} {413, 813}

\bibitem[\protect\citeauthoryear{{Cappellaro} et~al.,}{{Cappellaro}
  et~al.}{1997}]{cappellaro97}
{Cappellaro} E.,  et~al., 1997, \aap, \href
  {https://ui.adsabs.harvard.edu/abs/1997A&A...328..203C} {328, 203}

\bibitem[\protect\citeauthoryear{{Cappellaro} et~al.,}{{Cappellaro}
  et~al.}{2001}]{cappellaro01}
{Cappellaro} E.,  et~al., 2001, \mn@doi [\apj] {10.1086/319178}, \href
  {https://ui.adsabs.harvard.edu/#abs/2001ApJ...549L.215C} {549, L215}

\bibitem[\protect\citeauthoryear{{Carnall}}{{Carnall}}{2017}]{spectresref}
{Carnall} A.~C.,  2017, preprint, \href
  {https://ui.adsabs.harvard.edu/#abs/2017arXiv170505165C} {p.
  arXiv:1705.05165} (\mn@eprint {arXiv} {1705.05165})

\bibitem[\protect\citeauthoryear{{Casper} et~al.,}{{Casper}
  et~al.}{2013}]{CBET3588}
{Casper} C.,  et~al., 2013, Central Bureau Electronic Telegrams, \href
  {https://ui.adsabs.harvard.edu/#abs/2013CBET.3588....1C} {3588, 1}

\bibitem[\protect\citeauthoryear{{Catinella} et~al.,}{{Catinella}
  et~al.}{2005}]{catinella05}
{Catinella} B.,  et~al., 2005, \mn@doi [\aj] {10.1086/432543}, \href
  {https://ui.adsabs.harvard.edu/#abs/2005AJ....130.1037C} {130, 1037}

\bibitem[\protect\citeauthoryear{{Cenko} et~al.,}{{Cenko}
  et~al.}{2011}]{ATel3583}
{Cenko} S.~B.,  et~al., 2011, The Astronomer's Telegram, \href
  {https://ui.adsabs.harvard.edu/#abs/2011ATel.3583....1C} {3583, 1}

\bibitem[\protect\citeauthoryear{{Cenko} et~al.,}{{Cenko}
  et~al.}{2012}]{CBET3014}
{Cenko} S.~B.,  et~al., 2012, Central Bureau Electronic Telegrams, \href
  {https://ui.adsabs.harvard.edu/\#abs/2012CBET.3014....1C} {3014, 1}

\bibitem[\protect\citeauthoryear{{Cepa}}{{Cepa}}{2010}]{OSIRISref}
{Cepa} J.,  2010, \mn@doi [Astrophysics and Space Science Proceedings]
  {10.1007/978-3-642-11250-8_2}, \href
  {https://ui.adsabs.harvard.edu/#abs/2010ASSP...14...15C} {14, 15}

\bibitem[\protect\citeauthoryear{{Challis} et~al.,}{{Challis} \&
  {Berlind}}{2009}]{CBET2025}
{Challis} P.,  et~al., 2009, Central Bureau Electronic Telegrams, \href
  {https://ui.adsabs.harvard.edu/\#abs/2009CBET.2025....1C} {2025, 1}

\bibitem[\protect\citeauthoryear{{Chambers} et~al.,}{{Chambers}
  et~al.}{2016}]{chambers16}
{Chambers} K.~C.,  et~al., 2016, preprint, \href
  {https://ui.adsabs.harvard.edu/#abs/2016arXiv161205560C} {p.
  arXiv:1612.05560} (\mn@eprint {arXiv} {1612.05560})

\bibitem[\protect\citeauthoryear{{Chen} et~al.,}{{Chen} et~al.}{2019}]{chen19}
{Chen} P.,  et~al., 2019, \mn@doi [\apj] {10.3847/1538-4357/ab2630}, \href
  {https://ui.adsabs.harvard.edu/abs/2019ApJ...880...35C} {880, 35}

\bibitem[\protect\citeauthoryear{{Childress} et~al.,}{{Childress}
  et~al.}{2015}]{childress15}
{Childress} M.~J.,  et~al., 2015, \mn@doi [\mnras] {10.1093/mnras/stv2173},
  \href {https://ui.adsabs.harvard.edu/#abs/2015MNRAS.454.3816C} {454, 3816}

\bibitem[\protect\citeauthoryear{{Childress} et~al.,}{{Childress}
  et~al.}{2016}]{childress16}
{Childress} M.~J.,  et~al., 2016, \mn@doi [Publications of the Astronomical
  Society of Australia] {10.1017/pasa.2016.47}, \href
  {https://ui.adsabs.harvard.edu/#abs/2016PASA...33...55C} {33, e055}

\bibitem[\protect\citeauthoryear{{Chomiuk} et~al.,}{{Chomiuk}
  et~al.}{2016}]{chomiuk16}
{Chomiuk} L.,  et~al., 2016, \mn@doi [\apj] {10.3847/0004-637X/821/2/119},
  \href {https://ui.adsabs.harvard.edu/\#abs/2016ApJ...821..119C} {821, 119}

\bibitem[\protect\citeauthoryear{{Chornock} et~al.,}{{Chornock}
  et~al.}{2000}]{IAUC7463}
{Chornock} R.,  et~al., 2000, International Astronomical Union Circular, \href
  {https://ui.adsabs.harvard.edu/#abs/2000IAUC.7463....1C} {7463, 1}

\bibitem[\protect\citeauthoryear{{Christensen} et~al.,}{{Christensen}
  et~al.}{2003}]{christensen03}
{Christensen} L.,  et~al., 2003, \mn@doi [\aap] {10.1051/0004-6361:20030015},
  \href {https://ui.adsabs.harvard.edu/#abs/2003A&A...401..479C} {401, 479}

\bibitem[\protect\citeauthoryear{{Chugai}}{{Chugai}}{1986}]{chugai86}
{Chugai} N.~N.,  1986, \sovast, \href
  {https://ui.adsabs.harvard.edu/\#abs/1986SvA....30..563C} {30, 563}

\bibitem[\protect\citeauthoryear{{Chugai}}{{Chugai}}{2008}]{chugai08}
{Chugai} N.~N.,  2008, \mn@doi [Astronomy Letters] {10.1134/S1063773708060030},
  \href {https://ui.adsabs.harvard.edu/abs/2008AstL...34..389C} {34, 389}

\bibitem[\protect\citeauthoryear{{Colless} et~al.,}{{Colless}
  et~al.}{2003}]{colless03}
{Colless} M.,  et~al., 2003, arXiv e-prints, \href
  {https://ui.adsabs.harvard.edu/\#abs/2003astro.ph..6581C} {pp
  astro--ph/0306581}

\bibitem[\protect\citeauthoryear{{Collobert} et~al.,}{{Collobert}
  et~al.}{2006}]{collobert06}
{Collobert} M.,  et~al., 2006, \mn@doi [\mnras]
  {10.1111/j.1365-2966.2006.10538.x}, \href
  {https://ui.adsabs.harvard.edu/\#abs/2006MNRAS.370.1213C} {370, 1213}

\bibitem[\protect\citeauthoryear{{Contreras} et~al.,}{{Contreras}
  et~al.}{2010}]{contreras10}
{Contreras} C.,  et~al., 2010, \mn@doi [\aj] {10.1088/0004-6256/139/2/519},
  \href {https://ui.adsabs.harvard.edu/#abs/2010AJ....139..519C} {139, 519}

\bibitem[\protect\citeauthoryear{{Corsini} et~al.,}{{Corsini}
  et~al.}{2003}]{corsini03}
{Corsini} E.~M.,  et~al., 2003, \mn@doi [\apj] {10.1086/381080}, \href
  {https://ui.adsabs.harvard.edu/\#abs/2003ApJ...599L..29C} {599, L29}

\bibitem[\protect\citeauthoryear{{Cortini} et~al.,}{{Cortini}
  et~al.}{2014}]{CBET3911}
{Cortini} G.,  et~al., 2014, Central Bureau Electronic Telegrams, \href
  {https://ui.adsabs.harvard.edu/\#abs/2014CBET.3911....1C} {3911, 1}

\bibitem[\protect\citeauthoryear{{Cousins}}{{Cousins}}{1972}]{cousins72}
{Cousins} A.~W.~J.,  1972, Information Bulletin on Variable Stars, \href
  {https://ui.adsabs.harvard.edu/\#abs/1972IBVS..700....1C} {700, 1}

\bibitem[\protect\citeauthoryear{{Cox} et~al.,}{{Cox} et~al.}{2010}]{CBET2612}
{Cox} L.,  et~al., 2010, Central Bureau Electronic Telegrams, \href
  {https://ui.adsabs.harvard.edu/\#abs/2010CBET.2612....1C} {2612, 1}

\bibitem[\protect\citeauthoryear{{Cox} et~al.,}{{Cox} et~al.}{2011}]{CBET2676}
{Cox} L.,  et~al., 2011, Central Bureau Electronic Telegrams, \href
  {https://ui.adsabs.harvard.edu/\#abs/2011CBET.2676....1C} {2676, 1}

\bibitem[\protect\citeauthoryear{{Cragg} et~al.,}{{Cragg}
  et~al.}{1981}]{IAUC3583}
{Cragg} T.,  et~al., 1981, International Astronomical Union Circular, \href
  {https://ui.adsabs.harvard.edu/\#abs/1981IAUC.3583....1C} {3583, 1}

\bibitem[\protect\citeauthoryear{{Cristiani} et~al.,}{{Cristiani}
  et~al.}{1992}]{cristiani92}
{Cristiani} S.,  et~al., 1992, \aap, \href
  {https://ui.adsabs.harvard.edu/\#abs/1992A&A...259...63C} {259, 63}

\bibitem[\protect\citeauthoryear{{Cumming} et~al.,}{{Cumming}
  et~al.}{1994}]{IAUC5951}
{Cumming} R.~J.,  et~al., 1994, International Astronomical Union Circular,
  \href {https://ui.adsabs.harvard.edu/\#abs/1994IAUC.5951....1C} {5951, 1}

\bibitem[\protect\citeauthoryear{{D'Odorico}}{{D'Odorico}}{1990}]{EMMIref}
{D'Odorico} S.,  1990, The Messenger, \href
  {https://ui.adsabs.harvard.edu/\#abs/1990Msngr..61...51D} {61, 51}

\bibitem[\protect\citeauthoryear{{D'Onofrio} et~al.,}{{D'Onofrio}
  et~al.}{1995}]{donofrio95}
{D'Onofrio} M.,  et~al., 1995, \aap, \href
  {https://ui.adsabs.harvard.edu/#abs/1995A&A...296..319D} {296, 319}

\bibitem[\protect\citeauthoryear{{Delgado} et~al.,}{{Delgado}
  et~al.}{2016}]{TNSTR485}
{Delgado} A.,  et~al., 2016, Transient Name Server Discovery Report, \href
  {https://ui.adsabs.harvard.edu/#abs/2016TNSTR.485....1D} {2016-485, 1}

\bibitem[\protect\citeauthoryear{{Di Stefano} et~al.,}{{Di Stefano}
  et~al.}{2011}]{distefano11}
{Di Stefano} R.,  et~al., 2011, \mn@doi [\apj] {10.1088/2041-8205/738/1/L1},
  \href {https://ui.adsabs.harvard.edu/\#abs/2011ApJ...738L...1D} {738, L1}

\bibitem[\protect\citeauthoryear{{Dilday} et~al.,}{{Dilday}
  et~al.}{2012}]{dilday12}
{Dilday} B.,  et~al., 2012, \mn@doi [Science] {10.1126/science.1219164}, \href
  {https://ui.adsabs.harvard.edu/\#abs/2012Sci...337..942D} {337, 942}

\bibitem[\protect\citeauthoryear{{Dimitriadis} et~al.,}{{Dimitriadis}
  et~al.}{2014}]{ATel6749}
{Dimitriadis} G.,  et~al., 2014, The Astronomer's Telegram, \href
  {https://ui.adsabs.harvard.edu/#abs/2014ATel.6749....1D} {6749, 1}

\bibitem[\protect\citeauthoryear{{Dimitriadis} et~al.,}{{Dimitriadis}
  et~al.}{2019a}]{dimitriadis19}
{Dimitriadis} G.,  et~al., 2019a, \mn@doi [\apj] {10.3847/2041-8213/aaf9b1},
  \href {https://ui.adsabs.harvard.edu/\#abs/2019ApJ...870L..14D} {870, L14}

\bibitem[\protect\citeauthoryear{{Dimitriadis} et~al.,}{{Dimitriadis}
  et~al.}{2019b}]{dimitriadis18}
{Dimitriadis} G.,  et~al., 2019b, \mn@doi [\apjl] {10.3847/2041-8213/aaedb0},
  \href {https://ui.adsabs.harvard.edu/abs/2019ApJ...870L...1D} {870, L1}

\bibitem[\protect\citeauthoryear{{Dong} et~al.,}{{Dong} et~al.}{2015a}]{dong15}
{Dong} S.,  et~al., 2015a, \mn@doi [\mnras] {10.1093/mnrasl/slv129}, \href
  {https://ui.adsabs.harvard.edu/#abs/2015MNRAS.454L..61D} {454, L61}

\bibitem[\protect\citeauthoryear{{Dong} et~al.,}{{Dong}
  et~al.}{2015b}]{ATel7447}
{Dong} S.,  et~al., 2015b, The Astronomer's Telegram, \href
  {https://ui.adsabs.harvard.edu/\#abs/2015ATel.7447....1D} {7447, 1}

\bibitem[\protect\citeauthoryear{{Dong} et~al.,}{{Dong} et~al.}{2018a}]{dong18}
{Dong} S.,  et~al., 2018a, \mn@doi [\mnras] {10.1093/mnrasl/sly098}, \href
  {https://ui.adsabs.harvard.edu/\#abs/2018MNRAS.479L..70D} {479, L70}

\bibitem[\protect\citeauthoryear{{Dong} et~al.,}{{Dong}
  et~al.}{2018b}]{ATel11346}
{Dong} S.,  et~al., 2018b, The Astronomer's Telegram, \href
  {https://ui.adsabs.harvard.edu/\#abs/2018ATel11346....1D} {11346, 1}

\bibitem[\protect\citeauthoryear{{Dopita} et~al.,}{{Dopita}
  et~al.}{2007}]{WiFeSref1}
{Dopita} M.,  et~al., 2007, \mn@doi [\apss] {10.1007/s10509-007-9510-z}, \href
  {https://ui.adsabs.harvard.edu/\#abs/2007Ap&SS.310..255D} {310, 255}

\bibitem[\protect\citeauthoryear{{Dopita} et~al.,}{{Dopita}
  et~al.}{2010}]{WiFeSref2}
{Dopita} M.,  et~al., 2010, \mn@doi [\apss] {10.1007/s10509-010-0335-9}, \href
  {https://ui.adsabs.harvard.edu/\#abs/2010Ap&SS.327..245D} {327, 245}

\bibitem[\protect\citeauthoryear{{Downes} et~al.,}{{Downes}
  et~al.}{1993}]{downes93}
{Downes} D.,  et~al., 1993, \mn@doi [\apj] {10.1086/186984}, \href
  {https://ui.adsabs.harvard.edu/#abs/1993ApJ...414L..13D} {414, L13}

\bibitem[\protect\citeauthoryear{{Drake} et~al.,}{{Drake}
  et~al.}{2011a}]{CBET2636}
{Drake} A.~J.,  et~al., 2011a, Central Bureau Electronic Telegrams, \href
  {https://ui.adsabs.harvard.edu/\#abs/2011CBET.2636....1D} {2636, 1}

\bibitem[\protect\citeauthoryear{{Drake} et~al.,}{{Drake}
  et~al.}{2011b}]{CBET2703}
{Drake} A.~J.,  et~al., 2011b, Central Bureau Electronic Telegrams, \href
  {https://ui.adsabs.harvard.edu/#abs/2011CBET.2703....1D} {2703, 1}

\bibitem[\protect\citeauthoryear{{Drescher} et~al.,}{{Drescher}
  et~al.}{2012}]{CBET3346}
{Drescher} C.,  et~al., 2012, Central Bureau Electronic Telegrams, \href
  {https://ui.adsabs.harvard.edu/#abs/2012CBET.3346....1D} {3346, 1}

\bibitem[\protect\citeauthoryear{{Dressler} et~al.,}{{Dressler}
  et~al.}{2003}]{IAUC8198}
{Dressler} A.,  et~al., 2003, International Astronomical Union Circular, \href
  {https://ui.adsabs.harvard.edu/#abs/2003IAUC.8198....2D} {8198, 2}

\bibitem[\protect\citeauthoryear{{Dressler} et~al.,}{{Dressler}
  et~al.}{2011}]{IMACSref}
{Dressler} A.,  et~al., 2011, \mn@doi [Publications of the Astronomical Society
  of the Pacific] {10.1086/658908}, \href
  {https://ui.adsabs.harvard.edu/#abs/2011PASP..123..288D} {123, 288}

\bibitem[\protect\citeauthoryear{{Elias} et~al.,}{{Elias} \&
  {Frogel}}{1983}]{elias83}
{Elias} J.~H.,  et~al., 1983, \mn@doi [\apj] {10.1086/160993}, \href
  {https://ui.adsabs.harvard.edu/#abs/1983ApJ...268..718E} {268, 718}

\bibitem[\protect\citeauthoryear{{Elias-Rosa} et~al.,}{{Elias-Rosa}
  et~al.}{2005}]{IAUC8479}
{Elias-Rosa} N.,  et~al., 2005, International Astronomical Union Circular,
  \href {https://ui.adsabs.harvard.edu/\#abs/2005IAUC.8479....3E} {8479, 3}

\bibitem[\protect\citeauthoryear{{Elias-Rosa} et~al.,}{{Elias-Rosa}
  et~al.}{2006}]{eliasrosa06}
{Elias-Rosa} N.,  et~al., 2006, \mn@doi [\mnras]
  {10.1111/j.1365-2966.2006.10430.x}, \href
  {https://ui.adsabs.harvard.edu/#abs/2006MNRAS.369.1880E} {369, 1880}

\bibitem[\protect\citeauthoryear{{Epinat} et~al.,}{{Epinat}
  et~al.}{2008}]{epinat08}
{Epinat} B.,  et~al., 2008, \mn@doi [\mnras]
  {10.1111/j.1365-2966.2008.13422.x}, \href
  {https://ui.adsabs.harvard.edu/#abs/2008MNRAS.388..500E} {388, 500}

\bibitem[\protect\citeauthoryear{{Evans} et~al.,}{{Evans}
  et~al.}{1986}]{IAUC4208}
{Evans} R.,  et~al., 1986, International Astronomical Union Circular, \href
  {https://ui.adsabs.harvard.edu/#abs/1986IAUC.4208....1E} {4208, 1}

\bibitem[\protect\citeauthoryear{{Evans} et~al.,}{{Evans}
  et~al.}{2003}]{IAUC8171}
{Evans} R.,  et~al., 2003, International Astronomical Union Circular, \href
  {https://ui.adsabs.harvard.edu/\#abs/2003IAUC.8171....1E} {8171, 1}

\bibitem[\protect\citeauthoryear{{Evans} et~al.,}{{Evans}
  et~al.}{2018}]{evans18}
{Evans} D.~W.,  et~al., 2018, preprint, \href
  {https://ui.adsabs.harvard.edu/#abs/2018arXiv180409368E} {p.
  arXiv:1804.09368} (\mn@eprint {arXiv} {1804.09368})

\bibitem[\protect\citeauthoryear{{Everson} et~al.,}{{Everson}
  et~al.}{2017}]{ATel10518}
{Everson} R.~W.,  et~al., 2017, The Astronomer's Telegram, \href
  {https://ui.adsabs.harvard.edu/#abs/2017ATel10518....1E} {10518, 1}

\bibitem[\protect\citeauthoryear{{Faber} et~al.,}{{Faber}
  et~al.}{2003}]{DEIMOSref}
{Faber} S.~M.,  et~al., 2003, in Instrument Design and Performance for
  Optical/Infrared Ground-based Telescopes. Edited by Iye, Masanori; Moorwood,
  Alan F. M. Proceedings of the SPIE, Volume 4841, pp. 1657-1669 (2003).. pp
  1657--1669, \mn@doi{10.1117/12.460346}

\bibitem[\protect\citeauthoryear{{Fabricant} et~al.,}{{Fabricant}
  et~al.}{1998}]{FASTref}
{Fabricant} D.,  et~al., 1998, \mn@doi [Publications of the Astronomical
  Society of the Pacific] {10.1086/316111}, \href
  {https://ui.adsabs.harvard.edu/#abs/1998PASP..110...79F} {110, 79}

\bibitem[\protect\citeauthoryear{{Falco} et~al.,}{{Falco}
  et~al.}{1999}]{falco99}
{Falco} E.~E.,  et~al., 1999, \mn@doi [Publications of the Astronomical Society
  of the Pacific] {10.1086/316343}, \href
  {https://ui.adsabs.harvard.edu/#abs/1999PASP..111..438F} {111, 438}

\bibitem[\protect\citeauthoryear{{Feast} et~al.,}{{Feast}
  et~al.}{1986}]{IAUC4210}
{Feast} M.~W.,  et~al., 1986, International Astronomical Union Circular, \href
  {https://ui.adsabs.harvard.edu/#abs/1986IAUC.4210....1F} {4210, 1}

\bibitem[\protect\citeauthoryear{{Filippenko} et~al.,}{{Filippenko}
  et~al.}{1999}]{IAUC7108-1}
{Filippenko} A.~V.,  et~al., 1999, International Astronomical Union Circular,
  \href {https://ui.adsabs.harvard.edu/#abs/1999IAUC.7108....2F} {7108, 2}

\bibitem[\protect\citeauthoryear{{Filippenko} et~al.,}{{Filippenko}
  et~al.}{2007}]{CBET1101}
{Filippenko} A.~V.,  et~al., 2007, Central Bureau Electronic Telegrams, \href
  {https://ui.adsabs.harvard.edu/#abs/2007CBET.1101....1F} {1101, 1}

\bibitem[\protect\citeauthoryear{{Fink} et~al.,}{{Fink} et~al.}{2010}]{fink10}
{Fink} M.,  et~al., 2010, \mn@doi [\aap] {10.1051/0004-6361/200913892}, \href
  {https://ui.adsabs.harvard.edu/\#abs/2010A&A...514A..53F} {514, A53}

\bibitem[\protect\citeauthoryear{{Firth} et~al.,}{{Firth}
  et~al.}{2015}]{firth15}
{Firth} R.~E.,  et~al., 2015, \mn@doi [\mnras] {10.1093/mnras/stu2314}, \href
  {https://ui.adsabs.harvard.edu/#abs/2015MNRAS.446.3895F} {446, 3895}

\bibitem[\protect\citeauthoryear{{Fitzpatrick}}{{Fitzpatrick}}{1999}]{fitzpatrick99}
{Fitzpatrick} E.~L.,  1999, \mn@doi [Publications of the Astronomical Society
  of the Pacific] {10.1086/316293}, \href
  {https://ui.adsabs.harvard.edu/\#abs/1999PASP..111...63F} {111, 63}

\bibitem[\protect\citeauthoryear{{Flewelling} et~al.,}{{Flewelling}
  et~al.}{2016}]{flewelling16}
{Flewelling} H.~A.,  et~al., 2016, preprint, \href
  {https://ui.adsabs.harvard.edu/#abs/2016arXiv161205243F} {p.
  arXiv:1612.05243} (\mn@eprint {arXiv} {1612.05243})

\bibitem[\protect\citeauthoryear{{Folatelli} et~al.,}{{Folatelli}
  et~al.}{2010a}]{folatelli10}
{Folatelli} G.,  et~al., 2010a, \mn@doi [\aj] {10.1088/0004-6256/139/1/120},
  \href {https://ui.adsabs.harvard.edu/#abs/2010AJ....139..120F} {139, 120}

\bibitem[\protect\citeauthoryear{{Folatelli} et~al.,}{{Folatelli}
  et~al.}{2010b}]{CBET2390}
{Folatelli} G.,  et~al., 2010b, Central Bureau Electronic Telegrams, \href
  {https://ui.adsabs.harvard.edu/#abs/2010CBET.2390....1F} {2390, 1}

\bibitem[\protect\citeauthoryear{{Folatelli} et~al.,}{{Folatelli}
  et~al.}{2013}]{folatelli13}
{Folatelli} G.,  et~al., 2013, \mn@doi [\apj] {10.1088/0004-637X/773/1/53},
  \href {https://ui.adsabs.harvard.edu/#abs/2013ApJ...773...53F} {773}

\bibitem[\protect\citeauthoryear{{Foley} et~al.,}{{Foley}
  et~al.}{2013}]{foley13}
{Foley} R.~J.,  et~al., 2013, \mn@doi [\apj] {10.1088/0004-637X/767/1/57},
  \href {https://ui.adsabs.harvard.edu/\#abs/2013ApJ...767...57F} {767, 57}

\bibitem[\protect\citeauthoryear{{Foley} et~al.,}{{Foley}
  et~al.}{2014}]{foley14}
{Foley} R.~J.,  et~al., 2014, \mn@doi [\mnras] {10.1093/mnras/stu1378}, \href
  {https://ui.adsabs.harvard.edu/#abs/2014MNRAS.443.2887F} {443, 2887}

\bibitem[\protect\citeauthoryear{{Foley} et~al.,}{{Foley}
  et~al.}{2015}]{foley15}
{Foley} R.~J.,  et~al., 2015, \mn@doi [\apj] {10.1088/2041-8205/798/2/L37},
  \href {https://ui.adsabs.harvard.edu/\#abs/2015ApJ...798L..37F} {798, L37}

\bibitem[\protect\citeauthoryear{{Foley} et~al.,}{{Foley}
  et~al.}{2016}]{foley16}
{Foley} R.~J.,  et~al., 2016, \mn@doi [\mnras] {10.1093/mnras/stw1320}, \href
  {https://ui.adsabs.harvard.edu/\#abs/2016MNRAS.461..433F} {461, 433}

\bibitem[\protect\citeauthoryear{{Foley} et~al.,}{{Foley}
  et~al.}{2018}]{foley18}
{Foley} R.~J.,  et~al., 2018, \mn@doi [\mnras] {10.1093/mnras/stx3136}, \href
  {https://ui.adsabs.harvard.edu/\#abs/2018MNRAS.475..193F} {475, 193}

\bibitem[\protect\citeauthoryear{{Ford} et~al.,}{{Ford} et~al.}{1993}]{ford93}
{Ford} C.~H.,  et~al., 1993, \mn@doi [\aj] {10.1086/116708}, \href
  {https://ui.adsabs.harvard.edu/#abs/1993AJ....106.1101F} {106, 1101}

\bibitem[\protect\citeauthoryear{{Fossey} et~al.,}{{Fossey}
  et~al.}{2014}]{CBET3792}
{Fossey} S.~J.,  et~al., 2014, Central Bureau Electronic Telegrams, \href
  {https://ui.adsabs.harvard.edu/#abs/2014CBET.3792....1F} {3792, 1}

\bibitem[\protect\citeauthoryear{{Freedman} et~al.,}{{Freedman}
  et~al.}{2001}]{freedman01}
{Freedman} W.~L.,  et~al., 2001, \mn@doi [\apj] {10.1086/320638}, \href
  {https://ui.adsabs.harvard.edu/#abs/2001ApJ...553...47F} {553, 47}

\bibitem[\protect\citeauthoryear{{Freudling} et~al.,}{{Freudling}
  et~al.}{2013}]{esorexref}
{Freudling} W.,  et~al., 2013, \mn@doi [\aap] {10.1051/0004-6361/201322494},
  \href {http://adsabs.harvard.edu/abs/2013A%26A...559A..96F} {559, A96}

\bibitem[\protect\citeauthoryear{{Friedman} et~al.,}{{Friedman}
  et~al.}{2015}]{friedman15}
{Friedman} A.~S.,  et~al., 2015, \mn@doi [The Astrophysical Journal Supplement
  Series] {10.1088/0067-0049/220/1/9}, \href
  {https://ui.adsabs.harvard.edu/#abs/2015ApJS..220....9F} {220}

\bibitem[\protect\citeauthoryear{{Frieman} et~al.,}{{Frieman}
  et~al.}{2006}]{IAUC8754}
{Frieman} J.,  et~al., 2006, International Astronomical Union Circular, \href
  {https://ui.adsabs.harvard.edu/#abs/2006IAUC.8754....1F} {8754, 1}

\bibitem[\protect\citeauthoryear{{Frohmaier} et~al.,}{{Frohmaier}
  et~al.}{2015}]{ATel7452}
{Frohmaier} C.,  et~al., 2015, The Astronomer's Telegram, \href
  {https://ui.adsabs.harvard.edu/\#abs/2015ATel.7452....1F} {7452, 1}

\bibitem[\protect\citeauthoryear{{Frye} et~al.,}{{Frye}
  et~al.}{1972}]{IAUC2398}
{Frye} R.,  et~al., 1972, International Astronomical Union Circular, \href
  {https://ui.adsabs.harvard.edu/\#abs/1972IAUC.2398....1F} {2398, 1}

\bibitem[\protect\citeauthoryear{{Fukugita} et~al.,}{{Fukugita}
  et~al.}{1996}]{fukugita96}
{Fukugita} M.,  et~al., 1996, \mn@doi [\aj] {10.1086/117915}, \href
  {https://ui.adsabs.harvard.edu/#abs/1996AJ....111.1748F} {111, 1748}

\bibitem[\protect\citeauthoryear{{Gagliano} et~al.,}{{Gagliano}
  et~al.}{2016}]{2016TNSTR761}
{Gagliano} R.,  et~al., 2016, Transient Name Server Discovery Report, \href
  {https://ui.adsabs.harvard.edu/\#abs/2016TNSTR.761....1G} {2016-761, 1}

\bibitem[\protect\citeauthoryear{{Gagliano} et~al.,}{{Gagliano}
  et~al.}{2017}]{2017TNSTR961}
{Gagliano} R.,  et~al., 2017, Transient Name Server Discovery Report, \href
  {https://ui.adsabs.harvard.edu/\#abs/2017TNSTR.961....1G} {2017-961, 1}

\bibitem[\protect\citeauthoryear{{Gaia Collaboration} et~al.,}{{Gaia
  Collaboration} et~al.}{2016}]{Gaiaref}
{Gaia Collaboration} et~al., 2016, \mn@doi [\aap]
  {10.1051/0004-6361/201629272}, \href
  {https://ui.adsabs.harvard.edu/#abs/2016A&A...595A...1G} {595, A1}

\bibitem[\protect\citeauthoryear{{Gaia Collaboration} et~al.,}{{Gaia
  Collaboration} et~al.}{2018}]{GaiaDR2ref}
{Gaia Collaboration} et~al., 2018, preprint, \href
  {https://ui.adsabs.harvard.edu/#abs/2018arXiv180409365G} {p.
  arXiv:1804.09365} (\mn@eprint {arXiv} {1804.09365})

\bibitem[\protect\citeauthoryear{{Galbany} et~al.,}{{Galbany}
  et~al.}{2014}]{galbany14}
{Galbany} L.,  et~al., 2014, \mn@doi [\aap] {10.1051/0004-6361/201424717},
  \href {https://ui.adsabs.harvard.edu/#abs/2014A&A...572A..38G} {572, A38}

\bibitem[\protect\citeauthoryear{{Galbany} et~al.,}{{Galbany}
  et~al.}{2016}]{galbany16}
{Galbany} L.,  et~al., 2016, \mn@doi [\mnras] {10.1093/mnras/stw026}, \href
  {https://ui.adsabs.harvard.edu/#abs/2016MNRAS.457..525G} {457, 525}

\bibitem[\protect\citeauthoryear{{Gall} et~al.,}{{Gall} et~al.}{2018}]{gall18}
{Gall} C.,  et~al., 2018, \mn@doi [\aap] {10.1051/0004-6361/201730886}, \href
  {https://ui.adsabs.harvard.edu/\#abs/2018A&A...611A..58G} {611, A58}

\bibitem[\protect\citeauthoryear{{Ganeshalingam} et~al.,}{{Ganeshalingam}
  et~al.}{2010}]{ganeshalingam10}
{Ganeshalingam} M.,  et~al., 2010, \mn@doi [The Astrophysical Journal
  Supplement Series] {10.1088/0067-0049/190/2/418}, \href
  {https://ui.adsabs.harvard.edu/\#abs/2010ApJS..190..418G} {190, 418}

\bibitem[\protect\citeauthoryear{{Gao} et~al.,}{{Gao} et~al.}{2015}]{gao15}
{Gao} J.,  et~al., 2015, \mn@doi [\apj] {10.1088/2041-8205/807/2/L26}, \href
  {https://ui.adsabs.harvard.edu/\#abs/2015ApJ...807L..26G} {807, L26}

\bibitem[\protect\citeauthoryear{{Garnavich} et~al.,}{{Garnavich}
  et~al.}{2004}]{garnavich04}
{Garnavich} P.~M.,  et~al., 2004, \mn@doi [\apj] {10.1086/422986}, \href
  {https://ui.adsabs.harvard.edu/#abs/2004ApJ...613.1120G} {613, 1120}

\bibitem[\protect\citeauthoryear{{Garradd} et~al.,}{{Garradd}
  et~al.}{1996}]{IAUC6380}
{Garradd} G.~J.,  et~al., 1996, International Astronomical Union Circular,
  \href {https://ui.adsabs.harvard.edu/#abs/1996IAUC.6380....1G} {6380, 1}

\bibitem[\protect\citeauthoryear{{Gaskell} et~al.,}{{Gaskell}
  et~al.}{1989}]{IAUC4761}
{Gaskell} C.~M.,  et~al., 1989, International Astronomical Union Circular,
  \href {https://ui.adsabs.harvard.edu/\#abs/1989IAUC.4761....2G} {4761, 2}

\bibitem[\protect\citeauthoryear{{Gerardy} et~al.,}{{Gerardy} \&
  {Fesen}}{1999}]{IAUC7158}
{Gerardy} C.,  et~al., 1999, International Astronomical Union Circular, \href
  {https://ui.adsabs.harvard.edu/#abs/1999IAUC.7158....2G} {7158, 2}

\bibitem[\protect\citeauthoryear{{Gilfanov} et~al.,}{{Gilfanov} \&
  {Bogd{\'a}n}}{2010}]{gilfanov10}
{Gilfanov} M.,  et~al., 2010, \mn@doi [\nat] {10.1038/nature08685}, \href
  {https://ui.adsabs.harvard.edu/#abs/2010Natur.463..924G} {463, 924}

\bibitem[\protect\citeauthoryear{{Gilmore}}{{Gilmore}}{1991}]{IAUC5309}
{Gilmore} A.~C.,  1991, International Astronomical Union Circular, \href
  {https://ui.adsabs.harvard.edu/#abs/1991IAUC.5309....3G} {5309, 3}

\bibitem[\protect\citeauthoryear{{Ginsburg} et~al.,}{{Ginsburg}
  et~al.}{2019}]{astroqueryref}
{Ginsburg} A.,  et~al., 2019, arXiv e-prints, \href
  {https://ui.adsabs.harvard.edu/\#abs/2019arXiv190104520G} {p.
  arXiv:1901.04520}

\bibitem[\protect\citeauthoryear{{Giovanelli} et~al.,}{{Giovanelli}
  et~al.}{1997}]{giovanelli97}
{Giovanelli} R.,  et~al., 1997, \mn@doi [\aj] {10.1086/118233}, \href
  {https://ui.adsabs.harvard.edu/\#abs/1997AJ....113...22G} {113, 22}

\bibitem[\protect\citeauthoryear{{Gomez} et~al.,}{{Gomez}
  et~al.}{1996}]{gomez96}
{Gomez} G.,  et~al., 1996, \mn@doi [\aj] {10.1086/118166}, \href
  {https://ui.adsabs.harvard.edu/#abs/1996AJ....112.2094G} {112, 2094}

\bibitem[\protect\citeauthoryear{{Gonzalez} et~al.,}{{Gonzalez}
  et~al.}{2004}]{IAUC8409}
{Gonzalez} S.,  et~al., 2004, International Astronomical Union Circular, \href
  {https://ui.adsabs.harvard.edu/#abs/2004IAUC.8409....2G} {8409, 2}

\bibitem[\protect\citeauthoryear{{Goobar} et~al.,}{{Goobar}
  et~al.}{2014}]{goobar14}
{Goobar} A.,  et~al., 2014, \mn@doi [\apj] {10.1088/2041-8205/784/1/L12}, \href
  {https://ui.adsabs.harvard.edu/#abs/2014ApJ...784L..12G} {784}

\bibitem[\protect\citeauthoryear{{Graham} et~al.,}{{Graham}
  et~al.}{1978}]{graham78}
{Graham} D.~A.,  et~al., 1978, \aap, \href
  {https://ui.adsabs.harvard.edu/#abs/1978A&A....70L..69G} {70, L69}

\bibitem[\protect\citeauthoryear{{Graham} et~al.,}{{Graham}
  et~al.}{1998}]{graham98}
{Graham} A.~W.,  et~al., 1998, \mn@doi [Astronomy and Astrophysics Supplement
  Series] {10.1051/aas:1998325}, \href
  {https://ui.adsabs.harvard.edu/#abs/1998A&AS..133..325G} {133, 325}

\bibitem[\protect\citeauthoryear{{Graham} et~al.,}{{Graham}
  et~al.}{2017}]{graham17}
{Graham} M.~L.,  et~al., 2017, \mn@doi [\mnras] {10.1093/mnras/stx2224}, \href
  {https://ui.adsabs.harvard.edu/#abs/2017MNRAS.472.3437G} {472, 3437}

\bibitem[\protect\citeauthoryear{{Graham} et~al.,}{{Graham}
  et~al.}{2019}]{graham19}
{Graham} M.~L.,  et~al., 2019, \mn@doi [\apj] {10.3847/1538-4357/aaf41e}, \href
  {https://ui.adsabs.harvard.edu/\#abs/2019ApJ...871...62G} {871, 62}

\bibitem[\protect\citeauthoryear{{Grogin} et~al.,}{{Grogin}
  et~al.}{1998}]{grogin98}
{Grogin} N.~A.,  et~al., 1998, \mn@doi [The Astrophysical Journal Supplement
  Series] {10.1086/313164}, \href
  {https://ui.adsabs.harvard.edu/#abs/1998ApJS..119..277G} {119, 277}

\bibitem[\protect\citeauthoryear{{Guillochon} et~al.,}{{Guillochon}
  et~al.}{2017}]{OSCref}
{Guillochon} J.,  et~al., 2017, \mn@doi [\apj] {10.3847/1538-4357/835/1/64},
  \href {https://ui.adsabs.harvard.edu/#abs/2017ApJ...835...64G} {835}

\bibitem[\protect\citeauthoryear{{Guthrie} et~al.,}{{Guthrie} \&
  {Napier}}{1996}]{guthrie96}
{Guthrie} B.~N.~G.,  et~al., 1996, \aap, \href
  {https://ui.adsabs.harvard.edu/#abs/1996A&A...310..353G} {310, 353}

\bibitem[\protect\citeauthoryear{{Guti{\'e}rrez} et~al.,}{{Guti{\'e}rrez}
  et~al.}{2016}]{gutierrez16}
{Guti{\'e}rrez} C.~P.,  et~al., 2016, \mn@doi [\aap]
  {10.1051/0004-6361/201527228}, \href
  {https://ui.adsabs.harvard.edu/\#abs/2016A&A...590A...5G} {590, A5}

\bibitem[\protect\citeauthoryear{{Halevi} et~al.,}{{Halevi}
  et~al.}{2016}]{ATel9309}
{Halevi} G.,  et~al., 2016, The Astronomer's Telegram, \href
  {https://ui.adsabs.harvard.edu/#abs/2016ATel.9309....1H} {9309, 1}

\bibitem[\protect\citeauthoryear{{Hamuy} et~al.,}{{Hamuy}
  et~al.}{1991}]{hamuy91}
{Hamuy} M.,  et~al., 1991, \mn@doi [\aj] {10.1086/115867}, \href
  {https://ui.adsabs.harvard.edu/\#abs/1991AJ....102..208H} {102, 208}

\bibitem[\protect\citeauthoryear{{Hamuy} et~al.,}{{Hamuy}
  et~al.}{1996}]{hamuy96}
{Hamuy} M.,  et~al., 1996, \mn@doi [\aj] {10.1086/118192}, \href
  {https://ui.adsabs.harvard.edu/\#abs/1996AJ....112.2408H} {112, 2408}

\bibitem[\protect\citeauthoryear{{Hamuy} et~al.,}{{Hamuy}
  et~al.}{2006}]{CSPref}
{Hamuy} M.,  et~al., 2006, \mn@doi [Publications of the Astronomical Society of
  the Pacific] {10.1086/500228}, \href
  {https://ui.adsabs.harvard.edu/#abs/2006PASP..118....2H} {118, 2}

\bibitem[\protect\citeauthoryear{{Han} et~al.,}{{Han} \&
  {Podsiadlowski}}{2006}]{han06}
{Han} Z.,  et~al., 2006, \mn@doi [\mnras] {10.1111/j.1365-2966.2006.10185.x},
  \href {https://ui.adsabs.harvard.edu/abs/2006MNRAS.368.1095H} {368, 1095}

\bibitem[\protect\citeauthoryear{{Harutyunyan} et~al.,}{{Harutyunyan}
  et~al.}{2007}]{CBET1021}
{Harutyunyan} A.,  et~al., 2007, Central Bureau Electronic Telegrams, \href
  {https://ui.adsabs.harvard.edu/abs/2007CBET.1021....1H} {1021, 1}

\bibitem[\protect\citeauthoryear{{Harutyunyan} et~al.,}{{Harutyunyan}
  et~al.}{2009}]{CBET1768}
{Harutyunyan} A.,  et~al., 2009, Central Bureau Electronic Telegrams, \href
  {https://ui.adsabs.harvard.edu/#abs/2009CBET.1768....1H} {1768, 1}

\bibitem[\protect\citeauthoryear{{Hayden} et~al.,}{{Hayden}
  et~al.}{2010}]{hayden10}
{Hayden} B.~T.,  et~al., 2010, \mn@doi [\apj] {10.1088/0004-637X/722/2/1691},
  \href {https://ui.adsabs.harvard.edu/\#abs/2010ApJ...722.1691H} {722, 1691}

\bibitem[\protect\citeauthoryear{{Heraudeau} et~al.,}{{Heraudeau}
  et~al.}{1994}]{IAUC5952}
{Heraudeau} P.,  et~al., 1994, International Astronomical Union Circular, \href
  {https://ui.adsabs.harvard.edu/\#abs/1994IAUC.5952....3H} {5952, 3}

\bibitem[\protect\citeauthoryear{{Herbig} et~al.,}{{Herbig}
  et~al.}{1972}]{IAUC2407}
{Herbig} G.~H.,  et~al., 1972, International Astronomical Union Circular, \href
  {https://ui.adsabs.harvard.edu/#abs/1972IAUC.2407R...1H} {2407, 1}

\bibitem[\protect\citeauthoryear{{Hernandez} et~al.,}{{Hernandez}
  et~al.}{2000}]{hernandez00}
{Hernandez} M.,  et~al., 2000, \mn@doi [\mnras]
  {10.1046/j.1365-8711.2000.03841.x}, \href
  {https://ui.adsabs.harvard.edu/#abs/2000MNRAS.319..223H} {319, 223}

\bibitem[\protect\citeauthoryear{{Hicken} et~al.,}{{Hicken}
  et~al.}{2007}]{hicken07}
{Hicken} M.,  et~al., 2007, \mn@doi [\apj] {10.1086/523301}, \href
  {https://ui.adsabs.harvard.edu/#abs/2007ApJ...669L..17H} {669, L17}

\bibitem[\protect\citeauthoryear{{Hicken} et~al.,}{{Hicken}
  et~al.}{2009}]{hicken09}
{Hicken} M.,  et~al., 2009, \mn@doi [\apj] {10.1088/0004-637X/700/1/331}, \href
  {https://ui.adsabs.harvard.edu/#abs/2009ApJ...700..331H} {700, 331}

\bibitem[\protect\citeauthoryear{{Hicken} et~al.,}{{Hicken}
  et~al.}{2012}]{hicken12}
{Hicken} M.,  et~al., 2012, \mn@doi [The Astrophysical Journal Supplement
  Series] {10.1088/0067-0049/200/2/12}, \href
  {https://ui.adsabs.harvard.edu/#abs/2012ApJS..200...12H} {200}

\bibitem[\protect\citeauthoryear{{Ho} et~al.,}{{Ho} et~al.}{2001}]{ho01}
{Ho} W. C.~G.,  et~al., 2001, \mn@doi [Publications of the Astronomical Society
  of the Pacific] {10.1086/323970}, \href
  {https://ui.adsabs.harvard.edu/#abs/2001PASP..113.1349H} {113, 1349}

\bibitem[\protect\citeauthoryear{{H{\"o}flich} et~al.,}{{H{\"o}flich}
  et~al.}{2004}]{hoflich04}
{H{\"o}flich} P.,  et~al., 2004, \mn@doi [\apj] {10.1086/425571}, \href
  {https://ui.adsabs.harvard.edu/\#abs/2004ApJ...617.1258H} {617, 1258}

\bibitem[\protect\citeauthoryear{{Holberg} et~al.,}{{Holberg}
  et~al.}{1991}]{IAUC5270}
{Holberg} J.,  et~al., 1991, International Astronomical Union Circular, \href
  {https://ui.adsabs.harvard.edu/#abs/1991IAUC.5270....3H} {5270, 3}

\bibitem[\protect\citeauthoryear{{Holmbo} et~al.,}{{Holmbo}
  et~al.}{2018}]{holmbo18}
{Holmbo} S.,  et~al., 2018, arXiv e-prints, \href
  {https://ui.adsabs.harvard.edu/\#abs/2018arXiv180901359H} {p.
  arXiv:1809.01359}

\bibitem[\protect\citeauthoryear{{Holoien} et~al.,}{{Holoien}
  et~al.}{2014}]{ATel6637}
{Holoien} T.~W.~S.,  et~al., 2014, The Astronomer's Telegram, \href
  {https://ui.adsabs.harvard.edu/#abs/2014ATel.6637....1H} {6637, 1}

\bibitem[\protect\citeauthoryear{{Holoien} et~al.,}{{Holoien}
  et~al.}{2017a}]{ASASSNcat14}
{Holoien} T.~W.~S.,  et~al., 2017a, \mn@doi [\mnras] {10.1093/mnras/stw2273},
  \href {https://ui.adsabs.harvard.edu/\#abs/2017MNRAS.464.2672H} {464, 2672}

\bibitem[\protect\citeauthoryear{{Holoien} et~al.,}{{Holoien}
  et~al.}{2017b}]{ASASSNcat15}
{Holoien} T.~W.~S.,  et~al., 2017b, \mn@doi [\mnras] {10.1093/mnras/stx057},
  \href {https://ui.adsabs.harvard.edu/\#abs/2017MNRAS.467.1098H} {467, 1098}

\bibitem[\protect\citeauthoryear{{Holoien} et~al.,}{{Holoien}
  et~al.}{2017c}]{ASASSNcat16}
{Holoien} T.~W.~S.,  et~al., 2017c, \mn@doi [\mnras] {10.1093/mnras/stx1544},
  \href {https://ui.adsabs.harvard.edu/\#abs/2017MNRAS.471.4966H} {471, 4966}

\bibitem[\protect\citeauthoryear{{Holoien} et~al.,}{{Holoien}
  et~al.}{2019}]{ASASSNcat17}
{Holoien} T.~W.~S.,  et~al., 2019, \mn@doi [\mnras] {10.1093/mnras/stz073},
  \href {https://ui.adsabs.harvard.edu/\#abs/2019MNRAS.tmp...93H} {p.~93}

\bibitem[\protect\citeauthoryear{{Holtzman} et~al.,}{{Holtzman}
  et~al.}{2008}]{holtzman08}
{Holtzman} J.~A.,  et~al., 2008, \mn@doi [\aj] {10.1088/0004-6256/136/6/2306},
  \href {https://ui.adsabs.harvard.edu/\#abs/2008AJ....136.2306H} {136, 2306}

\bibitem[\protect\citeauthoryear{{Hook} et~al.,}{{Hook} et~al.}{2004}]{GMOSref}
{Hook} I.~M.,  et~al., 2004, \mn@doi [Publications of the Astronomical Society
  of the Pacific] {10.1086/383624}, \href
  {https://ui.adsabs.harvard.edu/#abs/2004PASP..116..425H} {116, 425}

\bibitem[\protect\citeauthoryear{{Hosseinzadeh} et~al.,}{{Hosseinzadeh}
  et~al.}{2017a}]{ATel10164}
{Hosseinzadeh} G.,  et~al., 2017a, The Astronomer's Telegram, \href
  {https://ui.adsabs.harvard.edu/#abs/2017ATel10164....1H} {10164, 1}

\bibitem[\protect\citeauthoryear{{Hosseinzadeh} et~al.,}{{Hosseinzadeh}
  et~al.}{2017b}]{ATel10639}
{Hosseinzadeh} G.,  et~al., 2017b, The Astronomer's Telegram, \href
  {https://ui.adsabs.harvard.edu/#abs/2017ATel10639....1H} {10639, 1}

\bibitem[\protect\citeauthoryear{{Howell}}{{Howell}}{2001}]{howell01}
{Howell} D.~A.,  2001, \mn@doi [\apjl] {10.1086/321702}, \href
  {https://ui.adsabs.harvard.edu/abs/2001ApJ...554L.193H} {554, L193}

\bibitem[\protect\citeauthoryear{{Howell} et~al.,}{{Howell}
  et~al.}{2006}]{howell06}
{Howell} D.~A.,  et~al., 2006, \mn@doi [\nat] {10.1038/nature05103}, \href
  {https://ui.adsabs.harvard.edu/\#abs/2006Natur.443..308H} {443, 308}

\bibitem[\protect\citeauthoryear{{Howerton} et~al.,}{{Howerton}
  et~al.}{2011}]{CBET2658}
{Howerton} S.,  et~al., 2011, Central Bureau Electronic Telegrams, \href
  {https://ui.adsabs.harvard.edu/\#abs/2011CBET.2658....1H} {2658, 1}

\bibitem[\protect\citeauthoryear{{Howerton} et~al.,}{{Howerton}
  et~al.}{2013}]{CBET3533}
{Howerton} S.,  et~al., 2013, Central Bureau Electronic Telegrams, \href
  {https://ui.adsabs.harvard.edu/#abs/2013CBET.3533....1H} {3533, 1}

\bibitem[\protect\citeauthoryear{{Hoyle} et~al.,}{{Hoyle} \&
  {Fowler}}{1960}]{hoyle60}
{Hoyle} F.,  et~al., 1960, \mn@doi [\apj] {10.1086/146963}, \href
  {https://ui.adsabs.harvard.edu/\#abs/1960ApJ...132..565H} {132, 565}

\bibitem[\protect\citeauthoryear{{Huang} et~al.,}{{Huang}
  et~al.}{2017}]{huang17}
{Huang} X.,  et~al., 2017, \mn@doi [\apj] {10.3847/1538-4357/836/2/157}, \href
  {https://ui.adsabs.harvard.edu/#abs/2017ApJ...836..157H} {836, 157}

\bibitem[\protect\citeauthoryear{{Huchra} et~al.,}{{Huchra}
  et~al.}{1999}]{huchra99}
{Huchra} J.~P.,  et~al., 1999, \mn@doi [The Astrophysical Journal Supplement
  Series] {10.1086/313194}, \href
  {https://ui.adsabs.harvard.edu/#abs/1999ApJS..121..287H} {121, 287}

\bibitem[\protect\citeauthoryear{{Hurst} et~al.,}{{Hurst}
  et~al.}{1998}]{IAUC6875}
{Hurst} G.~M.,  et~al., 1998, International Astronomical Union Circular, \href
  {https://ui.adsabs.harvard.edu/#abs/1998IAUC.6875....1H} {6875, 1}

\bibitem[\protect\citeauthoryear{{Hutchings} et~al.,}{{Hutchings} \&
  {Li}}{2002}]{IAUC7918}
{Hutchings} D.,  et~al., 2002, International Astronomical Union Circular, \href
  {https://ui.adsabs.harvard.edu/#abs/2002IAUC.7918....1H} {7918, 1}

\bibitem[\protect\citeauthoryear{{Iben} et~al.,}{{Iben} \&
  {Tutukov}}{1984}]{iben84}
{Iben} I.,  et~al., 1984, \mn@doi [The Astrophysical Journal Supplement Series]
  {10.1086/190932}, \href
  {https://ui.adsabs.harvard.edu/#abs/1984ApJS...54..335I} {54, 335}

\bibitem[\protect\citeauthoryear{{Iijima} et~al.,}{{Iijima}
  et~al.}{1994}]{IAUC6108}
{Iijima} T.,  et~al., 1994, International Astronomical Union Circular, \href
  {https://ui.adsabs.harvard.edu/#abs/1994IAUC.6108....1I} {6108, 1}

\bibitem[\protect\citeauthoryear{{Inserra} et~al.,}{{Inserra}
  et~al.}{2013}]{inserra13}
{Inserra} C.,  et~al., 2013, \mn@doi [\apj] {10.1088/0004-637X/770/2/128},
  \href {https://ui.adsabs.harvard.edu/#abs/2013ApJ...770..128I} {770}

\bibitem[\protect\citeauthoryear{{Itagaki} et~al.,}{{Itagaki}
  et~al.}{2014}]{CBET3792-1}
{Itagaki} K.,  et~al., 2014, Central Bureau Electronic Telegrams, \href
  {https://ui.adsabs.harvard.edu/#abs/2014CBET.3792....2I} {3792, 2}

\bibitem[\protect\citeauthoryear{{Jha} et~al.,}{{Jha} et~al.}{1999}]{jha99}
{Jha} S.,  et~al., 1999, \mn@doi [The Astrophysical Journal Supplement Series]
  {10.1086/313275}, \href
  {https://ui.adsabs.harvard.edu/#abs/1999ApJS..125...73J} {125, 73}

\bibitem[\protect\citeauthoryear{{Jha} et~al.,}{{Jha} et~al.}{2006}]{jha06}
{Jha} S.,  et~al., 2006, \mn@doi [\aj] {10.1086/497989}, \href
  {https://ui.adsabs.harvard.edu/#abs/2006AJ....131..527J} {131, 527}

\bibitem[\protect\citeauthoryear{{Jha} et~al.,}{{Jha} et~al.}{2017}]{ATel10261}
{Jha} S.~W.,  et~al., 2017, The Astronomer's Telegram, \href
  {https://ui.adsabs.harvard.edu/#abs/2017ATel10261....1J} {10261, 1}

\bibitem[\protect\citeauthoryear{{Jha} et~al.,}{{Jha} et~al.}{2019}]{jha19}
{Jha} S.~W.,  et~al., 2019, \mn@doi [Nature Astronomy]
  {10.1038/s41550-019-0858-0}, \href
  {https://ui.adsabs.harvard.edu/abs/2019NatAs...3..706J} {3, 706}

\bibitem[\protect\citeauthoryear{{Jones} et~al.,}{{Jones}
  et~al.}{2006}]{jones06}
{Jones} D.~H.,  et~al., 2006, \mn@doi [\mnras]
  {10.1111/j.1365-2966.2006.10291.x}, \href
  {https://ui.adsabs.harvard.edu/\#abs/2006MNRAS.369...25J} {369, 25}

\bibitem[\protect\citeauthoryear{{Jones} et~al.,}{{Jones}
  et~al.}{2009}]{jones09}
{Jones} D.~H.,  et~al., 2009, \mn@doi [\mnras]
  {10.1111/j.1365-2966.2009.15338.x}, \href
  {https://ui.adsabs.harvard.edu/#abs/2009MNRAS.399..683J} {399, 683}

\bibitem[\protect\citeauthoryear{{Jones} et~al.,}{{Jones}
  et~al.}{2017}]{ATel11092}
{Jones} S.,  et~al., 2017, The Astronomer's Telegram, \href
  {https://ui.adsabs.harvard.edu/\#abs/2017ATel11092....1J} {11092, 1}

\bibitem[\protect\citeauthoryear{{Jorden}}{{Jorden}}{1990}]{ISISref}
{Jorden} P.~R.,  1990, in Proc. SPIE Vol. 1235, p. 790-798, Instrumentation in
  Astronomy VII, David L. Crawford; Ed.. pp 790--798, \mn@doi{10.1117/12.19163}

\bibitem[\protect\citeauthoryear{{Justham}}{{Justham}}{2011}]{justham11}
{Justham} S.,  2011, \mn@doi [\apjl] {10.1088/2041-8205/730/2/L34}, \href
  {https://ui.adsabs.harvard.edu/abs/2011ApJ...730L..34J} {730, L34}

\bibitem[\protect\citeauthoryear{{Kandrashoff} et~al.,}{{Kandrashoff}
  et~al.}{2012}]{CBET3111}
{Kandrashoff} M.,  et~al., 2012, Central Bureau Electronic Telegrams, \href
  {https://ui.adsabs.harvard.edu/#abs/2012CBET.3111....1K} {3111, 1}

\bibitem[\protect\citeauthoryear{{Karamehmetoglu} et~al.,}{{Karamehmetoglu}
  et~al.}{2015}]{ATel7476}
{Karamehmetoglu} E.,  et~al., 2015, The Astronomer's Telegram, \href
  {https://ui.adsabs.harvard.edu/\#abs/2015ATel.7476....1K} {7476, 1}

\bibitem[\protect\citeauthoryear{{Kasen}}{{Kasen}}{2010}]{kasen10}
{Kasen} D.,  2010, \mn@doi [\apj] {10.1088/0004-637X/708/2/1025}, \href
  {https://ui.adsabs.harvard.edu/#abs/2010ApJ...708.1025K} {708, 1025}

\bibitem[\protect\citeauthoryear{{Kashikawa} et~al.,}{{Kashikawa}
  et~al.}{2002}]{FOCASref}
{Kashikawa} N.,  et~al., 2002, \mn@doi [Publications of the Astronomical
  Society of Japan] {10.1093/pasj/54.6.819}, \href
  {https://ui.adsabs.harvard.edu/#abs/2002PASJ...54..819K} {54, 819}

\bibitem[\protect\citeauthoryear{{Katz} et~al.,}{{Katz} \&
  {Dong}}{2012}]{katz12}
{Katz} B.,  et~al., 2012, preprint, \href
  {https://ui.adsabs.harvard.edu/#abs/2012arXiv1211.4584K} {p. arXiv:1211.4584}
  (\mn@eprint {arXiv} {1211.4584})

\bibitem[\protect\citeauthoryear{{Kawakita} et~al.,}{{Kawakita}
  et~al.}{2002}]{IAUC7848}
{Kawakita} H.,  et~al., 2002, International Astronomical Union Circular, \href
  {https://ui.adsabs.harvard.edu/#abs/2002IAUC.7848....2K} {7848, 2}

\bibitem[\protect\citeauthoryear{{Kelson}}{{Kelson}}{2003}]{kelson03}
{Kelson} D.~D.,  2003, \mn@doi [Publications of the Astronomical Society of the
  Pacific] {10.1086/375502}, \href
  {https://ui.adsabs.harvard.edu/#abs/2003PASP..115..688K} {115, 688}

\bibitem[\protect\citeauthoryear{{Kelson} et~al.,}{{Kelson}
  et~al.}{2000}]{kelson00}
{Kelson} D.~D.,  et~al., 2000, \mn@doi [\apj] {10.1086/308445}, \href
  {https://ui.adsabs.harvard.edu/#abs/2000ApJ...531..159K} {531, 159}

\bibitem[\protect\citeauthoryear{{Kent} et~al.,}{{Kent} et~al.}{2008}]{kent08}
{Kent} B.~R.,  et~al., 2008, \mn@doi [\aj] {10.1088/0004-6256/136/2/713}, \href
  {https://ui.adsabs.harvard.edu/#abs/2008AJ....136..713K} {136, 713}

\bibitem[\protect\citeauthoryear{{Khan} et~al.,}{{Khan} et~al.}{2011}]{khan11}
{Khan} R.,  et~al., 2011, \mn@doi [\apj] {10.1088/0004-637X/726/2/106}, \href
  {https://ui.adsabs.harvard.edu/\#abs/2011ApJ...726..106K} {726, 106}

\bibitem[\protect\citeauthoryear{{Kim} et~al.,}{{Kim} et~al.}{2013}]{CBET3743}
{Kim} H.,  et~al., 2013, Central Bureau Electronic Telegrams, \href
  {https://ui.adsabs.harvard.edu/\#abs/2013CBET.3743....1K} {3743, 1}

\bibitem[\protect\citeauthoryear{{Kirshner} et~al.,}{{Kirshner} \&
  {Oke}}{1975}]{kirshner75}
{Kirshner} R.~P.,  et~al., 1975, \mn@doi [\apj] {10.1086/153824}, \href
  {https://ui.adsabs.harvard.edu/#abs/1975ApJ...200..574K} {200, 574}

\bibitem[\protect\citeauthoryear{{Kirshner} et~al.,}{{Kirshner}
  et~al.}{1991}]{IAUC5403}
{Kirshner} R.,  et~al., 1991, International Astronomical Union Circular, \href
  {https://ui.adsabs.harvard.edu/#abs/1991IAUC.5403....2K} {5403, 2}

\bibitem[\protect\citeauthoryear{{Kirshner} et~al.,}{{Kirshner}
  et~al.}{1993}]{kirshner93}
{Kirshner} R.~P.,  et~al., 1993, \mn@doi [\apj] {10.1086/173188}, \href
  {https://ui.adsabs.harvard.edu/#abs/1993ApJ...415..589K} {415, 589}

\bibitem[\protect\citeauthoryear{{Kiyota} et~al.,}{{Kiyota}
  et~al.}{2014a}]{ATel6594}
{Kiyota} S.,  et~al., 2014a, The Astronomer's Telegram, \href
  {https://ui.adsabs.harvard.edu/\#abs/2014ATel.6594....1K} {6594, 1}

\bibitem[\protect\citeauthoryear{{Kiyota} et~al.,}{{Kiyota}
  et~al.}{2014b}]{ATel6683}
{Kiyota} S.,  et~al., 2014b, The Astronomer's Telegram, \href
  {https://ui.adsabs.harvard.edu/#abs/2014ATel.6683....1K} {6683, 1}

\bibitem[\protect\citeauthoryear{{Kiyota} et~al.,}{{Kiyota}
  et~al.}{2014c}]{ATel6802}
{Kiyota} S.,  et~al., 2014c, The Astronomer's Telegram, \href
  {https://ui.adsabs.harvard.edu/\#abs/2014ATel.6802....1K} {6802, 1}

\bibitem[\protect\citeauthoryear{{Kiyota} et~al.,}{{Kiyota}
  et~al.}{2014d}]{ATel6809}
{Kiyota} S.,  et~al., 2014d, The Astronomer's Telegram, \href
  {https://ui.adsabs.harvard.edu/\#abs/2014ATel.6809....1K} {6809, 1}

\bibitem[\protect\citeauthoryear{{Kleiser} et~al.,}{{Kleiser}
  et~al.}{2009}]{CBET1918}
{Kleiser} I.,  et~al., 2009, Central Bureau Electronic Telegrams, \href
  {https://ui.adsabs.harvard.edu/#abs/2009CBET.1918....1K} {1918, 1}

\bibitem[\protect\citeauthoryear{{Klotz} et~al.,}{{Klotz}
  et~al.}{2012}]{CBET3277}
{Klotz} A.,  et~al., 2012, Central Bureau Electronic Telegrams, \href
  {https://ui.adsabs.harvard.edu/#abs/2012CBET.3277....2K} {3277, 2}

\bibitem[\protect\citeauthoryear{{Kollmeier} et~al.,}{{Kollmeier}
  et~al.}{2019}]{kollmeier19}
{Kollmeier} J.~A.,  et~al., 2019, arXiv e-prints, \href
  {https://ui.adsabs.harvard.edu/\#abs/2019arXiv190202251K} {p.
  arXiv:1902.02251}

\bibitem[\protect\citeauthoryear{{Koribalski} et~al.,}{{Koribalski}
  et~al.}{2004}]{koribalski04}
{Koribalski} B.~S.,  et~al., 2004, \mn@doi [\aj] {10.1086/421744}, \href
  {https://ui.adsabs.harvard.edu/#abs/2004AJ....128...16K} {128, 16}

\bibitem[\protect\citeauthoryear{{Kosai} et~al.,}{{Kosai}
  et~al.}{1991}]{IAUC5400}
{Kosai} H.,  et~al., 1991, International Astronomical Union Circular, \href
  {https://ui.adsabs.harvard.edu/#abs/1991IAUC.5400....1K} {5400, 1}

\bibitem[\protect\citeauthoryear{{Kotak} et~al.,}{{Kotak}
  et~al.}{2003a}]{IAUC8099}
{Kotak} R.,  et~al., 2003a, International Astronomical Union Circular, \href
  {https://ui.adsabs.harvard.edu/#abs/2003IAUC.8099....1K} {8099, 1}

\bibitem[\protect\citeauthoryear{{Kotak} et~al.,}{{Kotak}
  et~al.}{2003b}]{IAUC8122}
{Kotak} R.,  et~al., 2003b, International Astronomical Union Circular, \href
  {https://ui.adsabs.harvard.edu/#abs/2003IAUC.8122....3K} {8122, 3}

\bibitem[\protect\citeauthoryear{{Kotak} et~al.,}{{Kotak}
  et~al.}{2005}]{kotak05}
{Kotak} R.,  et~al., 2005, \mn@doi [\aap] {10.1051/0004-6361:20052756}, \href
  {https://ui.adsabs.harvard.edu/#abs/2005A&A...436.1021K} {436, 1021}

\bibitem[\protect\citeauthoryear{{Kowal}}{{Kowal}}{1972}]{IAUC2405}
{Kowal} C.~T.,  1972, International Astronomical Union Circular, \href
  {https://ui.adsabs.harvard.edu/#abs/1972IAUC.2405....1K} {2405, 1}

\bibitem[\protect\citeauthoryear{{Kowalski} et~al.,}{{Kowalski}
  et~al.}{2008}]{kowalski08}
{Kowalski} M.,  et~al., 2008, \mn@doi [\apj] {10.1086/589937}, \href
  {https://ui.adsabs.harvard.edu/#abs/2008ApJ...686..749K} {686}

\bibitem[\protect\citeauthoryear{{Krisciunas} et~al.,}{{Krisciunas}
  et~al.}{2000}]{kriciunas00}
{Krisciunas} K.,  et~al., 2000, \mn@doi [\apj] {10.1086/309263}, \href
  {https://ui.adsabs.harvard.edu/#abs/2000ApJ...539..658K} {539, 658}

\bibitem[\protect\citeauthoryear{{Krisciunas} et~al.,}{{Krisciunas}
  et~al.}{2003}]{krisciunas03}
{Krisciunas} K.,  et~al., 2003, \mn@doi [\aj] {10.1086/345571}, \href
  {https://ui.adsabs.harvard.edu/\#abs/2003AJ....125..166K} {125, 166}

\bibitem[\protect\citeauthoryear{{Krisciunas} et~al.,}{{Krisciunas}
  et~al.}{2004}]{krisciunas04}
{Krisciunas} K.,  et~al., 2004, \mn@doi [\aj] {10.1086/425629}, \href
  {https://ui.adsabs.harvard.edu/#abs/2004AJ....128.3034K} {128, 3034}

\bibitem[\protect\citeauthoryear{{Krisciunas} et~al.,}{{Krisciunas}
  et~al.}{2007}]{krisciunas07}
{Krisciunas} K.,  et~al., 2007, \mn@doi [\aj] {10.1086/509126}, \href
  {https://ui.adsabs.harvard.edu/#abs/2007AJ....133...58K} {133, 58}

\bibitem[\protect\citeauthoryear{{Krisciunas} et~al.,}{{Krisciunas}
  et~al.}{2009}]{krisciunas09}
{Krisciunas} K.,  et~al., 2009, \mn@doi [\aj] {10.1088/0004-6256/138/6/1584},
  \href {https://ui.adsabs.harvard.edu/\#abs/2009AJ....138.1584K} {138, 1584}

\bibitem[\protect\citeauthoryear{{Krisciunas} et~al.,}{{Krisciunas}
  et~al.}{2017}]{krisciunas17}
{Krisciunas} K.,  et~al., 2017, \mn@doi [\aj] {10.3847/1538-3881/aa8df0}, \href
  {https://ui.adsabs.harvard.edu/\#abs/2017AJ....154..211K} {154, 211}

\bibitem[\protect\citeauthoryear{{Kromer} et~al.,}{{Kromer}
  et~al.}{2010}]{kromer10}
{Kromer} M.,  et~al., 2010, \mn@doi [\apj] {10.1088/0004-637X/719/2/1067},
  \href {https://ui.adsabs.harvard.edu/\#abs/2010ApJ...719.1067K} {719, 1067}

\bibitem[\protect\citeauthoryear{{Krumm} et~al.,}{{Krumm} \&
  {Salpeter}}{1980}]{krumm80}
{Krumm} N.,  et~al., 1980, \mn@doi [\aj] {10.1086/112801}, \href
  {https://ui.adsabs.harvard.edu/#abs/1980AJ.....85.1312K} {85, 1312}

\bibitem[\protect\citeauthoryear{{Lair} et~al.,}{{Lair} et~al.}{2006}]{lair06}
{Lair} J.~C.,  et~al., 2006, \mn@doi [\aj] {10.1086/508322}, \href
  {https://ui.adsabs.harvard.edu/#abs/2006AJ....132.2024L} {132, 2024}

\bibitem[\protect\citeauthoryear{{Lanz} et~al.,}{{Lanz} et~al.}{2005}]{lanz05}
{Lanz} T.,  et~al., 2005, \mn@doi [\apj] {10.1086/426382}, \href
  {https://ui.adsabs.harvard.edu/#abs/2005ApJ...619..517L} {619, 517}

\bibitem[\protect\citeauthoryear{{Larsen} et~al.,}{{Larsen}
  et~al.}{2001}]{larsen01}
{Larsen} S.~S.,  et~al., 2001, \mn@doi [\aj] {10.1086/321081}, \href
  {https://ui.adsabs.harvard.edu/\#abs/2001AJ....121.2974L} {121, 2974}

\bibitem[\protect\citeauthoryear{{Lauberts} et~al.,}{{Lauberts} \&
  {Valentijn}}{1989}]{lauberts89}
{Lauberts} A.,  et~al., 1989, {The surface photometry catalogue of the
  ESO-Uppsala galaxies}

\bibitem[\protect\citeauthoryear{{Lavery}}{{Lavery}}{1989}]{IAUC4743}
{Lavery} R.~J.,  1989, International Astronomical Union Circular, \href
  {https://ui.adsabs.harvard.edu/\#abs/1989IAUC.4743....2L} {4743, 2}

\bibitem[\protect\citeauthoryear{{Leadbeater}}{{Leadbeater}}{2016}]{2016TNSCR793}
{Leadbeater} R.,  2016, Transient Name Server Classification Report, \href
  {https://ui.adsabs.harvard.edu/\#abs/2016TNSCR.793....1L} {2016-793, 1}

\bibitem[\protect\citeauthoryear{{Leadbeater}}{{Leadbeater}}{2018}]{TNSCR159}
{Leadbeater} R.,  2018, Transient Name Server Classification Report, \href
  {https://ui.adsabs.harvard.edu/\#abs/2018TNSCR.159....1L} {2018-159, 1}

\bibitem[\protect\citeauthoryear{{Lee} et~al.,}{{Lee} et~al.}{1972}]{lee72}
{Lee} T.~A.,  et~al., 1972, \mn@doi [\apj] {10.1086/181052}, \href
  {https://ui.adsabs.harvard.edu/\#abs/1972ApJ...177L..59L} {177, L59}

\bibitem[\protect\citeauthoryear{{Leibendgut} et~al.,}{{Leibendgut}
  et~al.}{1993}]{leibendgut93}
{Leibendgut} B.,  et~al., 1993, \mn@doi [\aj] {10.1086/116427}, \href
  {https://ui.adsabs.harvard.edu/#abs/1993AJ....105..301L} {105, 301}

\bibitem[\protect\citeauthoryear{{Leloudas} et~al.,}{{Leloudas}
  et~al.}{2009}]{leloudas09}
{Leloudas} G.,  et~al., 2009, \mn@doi [\aap] {10.1051/0004-6361/200912364},
  \href {https://ui.adsabs.harvard.edu/#abs/2009A&A...505..265L} {505, 265}

\bibitem[\protect\citeauthoryear{{Leonard}}{{Leonard}}{2007}]{leonard07}
{Leonard} D.~C.,  2007, \mn@doi [\apj] {10.1086/522367}, \href
  {https://ui.adsabs.harvard.edu/#abs/2007ApJ...670.1275L} {670, 1275}

\bibitem[\protect\citeauthoryear{{Leonard} et~al.,}{{Leonard} \&
  {Filippenko}}{2001}]{leonard01}
{Leonard} D.~C.,  et~al., 2001, \mn@doi [Publications of the Astronomical
  Society of the Pacific] {10.1086/322151}, \href
  {https://ui.adsabs.harvard.edu/#abs/2001PASP..113..920L} {113, 920}

\bibitem[\protect\citeauthoryear{{Li} et~al.,}{{Li} et~al.}{1999}]{li99}
{Li} W.~D.,  et~al., 1999, \mn@doi [\aj] {10.1086/300895}, \href
  {https://ui.adsabs.harvard.edu/\#abs/1999AJ....117.2709L} {117, 2709}

\bibitem[\protect\citeauthoryear{{Li} et~al.,}{{Li} et~al.}{2000}]{li00}
{Li} W.~D.,  et~al., 2000, in {Holt} S.~S.,  et~al., eds,  American Institute
  of Physics Conference Series Vol. 522, American Institute of Physics
  Conference Series. pp 103--106 (\mn@eprint {arXiv} {astro-ph/9912336}),
  \mn@doi{10.1063/1.1291702}

\bibitem[\protect\citeauthoryear{{Li} et~al.,}{{Li} et~al.}{2001}]{li01}
{Li} W.,  et~al., 2001, \mn@doi [Publications of the Astronomical Society of
  the Pacific] {10.1086/323355}, \href
  {https://ui.adsabs.harvard.edu/#abs/2001PASP..113.1178L} {113, 1178}

\bibitem[\protect\citeauthoryear{{Li} et~al.,}{{Li} et~al.}{2003a}]{li03}
{Li} W.,  et~al., 2003a, \mn@doi [Publications of the Astronomical Society of
  the Pacific] {10.1086/374200}, \href
  {https://ui.adsabs.harvard.edu/\#abs/2003PASP..115..453L} {115, 453}

\bibitem[\protect\citeauthoryear{{Li} et~al.,}{{Li} et~al.}{2003b}]{IAUC8245}
{Li} W.,  et~al., 2003b, International Astronomical Union Circular, \href
  {https://ui.adsabs.harvard.edu/#abs/2003IAUC.8245....1L} {8245, 1}

\bibitem[\protect\citeauthoryear{{Li} et~al.,}{{Li} et~al.}{2019}]{li18}
{Li} W.,  et~al., 2019, \mn@doi [\apj] {10.3847/1538-4357/aaec74}, \href
  {https://ui.adsabs.harvard.edu/\#abs/2019ApJ...870...12L} {870, 12}

\bibitem[\protect\citeauthoryear{{Liller} et~al.,}{{Liller} \&
  {Buta}}{1992}]{IAUC5431}
{Liller} W.,  et~al., 1992, International Astronomical Union Circular, \href
  {https://ui.adsabs.harvard.edu/#abs/1992IAUC.5431....3L} {5431, 3}

\bibitem[\protect\citeauthoryear{{Liller} et~al.,}{{Liller}
  et~al.}{1992}]{IAUC5428}
{Liller} W.,  et~al., 1992, International Astronomical Union Circular, \href
  {https://ui.adsabs.harvard.edu/#abs/1992IAUC.5428....1L} {5428, 1}

\bibitem[\protect\citeauthoryear{{Lira} et~al.,}{{Lira} et~al.}{1998}]{lira98}
{Lira} P.,  et~al., 1998, \mn@doi [\aj] {10.1086/300175}, \href
  {https://ui.adsabs.harvard.edu/#abs/1998AJ....115..234L} {115, 234}

\bibitem[\protect\citeauthoryear{{Liu} et~al.,}{{Liu} et~al.}{2012}]{liu12}
{Liu} Z.~W.,  et~al., 2012, \mn@doi [\aap] {10.1051/0004-6361/201219357}, \href
  {https://ui.adsabs.harvard.edu/#abs/2012A&A...548A...2L} {548, A2}

\bibitem[\protect\citeauthoryear{{Liu} et~al.,}{{Liu} et~al.}{2013a}]{liu13}
{Liu} Z.-W.,  et~al., 2013a, \mn@doi [\apj] {10.1088/0004-637X/774/1/37}, \href
  {https://ui.adsabs.harvard.edu/#abs/2013ApJ...774...37L} {774, 37}

\bibitem[\protect\citeauthoryear{{Liu} et~al.,}{{Liu} et~al.}{2013b}]{liu13Iax}
{Liu} Z.-W.,  et~al., 2013b, \mn@doi [\apj] {10.1088/0004-637X/778/2/121},
  \href {https://ui.adsabs.harvard.edu/#abs/2013ApJ...778..121L} {778, 121}

\bibitem[\protect\citeauthoryear{{Livio} et~al.,}{{Livio} \&
  {Mazzali}}{2018}]{livio18}
{Livio} M.,  et~al., 2018, \mn@doi [\physrep] {10.1016/j.physrep.2018.02.002},
  \href {https://ui.adsabs.harvard.edu/#abs/2018PhR...736....1L} {736, 1}

\bibitem[\protect\citeauthoryear{{Livio} et~al.,}{{Livio} \&
  {Riess}}{2003}]{livio03}
{Livio} M.,  et~al., 2003, \mn@doi [\apjl] {10.1086/378765}, \href
  {https://ui.adsabs.harvard.edu/abs/2003ApJ...594L..93L} {594, L93}

\bibitem[\protect\citeauthoryear{{Livne}}{{Livne}}{1990}]{livne90}
{Livne} E.,  1990, \mn@doi [\apj] {10.1086/185721}, \href
  {https://ui.adsabs.harvard.edu/#abs/1990ApJ...354L..53L} {354, L53}

\bibitem[\protect\citeauthoryear{{Livne} et~al.,}{{Livne} \&
  {Arnett}}{1995}]{livne95}
{Livne} E.,  et~al., 1995, \mn@doi [\apj] {10.1086/176279}, \href
  {https://ui.adsabs.harvard.edu/\#abs/1995ApJ...452...62L} {452, 62}

\bibitem[\protect\citeauthoryear{{Longhetti} et~al.,}{{Longhetti}
  et~al.}{1998}]{longhetti98}
{Longhetti} M.,  et~al., 1998, \mn@doi [Astronomy and Astrophysics Supplement
  Series] {10.1051/aas:1998410}, \href
  {https://ui.adsabs.harvard.edu/\#abs/1998A&AS..130..251L} {130, 251}

\bibitem[\protect\citeauthoryear{{Lundqvist} et~al.,}{{Lundqvist}
  et~al.}{2013}]{lundquist13}
{Lundqvist} P.,  et~al., 2013, \mn@doi [\mnras] {10.1093/mnras/stt1303}, \href
  {https://ui.adsabs.harvard.edu/#abs/2013MNRAS.435..329L} {435, 329}

\bibitem[\protect\citeauthoryear{{Lundqvist} et~al.,}{{Lundqvist}
  et~al.}{2015}]{lundquist15}
{Lundqvist} P.,  et~al., 2015, \mn@doi [\aap] {10.1051/0004-6361/201525719},
  \href {https://ui.adsabs.harvard.edu/#abs/2015A&A...577A..39L} {577, A39}

\bibitem[\protect\citeauthoryear{{Maeda} et~al.,}{{Maeda}
  et~al.}{2009}]{maeda09}
{Maeda} K.,  et~al., 2009, \mn@doi [\apj] {10.1088/0004-637X/690/2/1745}, \href
  {https://ui.adsabs.harvard.edu/#abs/2009ApJ...690.1745M} {690, 1745}

\bibitem[\protect\citeauthoryear{{Maguire} et~al.,}{{Maguire}
  et~al.}{2014}]{maguire14}
{Maguire} K.,  et~al., 2014, \mn@doi [\mnras] {10.1093/mnras/stu1607}, \href
  {https://ui.adsabs.harvard.edu/#abs/2014MNRAS.444.3258M} {444, 3258}

\bibitem[\protect\citeauthoryear{{Maguire} et~al.,}{{Maguire}
  et~al.}{2016}]{maguire16}
{Maguire} K.,  et~al., 2016, \mn@doi [\mnras] {10.1093/mnras/stv2991}, \href
  {https://ui.adsabs.harvard.edu/#abs/2016MNRAS.457.3254M} {457, 3254}

\bibitem[\protect\citeauthoryear{{Maguire} et~al.,}{{Maguire}
  et~al.}{2018}]{maguire18}
{Maguire} K.,  et~al., 2018, \mn@doi [\mnras] {10.1093/mnras/sty820}, \href
  {https://ui.adsabs.harvard.edu/\#abs/2018MNRAS.477.3567M} {477, 3567}

\bibitem[\protect\citeauthoryear{{Malesani} et~al.,}{{Malesani}
  et~al.}{2018}]{ATel11516}
{Malesani} D.,  et~al., 2018, The Astronomer's Telegram, \href
  {https://ui.adsabs.harvard.edu/\#abs/2018ATel11516....1M} {11516, 1}

\bibitem[\protect\citeauthoryear{{Maoz} et~al.,}{{Maoz} et~al.}{2014}]{maoz14}
{Maoz} D.,  et~al., 2014, \mn@doi [Annual Review of Astronomy and Astrophysics]
  {10.1146/annurev-astro-082812-141031}, \href
  {https://ui.adsabs.harvard.edu/#abs/2014ARA&A..52..107M} {52, 107}

\bibitem[\protect\citeauthoryear{{Marietta} et~al.,}{{Marietta}
  et~al.}{2000}]{marietta00}
{Marietta} E.,  et~al., 2000, \mn@doi [The Astrophysical Journal Supplement
  Series] {10.1086/313392}, \href
  {https://ui.adsabs.harvard.edu/#abs/2000ApJS..128..615M} {128, 615}

\bibitem[\protect\citeauthoryear{{Marion} et~al.,}{{Marion}
  et~al.}{2012}]{CBET3146}
{Marion} G.~H.,  et~al., 2012, Central Bureau Electronic Telegrams, \href
  {https://ui.adsabs.harvard.edu/#abs/2012CBET.3146....2M} {3146, 2}

\bibitem[\protect\citeauthoryear{{Marshall} et~al.,}{{Marshall}
  et~al.}{2008}]{MagEref}
{Marshall} J.~L.,  et~al., 2008, in Ground-based and Airborne Instrumentation
  for Astronomy II. Edited by McLean, Ian S.; Casali, Mark M. Proceedings of
  the SPIE, Volume 7014, article id. 701454, 10 pp. (2008).. ,
  \mn@doi{10.1117/12.789972}

\bibitem[\protect\citeauthoryear{{Martin} et~al.,}{{Martin} \&
  {Biggs}}{2004}]{IAUC8282}
{Martin} R.,  et~al., 2004, International Astronomical Union Circular, \href
  {https://ui.adsabs.harvard.edu/#abs/2004IAUC.8282....1M} {8282, 1}

\bibitem[\protect\citeauthoryear{{Martin} et~al.,}{{Martin}
  et~al.}{2005}]{IAUC8490}
{Martin} R.,  et~al., 2005, International Astronomical Union Circular, \href
  {https://ui.adsabs.harvard.edu/#abs/2005IAUC.8490....1M} {8490, 1}

\bibitem[\protect\citeauthoryear{{Matheson} et~al.,}{{Matheson}
  et~al.}{2002}]{IAUC7903}
{Matheson} T.,  et~al., 2002, International Astronomical Union Circular, \href
  {https://ui.adsabs.harvard.edu/#abs/2002IAUC.7903....2M} {7903, 2}

\bibitem[\protect\citeauthoryear{{Matheson} et~al.,}{{Matheson}
  et~al.}{2008}]{matheson08}
{Matheson} T.,  et~al., 2008, \mn@doi [\aj] {10.1088/0004-6256/135/4/1598},
  \href {https://ui.adsabs.harvard.edu/\#abs/2008AJ....135.1598M} {135, 1598}

\bibitem[\protect\citeauthoryear{{Mathewson} et~al.,}{{Mathewson}
  et~al.}{1992}]{mathewson92}
{Mathewson} D.~S.,  et~al., 1992, \mn@doi [The Astrophysical Journal Supplement
  Series] {10.1086/191700}, \href
  {https://ui.adsabs.harvard.edu/#abs/1992ApJS...81..413M} {81, 413}

\bibitem[\protect\citeauthoryear{{Mattila} et~al.,}{{Mattila}
  et~al.}{2005}]{mattila05}
{Mattila} S.,  et~al., 2005, \mn@doi [\aap] {10.1051/0004-6361:20052731}, \href
  {https://ui.adsabs.harvard.edu/#abs/2005A&A...443..649M} {443, 649}

\bibitem[\protect\citeauthoryear{{Mattila} et~al.,}{{Mattila}
  et~al.}{2016}]{ATel8992}
{Mattila} S.,  et~al., 2016, The Astronomer's Telegram, \href
  {https://ui.adsabs.harvard.edu/#abs/2016ATel.8992....1M} {8992, 1}

\bibitem[\protect\citeauthoryear{{Maury} et~al.,}{{Maury}
  et~al.}{1990}]{IAUC5039}
{Maury} A.,  et~al., 1990, International Astronomical Union Circular, \href
  {https://ui.adsabs.harvard.edu/#abs/1990IAUC.5039....1M} {5039, 1}

\bibitem[\protect\citeauthoryear{{Maza} et~al.,}{{Maza}
  et~al.}{2010}]{CBET2388}
{Maza} J.,  et~al., 2010, Central Bureau Electronic Telegrams, \href
  {https://ui.adsabs.harvard.edu/#abs/2010CBET.2388....1M} {2388, 1}

\bibitem[\protect\citeauthoryear{{Mazzali} et~al.,}{{Mazzali}
  et~al.}{2015}]{mazzali15}
{Mazzali} P.~A.,  et~al., 2015, \mn@doi [\mnras] {10.1093/mnras/stv761}, \href
  {https://ui.adsabs.harvard.edu/#abs/2015MNRAS.450.2631M} {450, 2631}

\bibitem[\protect\citeauthoryear{{McCully} et~al.,}{{McCully}
  et~al.}{2014}]{mccully14}
{McCully} C.,  et~al., 2014, \mn@doi [\nat] {10.1038/nature13615}, \href
  {https://ui.adsabs.harvard.edu/\#abs/2014Natur.512...54M} {512, 54}

\bibitem[\protect\citeauthoryear{{Meng} et~al.,}{{Meng} \&
  {Podsiadlowski}}{2013}]{meng13}
{Meng} X.,  et~al., 2013, \mn@doi [\apj] {10.1088/2041-8205/778/2/L35}, \href
  {https://ui.adsabs.harvard.edu/\#abs/2013ApJ...778L..35M} {778, L35}

\bibitem[\protect\citeauthoryear{{Menzies} et~al.,}{{Menzies}
  et~al.}{1991}]{IAUC5246}
{Menzies} J.,  et~al., 1991, International Astronomical Union Circular, \href
  {https://ui.adsabs.harvard.edu/#abs/1991IAUC.5246....2M} {5246, 2}

\bibitem[\protect\citeauthoryear{{Meyer} et~al.,}{{Meyer}
  et~al.}{2004}]{meyer04}
{Meyer} M.~J.,  et~al., 2004, \mn@doi [\mnras]
  {10.1111/j.1365-2966.2004.07710.x}, \href
  {https://ui.adsabs.harvard.edu/#abs/2004MNRAS.350.1195M} {350, 1195}

\bibitem[\protect\citeauthoryear{{Mikuz}}{{Mikuz}}{1994}]{IAUC5958}
{Mikuz} H.,  1994, International Astronomical Union Circular, \href
  {https://ui.adsabs.harvard.edu/\#abs/1994IAUC.5958....3M} {5958, 3}

\bibitem[\protect\citeauthoryear{{Mikuz}}{{Mikuz}}{1995}]{IAUC6166}
{Mikuz} H.,  1995, International Astronomical Union Circular, \href
  {https://ui.adsabs.harvard.edu/#abs/1995IAUC.6166....2M} {6166, 2}

\bibitem[\protect\citeauthoryear{{Misra} et~al.,}{{Misra}
  et~al.}{2005}]{misra05}
{Misra} K.,  et~al., 2005, \mn@doi [\mnras] {10.1111/j.1365-2966.2005.09059.x},
  \href {https://ui.adsabs.harvard.edu/#abs/2005MNRAS.360..662M} {360, 662}

\bibitem[\protect\citeauthoryear{{Modjaz} et~al.,}{{Modjaz}
  et~al.}{2005a}]{CBET160}
{Modjaz} M.,  et~al., 2005a, Central Bureau Electronic Telegrams, \href
  {https://ui.adsabs.harvard.edu/#abs/2005CBET..160....1M} {160, 1}

\bibitem[\protect\citeauthoryear{{Modjaz} et~al.,}{{Modjaz}
  et~al.}{2005b}]{IAUC8491}
{Modjaz} M.,  et~al., 2005b, International Astronomical Union Circular, \href
  {https://ui.adsabs.harvard.edu/#abs/2005IAUC.8491....2M} {8491, 2}

\bibitem[\protect\citeauthoryear{{Molinari} et~al.,}{{Molinari}
  et~al.}{1999}]{DOLORESref}
{Molinari} E.,  et~al., 1999, in Looking Deep in the Southern Sky, Proceedings
  of the ESO/Australia Workshop held at Sydney, Australia, 10-12 December 1997.
  Edited by Faffaella Morganti and Warrick J. Couch. Berlin: Springer- Verlag,
  1999. p. 157.. p.~157

\bibitem[\protect\citeauthoryear{{Monard} et~al.,}{{Monard} \&
  {Africa}}{2010}]{CBET2434}
{Monard} L.~A.~G.,  et~al., 2010, Central Bureau Electronic Telegrams, \href
  {https://ui.adsabs.harvard.edu/#abs/2010CBET.2434....1M} {2434, 1}

\bibitem[\protect\citeauthoryear{{Monard} et~al.,}{{Monard}
  et~al.}{2001}]{IAUC7720}
{Monard} A.~G.,  et~al., 2001, International Astronomical Union Circular, \href
  {https://ui.adsabs.harvard.edu/\#abs/2001IAUC.7720....1M} {7720, 1}

\bibitem[\protect\citeauthoryear{{Monard} et~al.,}{{Monard}
  et~al.}{2007}]{CBET1100}
{Monard} L.~A.~G.,  et~al., 2007, Central Bureau Electronic Telegrams, \href
  {https://ui.adsabs.harvard.edu/#abs/2007CBET.1100....1M} {1100, 1}

\bibitem[\protect\citeauthoryear{{Monard} et~al.,}{{Monard}
  et~al.}{2011}]{CBET2635}
{Monard} L.~A.~G.,  et~al., 2011, Central Bureau Electronic Telegrams, \href
  {https://ui.adsabs.harvard.edu/\#abs/2011CBET.2635....1M} {2635, 1}

\bibitem[\protect\citeauthoryear{{Monard} et~al.,}{{Monard}
  et~al.}{2015}]{CBET4081}
{Monard} L.~A.~G.,  et~al., 2015, Central Bureau Electronic Telegrams, \href
  {https://ui.adsabs.harvard.edu/\#abs/2015CBET.4081....1M} {4081, 1}

\bibitem[\protect\citeauthoryear{{Moorwood} et~al.,}{{Moorwood}
  et~al.}{1998}]{SOFIref}
{Moorwood} A.,  et~al., 1998, The Messenger, \href
  {https://ui.adsabs.harvard.edu/#abs/1998Msngr..91....9M} {91, 9}

\bibitem[\protect\citeauthoryear{{Morrell} et~al.,}{{Morrell} \&
  {Shappee}}{2016}]{ATel9170}
{Morrell} N.,  et~al., 2016, The Astronomer's Telegram, \href
  {https://ui.adsabs.harvard.edu/\#abs/2016ATel.9170....1M} {9170, 1}

\bibitem[\protect\citeauthoryear{{Morrell} et~al.,}{{Morrell}
  et~al.}{2007}]{CBET1131}
{Morrell} N.,  et~al., 2007, Central Bureau Electronic Telegrams, \href
  {https://ui.adsabs.harvard.edu/#abs/2007CBET.1131....1M} {1131, 1}

\bibitem[\protect\citeauthoryear{{Morrell} et~al.,}{{Morrell}
  et~al.}{2014a}]{ATel6508}
{Morrell} N.,  et~al., 2014a, The Astronomer's Telegram, \href
  {https://ui.adsabs.harvard.edu/#abs/2014ATel.6508....1M} {6508, 1}

\bibitem[\protect\citeauthoryear{{Morrell} et~al.,}{{Morrell}
  et~al.}{2014b}]{ATel6765}
{Morrell} N.,  et~al., 2014b, The Astronomer's Telegram, \href
  {https://ui.adsabs.harvard.edu/\#abs/2014ATel.6765....1M} {6765, 1}

\bibitem[\protect\citeauthoryear{{Morrell} et~al.,}{{Morrell}
  et~al.}{2015}]{ATel6988}
{Morrell} N.,  et~al., 2015, The Astronomer's Telegram, \href
  {https://ui.adsabs.harvard.edu/\#abs/2015ATel.6988....1M} {6988, 1}

\bibitem[\protect\citeauthoryear{{Motohara} et~al.,}{{Motohara}
  et~al.}{2002}]{CISCOref}
{Motohara} K.,  et~al., 2002, \mn@doi [Publications of the Astronomical Society
  of Japan] {10.1093/pasj/54.2.315}, \href
  {https://ui.adsabs.harvard.edu/#abs/2002PASJ...54..315M} {54, 315}

\bibitem[\protect\citeauthoryear{{Motohara} et~al.,}{{Motohara}
  et~al.}{2006}]{motohara06}
{Motohara} K.,  et~al., 2006, \mn@doi [\apj] {10.1086/509919}, \href
  {https://ui.adsabs.harvard.edu/\#abs/2006ApJ...652L.101M} {652, L101}

\bibitem[\protect\citeauthoryear{{Munari} et~al.,}{{Munari}
  et~al.}{2013}]{munari13}
{Munari} U.,  et~al., 2013, \mn@doi [\na] {10.1016/j.newast.2012.09.003}, \href
  {https://ui.adsabs.harvard.edu/#abs/2013NewA...20...30M} {20, 30}

\bibitem[\protect\citeauthoryear{{Munch} et~al.,}{{Munch}
  et~al.}{1986}]{IAUC4183}
{Munch} G.,  et~al., 1986, International Astronomical Union Circular, \href
  {https://ui.adsabs.harvard.edu/#abs/1986IAUC.4183....1M} {4183, 1}

\bibitem[\protect\citeauthoryear{{Nakano} et~al.,}{{Nakano} \&
  {Itagaki}}{2007}]{CBET863}
{Nakano} S.,  et~al., 2007, Central Bureau Electronic Telegrams, \href
  {https://ui.adsabs.harvard.edu/#abs/2007CBET..863....1N} {863, 1}

\bibitem[\protect\citeauthoryear{{Nakano} et~al.,}{{Nakano}
  et~al.}{1995}]{IAUC6134}
{Nakano} S.,  et~al., 1995, International Astronomical Union Circular, \href
  {https://ui.adsabs.harvard.edu/#abs/1995IAUC.6134....1N} {6134, 1}

\bibitem[\protect\citeauthoryear{{Nakano} et~al.,}{{Nakano}
  et~al.}{2003}]{IAUC8097}
{Nakano} S.,  et~al., 2003, International Astronomical Union Circular, \href
  {https://ui.adsabs.harvard.edu/#abs/2003IAUC.8097....1N} {8097, 1}

\bibitem[\protect\citeauthoryear{{Nakano} et~al.,}{{Nakano}
  et~al.}{2005}]{IAUC8475}
{Nakano} S.,  et~al., 2005, International Astronomical Union Circular, \href
  {https://ui.adsabs.harvard.edu/\#abs/2005IAUC.8475....1N} {8475, 1}

\bibitem[\protect\citeauthoryear{{Nakano} et~al.,}{{Nakano}
  et~al.}{2007}]{CBET1017}
{Nakano} S.,  et~al., 2007, Central Bureau Electronic Telegrams, \href
  {https://ui.adsabs.harvard.edu/abs/2007CBET.1017....1N} {1017, 1}

\bibitem[\protect\citeauthoryear{{Nakano} et~al.,}{{Nakano}
  et~al.}{2008}]{CBET1193}
{Nakano} S.,  et~al., 2008, Central Bureau Electronic Telegrams, \href
  {https://ui.adsabs.harvard.edu/\#abs/2008CBET.1193....1N} {1193, 1}

\bibitem[\protect\citeauthoryear{{Nakano} et~al.,}{{Nakano}
  et~al.}{2011}]{CBET2783}
{Nakano} S.,  et~al., 2011, Central Bureau Electronic Telegrams, \href
  {https://ui.adsabs.harvard.edu/\#abs/2011CBET.2783....1N} {2783, 1}

\bibitem[\protect\citeauthoryear{{Nakano} et~al.,}{{Nakano}
  et~al.}{2012}]{CBET3209}
{Nakano} S.,  et~al., 2012, Central Bureau Electronic Telegrams, \href
  {https://ui.adsabs.harvard.edu/#abs/2012CBET.3209....1N} {3209, 1}

\bibitem[\protect\citeauthoryear{{Nakano} et~al.,}{{Nakano}
  et~al.}{2015}]{CBET4106}
{Nakano} S.,  et~al., 2015, Central Bureau Electronic Telegrams, \href
  {https://ui.adsabs.harvard.edu/\#abs/2015CBET.4106....1N} {4106, 1}

\bibitem[\protect\citeauthoryear{{Nicolas} et~al.,}{{Nicolas}
  et~al.}{2014}]{ATel6500}
{Nicolas} J.,  et~al., 2014, The Astronomer's Telegram, \href
  {https://ui.adsabs.harvard.edu/#abs/2014ATel.6500....1N} {6500, 1}

\bibitem[\protect\citeauthoryear{{Nishiyama} et~al.,}{{Nishiyama}
  et~al.}{2012}]{CBET3349}
{Nishiyama} K.,  et~al., 2012, Central Bureau Electronic Telegrams, \href
  {https://ui.adsabs.harvard.edu/#abs/2012CBET.3349....1N} {3349, 1}

\bibitem[\protect\citeauthoryear{{Noguchi} et~al.,}{{Noguchi}
  et~al.}{2011a}]{CBET2940}
{Noguchi} T.,  et~al., 2011a, Central Bureau Electronic Telegrams, \href
  {https://ui.adsabs.harvard.edu/\#abs/2011CBET.2940....1N} {2940, 1}

\bibitem[\protect\citeauthoryear{{Noguchi} et~al.,}{{Noguchi}
  et~al.}{2011b}]{CBET2943}
{Noguchi} T.,  et~al., 2011b, Central Bureau Electronic Telegrams, \href
  {https://ui.adsabs.harvard.edu/\#abs/2011CBET.2943....1N} {2943, 1}

\bibitem[\protect\citeauthoryear{{Nomoto}}{{Nomoto}}{1982}]{nomoto82}
{Nomoto} K.,  1982, \mn@doi [\apj] {10.1086/159682}, \href
  {https://ui.adsabs.harvard.edu/#abs/1982ApJ...253..798N} {253, 798}

\bibitem[\protect\citeauthoryear{{Nucita} et~al.,}{{Nucita}
  et~al.}{2017}]{nucita17}
{Nucita} A.~A.,  et~al., 2017, \mn@doi [\apj] {10.3847/1538-4357/aa9481}, \href
  {https://ui.adsabs.harvard.edu/\#abs/2017ApJ...850..111N} {850, 111}

\bibitem[\protect\citeauthoryear{{Nugent} et~al.,}{{Nugent}
  et~al.}{2011a}]{nugent11}
{Nugent} P.~E.,  et~al., 2011a, \mn@doi [\nat] {10.1038/nature10644}, \href
  {https://ui.adsabs.harvard.edu/#abs/2011Natur.480..344N} {480, 344}

\bibitem[\protect\citeauthoryear{{Nugent} et~al.,}{{Nugent}
  et~al.}{2011b}]{ATel3581}
{Nugent} P.,  et~al., 2011b, The Astronomer's Telegram, \href
  {https://ui.adsabs.harvard.edu/#abs/2011ATel.3581....1N} {3581, 1}

\bibitem[\protect\citeauthoryear{{Nyholm} et~al.,}{{Nyholm}
  et~al.}{2017}]{ATel10131}
{Nyholm} A.,  et~al., 2017, The Astronomer's Telegram, \href
  {https://ui.adsabs.harvard.edu/#abs/2017ATel10131....1N} {10131, 1}

\bibitem[\protect\citeauthoryear{{Ochner} et~al.,}{{Ochner}
  et~al.}{2016}]{ATel9018}
{Ochner} P.,  et~al., 2016, The Astronomer's Telegram, \href
  {https://ui.adsabs.harvard.edu/\#abs/2016ATel.9018....1O} {9018, 1}

\bibitem[\protect\citeauthoryear{{Ogando} et~al.,}{{Ogando}
  et~al.}{2008}]{ogando08}
{Ogando} R. L.~C.,  et~al., 2008, \mn@doi [\aj] {10.1088/0004-6256/135/6/2424},
  \href {https://ui.adsabs.harvard.edu/#abs/2008AJ....135.2424O} {135, 2424}

\bibitem[\protect\citeauthoryear{{Oke} et~al.,}{{Oke} \&
  {Gunn}}{1982}]{DBSPref}
{Oke} J.~B.,  et~al., 1982, \mn@doi [Publications of the Astronomical Society
  of the Pacific] {10.1086/131027}, \href
  {https://ui.adsabs.harvard.edu/#abs/1982PASP...94..586O} {94, 586}

\bibitem[\protect\citeauthoryear{{Oke} et~al.,}{{Oke} et~al.}{1995}]{LRISref}
{Oke} J.~B.,  et~al., 1995, \mn@doi [Publications of the Astronomical Society
  of the Pacific] {10.1086/133562}, \href
  {https://ui.adsabs.harvard.edu/#abs/1995PASP..107..375O} {107, 375}

\bibitem[\protect\citeauthoryear{{Olszewski}}{{Olszewski}}{1982}]{olsz92}
{Olszewski} E.~W.,  1982, Information Bulletin on Variable Stars, \href
  {https://ui.adsabs.harvard.edu/\#abs/1982IBVS.2065....1O} {2065, 1}

\bibitem[\protect\citeauthoryear{{Osmer} et~al.,}{{Osmer}
  et~al.}{1972}]{osmer72}
{Osmer} P.~S.,  et~al., 1972, \mn@doi [Nature Physical Science]
  {10.1038/physci238021a0}, \href
  {https://ui.adsabs.harvard.edu/\#abs/1972NPhS..238...21O} {238, 21}

\bibitem[\protect\citeauthoryear{{Pakmor} et~al.,}{{Pakmor}
  et~al.}{2012}]{pakmor12}
{Pakmor} R.,  et~al., 2012, \mn@doi [\apj] {10.1088/2041-8205/747/1/L10}, \href
  {https://ui.adsabs.harvard.edu/\#abs/2012ApJ...747L..10P} {747, L10}

\bibitem[\protect\citeauthoryear{{Pan} et~al.,}{{Pan} et~al.}{2012}]{pan12}
{Pan} K.-C.,  et~al., 2012, \mn@doi [\apj] {10.1088/0004-637X/750/2/151}, \href
  {https://ui.adsabs.harvard.edu/#abs/2012ApJ...750..151P} {750, 151}

\bibitem[\protect\citeauthoryear{{Pan} et~al.,}{{Pan} et~al.}{2015}]{pan15}
{Pan} Y.~C.,  et~al., 2015, \mn@doi [\mnras] {10.1093/mnras/stv1605}, \href
  {https://ui.adsabs.harvard.edu/#abs/2015MNRAS.452.4307P} {452, 4307}

\bibitem[\protect\citeauthoryear{{Pan} et~al.,}{{Pan} et~al.}{2017}]{ATel10225}
{Pan} Y.~C.,  et~al., 2017, The Astronomer's Telegram, \href
  {https://ui.adsabs.harvard.edu/#abs/2017ATel10225....1P} {10225, 1}

\bibitem[\protect\citeauthoryear{{Parker}}{{Parker}}{2016}]{2016TNSTR304}
{Parker} S.,  2016, Transient Name Server Discovery Report, \href
  {https://ui.adsabs.harvard.edu/\#abs/2016TNSTR.304....1P} {2016-304, 1}

\bibitem[\protect\citeauthoryear{{Parker}}{{Parker}}{2017}]{ATRN12243}
{Parker} S.,  2017, Transient Name Server Discovery Report, \href
  {https://ui.adsabs.harvard.edu/#abs/2017TNSTR.700....1P} {2017-700, 1}

\bibitem[\protect\citeauthoryear{{Parker} et~al.,}{{Parker}
  et~al.}{2013a}]{CBET3416}
{Parker} S.,  et~al., 2013a, Central Bureau Electronic Telegrams, \href
  {https://ui.adsabs.harvard.edu/#abs/2013CBET.3416....1P} {3416, 1}

\bibitem[\protect\citeauthoryear{{Parker} et~al.,}{{Parker}
  et~al.}{2013b}]{CBET3539}
{Parker} P.,  et~al., 2013b, Central Bureau Electronic Telegrams, \href
  {https://ui.adsabs.harvard.edu/#abs/2013CBET.3539....1P} {3539, 1}

\bibitem[\protect\citeauthoryear{{Pastorello} et~al.,}{{Pastorello}
  et~al.}{2007}]{pastorello07}
{Pastorello} A.,  et~al., 2007, \mn@doi [\mnras]
  {10.1111/j.1365-2966.2007.11527.x}, \href
  {https://ui.adsabs.harvard.edu/#abs/2007MNRAS.376.1301P} {376, 1301}

\bibitem[\protect\citeauthoryear{{Patat} et~al.,}{{Patat}
  et~al.}{2007}]{patat07}
{Patat} F.,  et~al., 2007, \mn@doi [Science] {10.1126/science.1143005}, \href
  {https://ui.adsabs.harvard.edu/abs/2007Sci...317..924P} {317, 924}

\bibitem[\protect\citeauthoryear{{Paturel} et~al.,}{{Paturel}
  et~al.}{2003}]{paturel03}
{Paturel} G.,  et~al., 2003, \mn@doi [\aap] {10.1051/0004-6361:20031412}, \href
  {https://ui.adsabs.harvard.edu/#abs/2003A&A...412...57P} {412, 57}

\bibitem[\protect\citeauthoryear{{Pejcha} et~al.,}{{Pejcha}
  et~al.}{2013}]{pejcha13}
{Pejcha} O.,  et~al., 2013, \mn@doi [\mnras] {10.1093/mnras/stt1281}, \href
  {https://ui.adsabs.harvard.edu/#abs/2013MNRAS.435..943P} {435, 943}

\bibitem[\protect\citeauthoryear{{Pessa} et~al.,}{{Pessa}
  et~al.}{2018}]{pymuseref}
{Pessa} I.,  et~al., 2018, preprint, \href
  {https://ui.adsabs.harvard.edu/#abs/2018arXiv180305005P} {p.
  arXiv:1803.05005} (\mn@eprint {arXiv} {1803.05005})

\bibitem[\protect\citeauthoryear{{Petrushevska} et~al.,}{{Petrushevska}
  et~al.}{2016}]{ATel9049}
{Petrushevska} T.,  et~al., 2016, The Astronomer's Telegram, \href
  {https://ui.adsabs.harvard.edu/\#abs/2016ATel.9049....1P} {9049, 1}

\bibitem[\protect\citeauthoryear{{Phillips} et~al.,}{{Phillips}
  et~al.}{1987}]{phillips87}
{Phillips} M.~M.,  et~al., 1987, \mn@doi [Publications of the Astronomical
  Society of the Pacific] {10.1086/132020}, \href
  {https://ui.adsabs.harvard.edu/#abs/1987PASP...99..592P} {99, 592}

\bibitem[\protect\citeauthoryear{{Phillips} et~al.,}{{Phillips}
  et~al.}{2006}]{phillips06}
{Phillips} M.~M.,  et~al., 2006, \mn@doi [\aj] {10.1086/503108}, \href
  {https://ui.adsabs.harvard.edu/\#abs/2006AJ....131.2615P} {131, 2615}

\bibitem[\protect\citeauthoryear{{Phillips} et~al.,}{{Phillips}
  et~al.}{2007}]{phillips07}
{Phillips} M.~M.,  et~al., 2007, \mn@doi [Publications of the Astronomical
  Society of the Pacific] {10.1086/518372}, \href
  {https://ui.adsabs.harvard.edu/\#abs/2007PASP..119..360P} {119, 360}

\bibitem[\protect\citeauthoryear{{Phillips} et~al.,}{{Phillips}
  et~al.}{2013}]{phillips13}
{Phillips} M.~M.,  et~al., 2013, \mn@doi [\apj] {10.1088/0004-637X/779/1/38},
  \href {https://ui.adsabs.harvard.edu/\#abs/2013ApJ...779...38P} {779, 38}

\bibitem[\protect\citeauthoryear{{Phillips} et~al.,}{{Phillips}
  et~al.}{2019}]{phillips19}
{Phillips} M.~M.,  et~al., 2019, \mn@doi [Publications of the Astronomical
  Society of the Pacific] {10.1088/1538-3873/aae8bd}, \href
  {https://ui.adsabs.harvard.edu/\#abs/2019PASP..131a4001P} {131, 014001}

\bibitem[\protect\citeauthoryear{{Pignata} et~al.,}{{Pignata}
  et~al.}{2004}]{pignata04}
{Pignata} G.,  et~al., 2004, \mn@doi [\mnras]
  {10.1111/j.1365-2966.2004.08308.x}, \href
  {https://ui.adsabs.harvard.edu/#abs/2004MNRAS.355..178P} {355, 178}

\bibitem[\protect\citeauthoryear{{Pignata} et~al.,}{{Pignata}
  et~al.}{2008}]{pignata08}
{Pignata} G.,  et~al., 2008, \mn@doi [\mnras]
  {10.1111/j.1365-2966.2008.13434.x}, \href
  {https://ui.adsabs.harvard.edu/#abs/2008MNRAS.388..971P} {388, 971}

\bibitem[\protect\citeauthoryear{{Pignata} et~al.,}{{Pignata}
  et~al.}{2009}]{CBET2022}
{Pignata} G.,  et~al., 2009, Central Bureau Electronic Telegrams, \href
  {https://ui.adsabs.harvard.edu/\#abs/2009CBET.2022....1P} {2022, 1}

\bibitem[\protect\citeauthoryear{{Pignata} et~al.,}{{Pignata}
  et~al.}{2010}]{CBET2344}
{Pignata} G.,  et~al., 2010, Central Bureau Electronic Telegrams, \href
  {https://ui.adsabs.harvard.edu/\#abs/2010CBET.2344....1P} {2344, 1}

\bibitem[\protect\citeauthoryear{{Pisano} et~al.,}{{Pisano}
  et~al.}{2011}]{pisano11}
{Pisano} D.~J.,  et~al., 2011, \mn@doi [The Astrophysical Journal Supplement
  Series] {10.1088/0067-0049/197/2/28}, \href
  {https://ui.adsabs.harvard.edu/\#abs/2011ApJS..197...28P} {197, 28}

\bibitem[\protect\citeauthoryear{{Pogge} et~al.,}{{Pogge}
  et~al.}{2010}]{MODSref}
{Pogge} R.~W.,  et~al., 2010, in Proceedings of the SPIE, Volume 7735, id.
  77350A (2010).. , \mn@doi{10.1117/12.857215}

\bibitem[\protect\citeauthoryear{{Pollas} et~al.,}{{Pollas} \&
  {Klotz}}{2007}]{CBET1121}
{Pollas} C.,  et~al., 2007, Central Bureau Electronic Telegrams, \href
  {https://ui.adsabs.harvard.edu/#abs/2007CBET.1121....1P} {1121, 1}

\bibitem[\protect\citeauthoryear{{Pollas} et~al.,}{{Pollas}
  et~al.}{1989}]{IAUC4742}
{Pollas} C.,  et~al., 1989, International Astronomical Union Circular, \href
  {https://ui.adsabs.harvard.edu/\#abs/1989IAUC.4742....2P} {4742, 2}

\bibitem[\protect\citeauthoryear{{Ponticello} et~al.,}{{Ponticello}
  et~al.}{2006}]{IAUC8667}
{Ponticello} N.~J.,  et~al., 2006, International Astronomical Union Circular,
  \href {https://ui.adsabs.harvard.edu/#abs/2006IAUC.8667....1P} {8667, 1}

\bibitem[\protect\citeauthoryear{{Prentice} et~al.,}{{Prentice} \&
  {Ashall}}{2017}]{2017TNSCR978}
{Prentice} S.,  et~al., 2017, Transient Name Server Classification Report,
  \href {https://ui.adsabs.harvard.edu/\#abs/2017TNSCR.978....1P} {2017-978, 1}

\bibitem[\protect\citeauthoryear{{Press} et~al.,}{{Press}
  et~al.}{1992}]{press92}
{Press} W.~H.,  et~al., 1992, {Numerical recipes in FORTRAN. The art of
  scientific computing}

\bibitem[\protect\citeauthoryear{{Prieto} et~al.,}{{Prieto} \&
  {Morrell}}{2011}]{CBET2613}
{Prieto} J.~L.,  et~al., 2011, Central Bureau Electronic Telegrams, \href
  {https://ui.adsabs.harvard.edu/\#abs/2011CBET.2613....1P} {2613, 1}

\bibitem[\protect\citeauthoryear{{Prieto} et~al.,}{{Prieto}
  et~al.}{2006}]{CBET651}
{Prieto} J.~L.,  et~al., 2006, Central Bureau Electronic Telegrams, \href
  {https://ui.adsabs.harvard.edu/#abs/2006CBET..651....1P} {651, 1}

\bibitem[\protect\citeauthoryear{{Prieto} et~al.,}{{Prieto}
  et~al.}{2010}]{CBET2453}
{Prieto} J.~L.,  et~al., 2010, Central Bureau Electronic Telegrams, \href
  {https://ui.adsabs.harvard.edu/#abs/2010CBET.2453....1P} {2453, 1}

\bibitem[\protect\citeauthoryear{{Prieto} et~al.,}{{Prieto}
  et~al.}{2019}]{prieto19}
{Prieto} J.~L.,  et~al., 2019, arXiv e-prints, \href
  {https://ui.adsabs.harvard.edu/abs/2019arXiv190905267P} {p. arXiv:1909.05267}

\bibitem[\protect\citeauthoryear{{Przybylski}}{{Przybylski}}{1972}]{IAUC2434}
{Przybylski} A.,  1972, International Astronomical Union Circular, \href
  {https://ui.adsabs.harvard.edu/\#abs/1972IAUC.2434....1P} {2434, 1}

\bibitem[\protect\citeauthoryear{{Puckett} et~al.,}{{Puckett}
  et~al.}{2009}]{CBET1762}
{Puckett} T.,  et~al., 2009, Central Bureau Electronic Telegrams, \href
  {https://ui.adsabs.harvard.edu/#abs/2009CBET.1762....1P} {1762, 1}

\bibitem[\protect\citeauthoryear{{Pugh} et~al.,}{{Pugh} \&
  {Li}}{2005}]{CBET158}
{Pugh} H.,  et~al., 2005, Central Bureau Electronic Telegrams, \href
  {https://ui.adsabs.harvard.edu/#abs/2005CBET..158....1P} {158, 1}

\bibitem[\protect\citeauthoryear{{Quimby} et~al.,}{{Quimby}
  et~al.}{2005}]{IAUC8625}
{Quimby} R.,  et~al., 2005, International Astronomical Union Circular, \href
  {https://ui.adsabs.harvard.edu/\#abs/2005IAUC.8625....2Q} {8625, 2}

\bibitem[\protect\citeauthoryear{{Quimby} et~al.,}{{Quimby}
  et~al.}{2006a}]{CBET393}
{Quimby} R.,  et~al., 2006a, Central Bureau Electronic Telegrams, \href
  {https://ui.adsabs.harvard.edu/#abs/2006CBET..393....1Q} {393, 1}

\bibitem[\protect\citeauthoryear{{Quimby} et~al.,}{{Quimby}
  et~al.}{2006b}]{IAUC8723}
{Quimby} R.,  et~al., 2006b, International Astronomical Union Circular, \href
  {https://ui.adsabs.harvard.edu/\#abs/2006IAUC.8723....2Q} {8723, 2}

\bibitem[\protect\citeauthoryear{{Rand}}{{Rand}}{1995}]{rand95}
{Rand} R.~J.,  1995, \mn@doi [\aj] {10.1086/117462}, \href
  {https://ui.adsabs.harvard.edu/#abs/1995AJ....109.2444R} {109, 2444}

\bibitem[\protect\citeauthoryear{{Raskin} et~al.,}{{Raskin} \&
  {Kasen}}{2013}]{raskin13}
{Raskin} C.,  et~al., 2013, \mn@doi [\apj] {10.1088/0004-637X/772/1/1}, \href
  {https://ui.adsabs.harvard.edu/abs/2013ApJ...772....1R} {772, 1}

\bibitem[\protect\citeauthoryear{{Rebassa-Mansergas}
  et~al.,}{{Rebassa-Mansergas} et~al.}{2018}]{rebassa18}
{Rebassa-Mansergas} A.,  et~al., 2018, preprint, \href
  {https://ui.adsabs.harvard.edu/#abs/2018arXiv180907158R} {p.
  arXiv:1809.07158} (\mn@eprint {arXiv} {1809.07158})

\bibitem[\protect\citeauthoryear{{Rhee} et~al.,}{{Rhee} \& {van
  Albada}}{1996}]{rhee96}
{Rhee} M.~H.,  et~al., 1996, Astronomy and Astrophysics Supplement Series,
  \href {https://ui.adsabs.harvard.edu/#abs/1996A&AS..115..407R} {115, 407}

\bibitem[\protect\citeauthoryear{{Richmond} et~al.,}{{Richmond} \&
  {Smith}}{2012}]{richmond12}
{Richmond} M.~W.,  et~al., 2012, Journal of the American Association of
  Variable Star Observers (JAAVSO), \href
  {https://ui.adsabs.harvard.edu/#abs/2012JAVSO..40..872R} {40, 872}

\bibitem[\protect\citeauthoryear{{Richmond} et~al.,}{{Richmond}
  et~al.}{1995}]{richmond95}
{Richmond} M.~W.,  et~al., 1995, \mn@doi [\aj] {10.1086/117437}, \href
  {https://ui.adsabs.harvard.edu/\#abs/1995AJ....109.2121R} {109, 2121}

\bibitem[\protect\citeauthoryear{{Riello} et~al.,}{{Riello}
  et~al.}{2002}]{IAUC7919}
{Riello} M.,  et~al., 2002, International Astronomical Union Circular, \href
  {https://ui.adsabs.harvard.edu/#abs/2002IAUC.7919....2R} {7919, 2}

\bibitem[\protect\citeauthoryear{{Riello} et~al.,}{{Riello}
  et~al.}{2018}]{riello18}
{Riello} M.,  et~al., 2018, preprint, \href
  {https://ui.adsabs.harvard.edu/#abs/2018arXiv180409367R} {p.
  arXiv:1804.09367} (\mn@eprint {arXiv} {1804.09367})

\bibitem[\protect\citeauthoryear{{Riess} et~al.,}{{Riess}
  et~al.}{1999}]{riess99}
{Riess} A.~G.,  et~al., 1999, \mn@doi [\aj] {10.1086/300738}, \href
  {https://ui.adsabs.harvard.edu/#abs/1999AJ....117..707R} {117, 707}

\bibitem[\protect\citeauthoryear{{Riess} et~al.,}{{Riess}
  et~al.}{2005}]{riess05}
{Riess} A.~G.,  et~al., 2005, \mn@doi [\apj] {10.1086/430497}, \href
  {https://ui.adsabs.harvard.edu/#abs/2005ApJ...627..579R} {627, 579}

\bibitem[\protect\citeauthoryear{{Riess} et~al.,}{{Riess}
  et~al.}{2016}]{riess16}
{Riess} A.~G.,  et~al., 2016, \mn@doi [\apj] {10.3847/0004-637X/826/1/56},
  \href {https://ui.adsabs.harvard.edu/#abs/2016ApJ...826...56R} {826}

\bibitem[\protect\citeauthoryear{{Romero-Canizales} et~al.,}{{Romero-Canizales}
  et~al.}{2014}]{ATel6618}
{Romero-Canizales} C.,  et~al., 2014, The Astronomer's Telegram, \href
  {https://ui.adsabs.harvard.edu/\#abs/2014ATel.6618....1R} {6618, 1}

\bibitem[\protect\citeauthoryear{{R{\"o}pke} et~al.,}{{R{\"o}pke}
  et~al.}{2012}]{ropke12}
{R{\"o}pke} F.~K.,  et~al., 2012, \mn@doi [\apj] {10.1088/2041-8205/750/1/L19},
  \href {https://ui.adsabs.harvard.edu/#abs/2012ApJ...750L..19R} {750, L19}

\bibitem[\protect\citeauthoryear{{Rothberg} et~al.,}{{Rothberg} \&
  {Joseph}}{2006}]{rothberg06}
{Rothberg} B.,  et~al., 2006, \mn@doi [\aj] {10.1086/498452}, \href
  {https://ui.adsabs.harvard.edu/#abs/2006AJ....131..185R} {131, 185}

\bibitem[\protect\citeauthoryear{{Rudy} et~al.,}{{Rudy}
  et~al.}{2015}]{ATel7825}
{Rudy} R.~J.,  et~al., 2015, The Astronomer's Telegram, \href
  {https://ui.adsabs.harvard.edu/\#abs/2015ATel.7825....1R} {7825, 1}

\bibitem[\protect\citeauthoryear{{Ruffini} et~al.,}{{Ruffini} \&
  {Casey}}{2019}]{ruffini19}
{Ruffini} N.~J.,  et~al., 2019, \mn@doi [\mnras] {10.1093/mnras/stz2176}, \href
  {https://ui.adsabs.harvard.edu/abs/2019MNRAS.489..420R} {489, 420}

\bibitem[\protect\citeauthoryear{{Saha} et~al.,}{{Saha} et~al.}{2006}]{saha06}
{Saha} A.,  et~al., 2006, \mn@doi [The Astrophysical Journal Supplement Series]
  {10.1086/503800}, \href
  {https://ui.adsabs.harvard.edu/#abs/2006ApJS..165..108S} {165, 108}

\bibitem[\protect\citeauthoryear{{Sahu} et~al.,}{{Sahu} et~al.}{2008}]{sahu08}
{Sahu} D.~K.,  et~al., 2008, \mn@doi [\apj] {10.1086/587772}, \href
  {https://ui.adsabs.harvard.edu/\#abs/2008ApJ...680..580S} {680, 580}

\bibitem[\protect\citeauthoryear{{Sako} et~al.,}{{Sako} et~al.}{2014}]{sako14}
{Sako} M.,  et~al., 2014, preprint, \href
  {https://ui.adsabs.harvard.edu/#abs/2014arXiv1401.3317S} {} (\mn@eprint
  {arXiv} {1401.3317})

\bibitem[\protect\citeauthoryear{{Salgado} et~al.,}{{Salgado}
  et~al.}{2007}]{CBET865}
{Salgado} F.,  et~al., 2007, Central Bureau Electronic Telegrams, \href
  {https://ui.adsabs.harvard.edu/#abs/2007CBET..865....1S} {865, 1}

\bibitem[\protect\citeauthoryear{{Salvo} et~al.,}{{Salvo}
  et~al.}{2001}]{salvo01}
{Salvo} M.~E.,  et~al., 2001, \mn@doi [\mnras]
  {10.1046/j.1365-8711.2001.03995.x}, \href
  {https://ui.adsabs.harvard.edu/#abs/2001MNRAS.321..254S} {321, 254}

\bibitem[\protect\citeauthoryear{{Salvo} et~al.,}{{Salvo}
  et~al.}{2006}]{CBET557}
{Salvo} M.,  et~al., 2006, Central Bureau Electronic Telegrams, \href
  {https://ui.adsabs.harvard.edu/\#abs/2006CBET..557....1S} {557, 1}

\bibitem[\protect\citeauthoryear{{Sand} et~al.,}{{Sand}
  et~al.}{2017}]{ATel10569}
{Sand} D.~J.,  et~al., 2017, The Astronomer's Telegram, \href
  {https://ui.adsabs.harvard.edu/#abs/2017ATel10569....1S} {10569, 1}

\bibitem[\protect\citeauthoryear{{Sand} et~al.,}{{Sand} et~al.}{2018a}]{sand18}
{Sand} D.~J.,  et~al., 2018a, preprint, \href
  {https://ui.adsabs.harvard.edu/#abs/2018arXiv180403666S} {p.
  arXiv:1804.03666} (\mn@eprint {arXiv} {1804.03666})

\bibitem[\protect\citeauthoryear{{Sand} et~al.,}{{Sand}
  et~al.}{2018b}]{ATel11328}
{Sand} D.,  et~al., 2018b, The Astronomer's Telegram, \href
  {https://ui.adsabs.harvard.edu/\#abs/2018ATel11328....1S} {11328, 1}

\bibitem[\protect\citeauthoryear{{Sand} et~al.,}{{Sand}
  et~al.}{2018c}]{ATel11330}
{Sand} D.,  et~al., 2018c, The Astronomer's Telegram, \href
  {https://ui.adsabs.harvard.edu/\#abs/2018ATel11330....1S} {11330, 1}

\bibitem[\protect\citeauthoryear{{Sand} et~al.,}{{Sand}
  et~al.}{2018d}]{ATel11371}
{Sand} D.,  et~al., 2018d, The Astronomer's Telegram, \href
  {https://ui.adsabs.harvard.edu/\#abs/2018ATel11371....1S} {11371, 1}

\bibitem[\protect\citeauthoryear{{Sand} et~al.,}{{Sand} et~al.}{2019}]{sand19}
{Sand} D.~J.,  et~al., 2019, \mn@doi [\apjl] {10.3847/2041-8213/ab1eaf}, \href
  {https://ui.adsabs.harvard.edu/abs/2019ApJ...877L...4S} {877, L4}

\bibitem[\protect\citeauthoryear{{Saulder} et~al.,}{{Saulder}
  et~al.}{2016}]{saulder16}
{Saulder} C.,  et~al., 2016, \mn@doi [\aap] {10.1051/0004-6361/201526711},
  \href {https://ui.adsabs.harvard.edu/\#abs/2016A&A...596A..14S} {596, A14}

\bibitem[\protect\citeauthoryear{{Scalzo} et~al.,}{{Scalzo}
  et~al.}{2010}]{scalzo10}
{Scalzo} R.~A.,  et~al., 2010, \mn@doi [\apj] {10.1088/0004-637X/713/2/1073},
  \href {https://ui.adsabs.harvard.edu/\#abs/2010ApJ...713.1073S} {713, 1073}

\bibitem[\protect\citeauthoryear{{Scalzo} et~al.,}{{Scalzo}
  et~al.}{2014}]{scalzo14}
{Scalzo} R.~A.,  et~al., 2014, \mn@doi [\mnras] {10.1093/mnras/stu1808}, \href
  {https://ui.adsabs.harvard.edu/#abs/2014MNRAS.445.2535S} {445, 2535}

\bibitem[\protect\citeauthoryear{{Scalzo} et~al.,}{{Scalzo}
  et~al.}{2019}]{scalzo18}
{Scalzo} R.~A.,  et~al., 2019, \mn@doi [\mnras] {10.1093/mnras/sty3178}, \href
  {https://ui.adsabs.harvard.edu/\#abs/2019MNRAS.483..628S} {483, 628}

\bibitem[\protect\citeauthoryear{{Schaefer}}{{Schaefer}}{1987}]{IAUC4421}
{Schaefer} B.,  1987, International Astronomical Union Circular, \href
  {https://ui.adsabs.harvard.edu/#abs/1987IAUC.4421....1S} {4421, 1}

\bibitem[\protect\citeauthoryear{{Schlegel} et~al.,}{{Schlegel}
  et~al.}{1998}]{schlegel98}
{Schlegel} D.~J.,  et~al., 1998, \mn@doi [\apj] {10.1086/305772}, \href
  {https://ui.adsabs.harvard.edu/\#abs/1998ApJ...500..525S} {500, 525}

\bibitem[\protect\citeauthoryear{{Schmidt} et~al.,}{{Schmidt}
  et~al.}{1994}]{schmidt94}
{Schmidt} B.~P.,  et~al., 1994, \mn@doi [\apj] {10.1086/187562}, \href
  {https://ui.adsabs.harvard.edu/#abs/1994ApJ...434L..19S} {434, L19}

\bibitem[\protect\citeauthoryear{{Schneider} et~al.,}{{Schneider}
  et~al.}{1990}]{schneider90}
{Schneider} S.~E.,  et~al., 1990, \mn@doi [The Astrophysical Journal Supplement
  Series] {10.1086/191416}, \href
  {https://ui.adsabs.harvard.edu/\#abs/1990ApJS...72..245S} {72, 245}

\bibitem[\protect\citeauthoryear{{Schneider} et~al.,}{{Schneider}
  et~al.}{1992}]{schneider92}
{Schneider} S.~E.,  et~al., 1992, \mn@doi [The Astrophysical Journal Supplement
  Series] {10.1086/191684}, \href
  {https://ui.adsabs.harvard.edu/#abs/1992ApJS...81....5S} {81, 5}

\bibitem[\protect\citeauthoryear{{Schwartz} et~al.,}{{Schwartz} \&
  {Holvorcem}}{2003}]{IAUC8121}
{Schwartz} M.,  et~al., 2003, International Astronomical Union Circular, \href
  {https://ui.adsabs.harvard.edu/#abs/2003IAUC.8121....1S} {8121, 1}

\bibitem[\protect\citeauthoryear{{Serduke} et~al.,}{{Serduke}
  et~al.}{2005}]{CBET269}
{Serduke} F.~J.~D.,  et~al., 2005, Central Bureau Electronic Telegrams, \href
  {https://ui.adsabs.harvard.edu/\#abs/2005CBET..269....1S} {269, 1}

\bibitem[\protect\citeauthoryear{{Shappee} et~al.,}{{Shappee} \&
  {Thompson}}{2013}]{shappeekozai}
{Shappee} B.~J.,  et~al., 2013, \mn@doi [\apj] {10.1088/0004-637X/766/1/64},
  \href {https://ui.adsabs.harvard.edu/#abs/2013ApJ...766...64S} {766}

\bibitem[\protect\citeauthoryear{{Shappee} et~al.,}{{Shappee}
  et~al.}{2013a}]{shappee13}
{Shappee} B.~J.,  et~al., 2013a, \mn@doi [\apj] {10.1088/2041-8205/762/1/L5},
  \href {https://ui.adsabs.harvard.edu/#abs/2013ApJ...762L...5S} {762}

\bibitem[\protect\citeauthoryear{{Shappee} et~al.,}{{Shappee}
  et~al.}{2013b}]{shappee_overlum}
{Shappee} B.~J.,  et~al., 2013b, \mn@doi [\apj] {10.1088/0004-637X/765/2/150},
  \href {https://ui.adsabs.harvard.edu/#abs/2013ApJ...765..150S} {765, 150}

\bibitem[\protect\citeauthoryear{{Shappee} et~al.,}{{Shappee}
  et~al.}{2014a}]{shappee14}
{Shappee} B.~J.,  et~al., 2014a, \mn@doi [\apj] {10.1088/0004-637X/788/1/48},
  \href {https://ui.adsabs.harvard.edu/\#abs/2014ApJ...788...48S} {788, 48}

\bibitem[\protect\citeauthoryear{{Shappee} et~al.,}{{Shappee}
  et~al.}{2014b}]{ATel6812}
{Shappee} B.~J.,  et~al., 2014b, The Astronomer's Telegram, \href
  {https://ui.adsabs.harvard.edu/\#abs/2014ATel.6812....1S} {6812, 1}

\bibitem[\protect\citeauthoryear{{Shappee} et~al.,}{{Shappee}
  et~al.}{2015}]{ATel6882}
{Shappee} B.~J.,  et~al., 2015, The Astronomer's Telegram, \href
  {https://ui.adsabs.harvard.edu/\#abs/2015ATel.6882....1S} {6882, 1}

\bibitem[\protect\citeauthoryear{{Shappee} et~al.,}{{Shappee}
  et~al.}{2017}]{shappee17}
{Shappee} B.~J.,  et~al., 2017, \mn@doi [\apj] {10.3847/1538-4357/aa6eab},
  \href {https://ui.adsabs.harvard.edu/#abs/2017ApJ...841...48S} {841, 48}

\bibitem[\protect\citeauthoryear{{Shappee} et~al.,}{{Shappee}
  et~al.}{2018}]{shappee18}
{Shappee} B.~J.,  et~al., 2018, \mn@doi [\apj] {10.3847/1538-4357/aaa1e9},
  \href {https://ui.adsabs.harvard.edu/\#abs/2018ApJ...855....6S} {855, 6}

\bibitem[\protect\citeauthoryear{{Shappee} et~al.,}{{Shappee}
  et~al.}{2019}]{shappee18bt}
{Shappee} B.~J.,  et~al., 2019, \mn@doi [\apj] {10.3847/1538-4357/aaec79},
  \href {https://ui.adsabs.harvard.edu/abs/2019ApJ...870...13S} {870, 13}

\bibitem[\protect\citeauthoryear{{Sheinis} et~al.,}{{Sheinis}
  et~al.}{2002}]{ESIref}
{Sheinis} A.~I.,  et~al., 2002, \mn@doi [Publications of the Astronomical
  Society of the Pacific] {10.1086/341706}, \href
  {https://ui.adsabs.harvard.edu/#abs/2002PASP..114..851S} {114, 851}

\bibitem[\protect\citeauthoryear{{Shen} et~al.,}{{Shen} et~al.}{2018}]{shen18}
{Shen} K.~J.,  et~al., 2018, preprint, \href
  {https://ui.adsabs.harvard.edu/#abs/2018arXiv180411163S} {p.
  arXiv:1804.11163} (\mn@eprint {arXiv} {1804.11163})

\bibitem[\protect\citeauthoryear{{Silverman} et~al.,}{{Silverman}
  et~al.}{2011}]{silverman11}
{Silverman} J.~M.,  et~al., 2011, \mn@doi [\mnras]
  {10.1111/j.1365-2966.2010.17474.x}, \href
  {https://ui.adsabs.harvard.edu/\#abs/2011MNRAS.410..585S} {410, 585}

\bibitem[\protect\citeauthoryear{{Silverman} et~al.,}{{Silverman}
  et~al.}{2012}]{silverman12}
{Silverman} J.~M.,  et~al., 2012, \mn@doi [\mnras]
  {10.1111/j.1365-2966.2012.21270.x}, \href
  {https://ui.adsabs.harvard.edu/#abs/2012MNRAS.425.1789S} {425, 1789}

\bibitem[\protect\citeauthoryear{{Silverman} et~al.,}{{Silverman}
  et~al.}{2013}]{silverman13}
{Silverman} J.~M.,  et~al., 2013, \mn@doi [\mnras] {10.1093/mnras/sts674},
  \href {https://ui.adsabs.harvard.edu/#abs/2013MNRAS.430.1030S} {430, 1030}

\bibitem[\protect\citeauthoryear{{Simon} et~al.,}{{Simon}
  et~al.}{2009}]{simon09}
{Simon} J.~D.,  et~al., 2009, \mn@doi [\apj] {10.1088/0004-637X/702/2/1157},
  \href {https://ui.adsabs.harvard.edu/abs/2009ApJ...702.1157S} {702, 1157}

\bibitem[\protect\citeauthoryear{{Smartt} et~al.,}{{Smartt}
  et~al.}{2002}]{IAUC7961}
{Smartt} S.~J.,  et~al., 2002, International Astronomical Union Circular, \href
  {https://ui.adsabs.harvard.edu/#abs/2002IAUC.7961....2S} {7961, 2}

\bibitem[\protect\citeauthoryear{{Smartt} et~al.,}{{Smartt}
  et~al.}{2015a}]{smartt15}
{Smartt} S.~J.,  et~al., 2015a, \mn@doi [\aap] {10.1051/0004-6361/201425237},
  \href {https://ui.adsabs.harvard.edu/#abs/2015A&A...579A..40S} {579}

\bibitem[\protect\citeauthoryear{{Smartt} et~al.,}{{Smartt}
  et~al.}{2015b}]{PESSTO}
{Smartt} S.~J.,  et~al., 2015b, \mn@doi [\aap] {10.1051/0004-6361/201425237},
  \href {https://ui.adsabs.harvard.edu/\#abs/2015A&A...579A..40S} {579, A40}

\bibitem[\protect\citeauthoryear{{Smith} et~al.,}{{Smith}
  et~al.}{2000}]{smith00}
{Smith} R.~J.,  et~al., 2000, \mn@doi [\mnras]
  {10.1046/j.1365-8711.2000.03251.x}, \href
  {https://ui.adsabs.harvard.edu/\#abs/2000MNRAS.313..469S} {313, 469}

\bibitem[\protect\citeauthoryear{{Smoker} et~al.,}{{Smoker}
  et~al.}{2000}]{smoker00}
{Smoker} J.~V.,  et~al., 2000, \aap, \href
  {https://ui.adsabs.harvard.edu/#abs/2000A&A...361...19S} {361, 19}

\bibitem[\protect\citeauthoryear{{Sollerman} et~al.,}{{Sollerman}
  et~al.}{2001}]{IAUC7723}
{Sollerman} J.,  et~al., 2001, International Astronomical Union Circular, \href
  {https://ui.adsabs.harvard.edu/\#abs/2001IAUC.7723....2S} {7723, 2}

\bibitem[\protect\citeauthoryear{{Sollerman} et~al.,}{{Sollerman}
  et~al.}{2004}]{sollerman04}
{Sollerman} J.,  et~al., 2004, \mn@doi [\aap] {10.1051/0004-6361:20041320},
  \href {https://ui.adsabs.harvard.edu/#abs/2004A&A...428..555S} {428, 555}

\bibitem[\protect\citeauthoryear{{Springob} et~al.,}{{Springob}
  et~al.}{2014}]{springob14}
{Springob} C.~M.,  et~al., 2014, \mn@doi [\mnras] {10.1093/mnras/stu1743},
  \href {https://ui.adsabs.harvard.edu/\#abs/2014MNRAS.445.2677S} {445, 2677}

\bibitem[\protect\citeauthoryear{{Spyromilio} et~al.,}{{Spyromilio}
  et~al.}{2004}]{spyromilio04}
{Spyromilio} J.,  et~al., 2004, \mn@doi [\aap] {10.1051/0004-6361:20040570},
  \href {https://ui.adsabs.harvard.edu/#abs/2004A&A...426..547S} {426, 547}

\bibitem[\protect\citeauthoryear{{Srivastav} et~al.,}{{Srivastav}
  et~al.}{2016}]{srivastav16}
{Srivastav} S.,  et~al., 2016, \mn@doi [\mnras] {10.1093/mnras/stw039}, \href
  {https://ui.adsabs.harvard.edu/#abs/2016MNRAS.457.1000S} {457, 1000}

\bibitem[\protect\citeauthoryear{{Stanek}}{{Stanek}}{2018}]{2018TNSTR16974}
{Stanek} K.~Z.,  2018, Transient Name Server Discovery Report, \href
  {https://ui.adsabs.harvard.edu/\#abs/2018TNSTR.234....1S} {2018-234, 1}

\bibitem[\protect\citeauthoryear{{Stanek} et~al.,}{{Stanek}
  et~al.}{2014}]{ATel6830}
{Stanek} K.~Z.,  et~al., 2014, The Astronomer's Telegram, \href
  {https://ui.adsabs.harvard.edu/\#abs/2014ATel.6830....1S} {6830, 1}

\bibitem[\protect\citeauthoryear{{Stanishev} et~al.,}{{Stanishev} \&
  {Pursimo}}{2008}]{CBET1232}
{Stanishev} V.,  et~al., 2008, Central Bureau Electronic Telegrams, \href
  {https://ui.adsabs.harvard.edu/#abs/2008CBET.1232....1S} {1232, 1}

\bibitem[\protect\citeauthoryear{{Stanishev} et~al.,}{{Stanishev}
  et~al.}{2007}]{stanishev07}
{Stanishev} V.,  et~al., 2007, \mn@doi [\aap] {10.1051/0004-6361:20066020},
  \href {https://ui.adsabs.harvard.edu/#abs/2007A&A...469..645S} {469, 645}

\bibitem[\protect\citeauthoryear{{Sternberg} et~al.,}{{Sternberg}
  et~al.}{2011}]{sternberg11}
{Sternberg} A.,  et~al., 2011, \mn@doi [Science] {10.1126/science.1203836},
  \href {https://ui.adsabs.harvard.edu/abs/2011Sci...333..856S} {333, 856}

\bibitem[\protect\citeauthoryear{{Sternberg} et~al.,}{{Sternberg}
  et~al.}{2014}]{sternberg13}
{Sternberg} A.,  et~al., 2014, \mn@doi [\mnras] {10.1093/mnras/stu1202}, \href
  {https://ui.adsabs.harvard.edu/abs/2014MNRAS.443.1849S} {443, 1849}

\bibitem[\protect\citeauthoryear{{Stone} et~al.,}{{Stone}
  et~al.}{2018}]{ATel11343}
{Stone} G.,  et~al., 2018, The Astronomer's Telegram, \href
  {https://ui.adsabs.harvard.edu/\#abs/2018ATel11343....1S} {11343, 1}

\bibitem[\protect\citeauthoryear{{Strauss} et~al.,}{{Strauss}
  et~al.}{1992}]{strauss92}
{Strauss} M.~A.,  et~al., 1992, \mn@doi [The Astrophysical Journal Supplement
  Series] {10.1086/191730}, \href
  {https://ui.adsabs.harvard.edu/#abs/1992ApJS...83...29S} {83, 29}

\bibitem[\protect\citeauthoryear{{Stritzinger}}{{Stritzinger}}{2010}]{CBET2346}
{Stritzinger} M.,  2010, Central Bureau Electronic Telegrams, \href
  {https://ui.adsabs.harvard.edu/\#abs/2010CBET.2346....1S} {2346, 1}

\bibitem[\protect\citeauthoryear{{Stritzinger} et~al.,}{{Stritzinger}
  et~al.}{2006a}]{stritzinger06}
{Stritzinger} M.,  et~al., 2006a, \mn@doi [\aap] {10.1051/0004-6361:20065514},
  \href {https://ui.adsabs.harvard.edu/\#abs/2006A&A...460..793S} {460, 793}

\bibitem[\protect\citeauthoryear{{Stritzinger} et~al.,}{{Stritzinger}
  et~al.}{2006b}]{stritzinger06-ni}
{Stritzinger} M.,  et~al., 2006b, \mn@doi [\aap] {10.1051/0004-6361:20065514},
  \href {https://ui.adsabs.harvard.edu/abs/2006A&A...460..793S} {460, 793}

\bibitem[\protect\citeauthoryear{{Stritzinger} et~al.,}{{Stritzinger}
  et~al.}{2010}]{stritzinger10}
{Stritzinger} M.,  et~al., 2010, \mn@doi [\aj] {10.1088/0004-6256/140/6/2036},
  \href {https://ui.adsabs.harvard.edu/\#abs/2010AJ....140.2036S} {140, 2036}

\bibitem[\protect\citeauthoryear{{Stritzinger} et~al.,}{{Stritzinger}
  et~al.}{2011}]{stritzinger11}
{Stritzinger} M.~D.,  et~al., 2011, \mn@doi [\aj]
  {10.1088/0004-6256/142/5/156}, \href
  {https://ui.adsabs.harvard.edu/#abs/2011AJ....142..156S} {142}

\bibitem[\protect\citeauthoryear{{Stritzinger} et~al.,}{{Stritzinger}
  et~al.}{2015}]{stritzinger15}
{Stritzinger} M.~D.,  et~al., 2015, \mn@doi [\aap]
  {10.1051/0004-6361/201424168}, \href
  {https://ui.adsabs.harvard.edu/\#abs/2015A&A...573A...2S} {573, A2}

\bibitem[\protect\citeauthoryear{{Strohmayer} et~al.,}{{Strohmayer}
  et~al.}{1996}]{IAUC6484}
{Strohmayer} T.,  et~al., 1996, International Astronomical Union Circular,
  \href {https://ui.adsabs.harvard.edu/#abs/1996IAUC.6484....1S} {6484, 1}

\bibitem[\protect\citeauthoryear{{Suntzeff} et~al.,}{{Suntzeff}
  et~al.}{1989}]{IAUC4728}
{Suntzeff} N.,  et~al., 1989, International Astronomical Union Circular, \href
  {https://ui.adsabs.harvard.edu/\#abs/1989IAUC.4728....1S} {4728, 1}

\bibitem[\protect\citeauthoryear{{Suntzeff} et~al.,}{{Suntzeff}
  et~al.}{1992}]{IAUC5432}
{Suntzeff} N.,  et~al., 1992, International Astronomical Union Circular, \href
  {https://ui.adsabs.harvard.edu/#abs/1992IAUC.5432....2S} {5432, 2}

\bibitem[\protect\citeauthoryear{{Suntzeff} et~al.,}{{Suntzeff}
  et~al.}{1996}]{IAUC6381}
{Suntzeff} N.~B.,  et~al., 1996, International Astronomical Union Circular,
  \href {https://ui.adsabs.harvard.edu/#abs/1996IAUC.6381....1S} {6381, 1}

\bibitem[\protect\citeauthoryear{{Suntzeff} et~al.,}{{Suntzeff}
  et~al.}{1999}]{suntzeff99}
{Suntzeff} N.~B.,  et~al., 1999, \mn@doi [\aj] {10.1086/300771}, \href
  {https://ui.adsabs.harvard.edu/#abs/1999AJ....117.1175S} {117, 1175}

\bibitem[\protect\citeauthoryear{{Suntzeff} et~al.,}{{Suntzeff}
  et~al.}{2004}]{IAUC8283}
{Suntzeff} N.,  et~al., 2004, International Astronomical Union Circular, \href
  {https://ui.adsabs.harvard.edu/#abs/2004IAUC.8283....1S} {8283, 1}

\bibitem[\protect\citeauthoryear{{Szab{\'o}} et~al.,}{{Szab{\'o}}
  et~al.}{2003}]{szabo03}
{Szab{\'o}} G.~M.,  et~al., 2003, \mn@doi [\aap] {10.1051/0004-6361:20031008},
  \href {https://ui.adsabs.harvard.edu/\#abs/2003A&A...408..915S} {408, 915}

\bibitem[\protect\citeauthoryear{{Taam}}{{Taam}}{1980}]{taam80}
{Taam} R.~E.,  1980, \mn@doi [\apj] {10.1086/157852}, \href
  {https://ui.adsabs.harvard.edu/#abs/1980ApJ...237..142T} {237, 142}

\bibitem[\protect\citeauthoryear{{Tartaglia} et~al.,}{{Tartaglia}
  et~al.}{2017a}]{ATel10158}
{Tartaglia} L.,  et~al., 2017a, The Astronomer's Telegram, \href
  {https://ui.adsabs.harvard.edu/#abs/2017ATel10158....1T} {10158, 1}

\bibitem[\protect\citeauthoryear{{Tartaglia} et~al.,}{{Tartaglia}
  et~al.}{2017b}]{ATel10260}
{Tartaglia} L.,  et~al., 2017b, The Astronomer's Telegram, \href
  {https://ui.adsabs.harvard.edu/#abs/2017ATel10260....1T} {10260, 1}

\bibitem[\protect\citeauthoryear{{Tartaglia} et~al.,}{{Tartaglia}
  et~al.}{2017c}]{ATel10439}
{Tartaglia} L.,  et~al., 2017c, The Astronomer's Telegram, \href
  {https://ui.adsabs.harvard.edu/#abs/2017ATel10439....1T} {10439, 1}

\bibitem[\protect\citeauthoryear{{Tartaglia} et~al.,}{{Tartaglia}
  et~al.}{2017d}]{ATel10629}
{Tartaglia} L.,  et~al., 2017d, The Astronomer's Telegram, \href
  {https://ui.adsabs.harvard.edu/#abs/2017ATel10629....1T} {10629, 1}

\bibitem[\protect\citeauthoryear{{Taubenberger} et~al.,}{{Taubenberger}
  et~al.}{2013a}]{taubenberger13}
{Taubenberger} S.,  et~al., 2013a, \mn@doi [\mnras] {10.1093/mnras/stt668},
  \href {https://ui.adsabs.harvard.edu/#abs/2013MNRAS.432.3117T} {432, 3117}

\bibitem[\protect\citeauthoryear{{Taubenberger} et~al.,}{{Taubenberger}
  et~al.}{2013b}]{taubenberger10lp}
{Taubenberger} S.,  et~al., 2013b, \mn@doi [\apj]
  {10.1088/2041-8205/775/2/L43}, \href
  {https://ui.adsabs.harvard.edu/\#abs/2013ApJ...775L..43T} {775, L43}

\bibitem[\protect\citeauthoryear{{Taubenberger} et~al.,}{{Taubenberger}
  et~al.}{2015}]{taubenberger15}
{Taubenberger} S.,  et~al., 2015, \mn@doi [\mnras] {10.1093/mnrasl/slu201},
  \href {https://ui.adsabs.harvard.edu/#abs/2015MNRAS.448L..48T} {448, L48}

\bibitem[\protect\citeauthoryear{{Terreran} et~al.,}{{Terreran}
  et~al.}{2016}]{ATel9403}
{Terreran} G.,  et~al., 2016, The Astronomer's Telegram, \href
  {https://ui.adsabs.harvard.edu/\#abs/2016ATel.9403....1T} {9403, 1}

\bibitem[\protect\citeauthoryear{{The Astropy Collaboration} et~al.,}{{The
  Astropy Collaboration} et~al.}{2018}]{astropyref}
{The Astropy Collaboration} et~al., 2018, preprint, \href
  {https://ui.adsabs.harvard.edu/#abs/2018arXiv180102634T} {p.
  arXiv:1801.02634} (\mn@eprint {arXiv} {1801.02634})

\bibitem[\protect\citeauthoryear{{Theureau} et~al.,}{{Theureau}
  et~al.}{1998}]{theureau98}
{Theureau} G.,  et~al., 1998, \mn@doi [Astronomy and Astrophysics Supplement
  Series] {10.1051/aas:1998416}, \href
  {https://ui.adsabs.harvard.edu/#abs/1998A&AS..130..333T} {130, 333}

\bibitem[\protect\citeauthoryear{{Theureau} et~al.,}{{Theureau}
  et~al.}{2007}]{theureau07}
{Theureau} G.,  et~al., 2007, \mn@doi [\aap] {10.1051/0004-6361:20066187},
  \href {https://ui.adsabs.harvard.edu/#abs/2007A&A...465...71T} {465, 71}

\bibitem[\protect\citeauthoryear{{Thompson}}{{Thompson}}{2011}]{thompson11}
{Thompson} T.~A.,  2011, \mn@doi [\apj] {10.1088/0004-637X/741/2/82}, \href
  {https://ui.adsabs.harvard.edu/#abs/2011ApJ...741...82T} {741, 82}

\bibitem[\protect\citeauthoryear{{Tinella}}{{Tinella}}{2016}]{2016TNSTR305}
{Tinella} V.,  2016, Transient Name Server Discovery Report, \href
  {https://ui.adsabs.harvard.edu/\#abs/2016TNSTR.305....1T} {2016-305, 1}

\bibitem[\protect\citeauthoryear{{Tonry} et~al.,}{{Tonry}
  et~al.}{2001}]{tonry01}
{Tonry} J.~L.,  et~al., 2001, \mn@doi [\apj] {10.1086/318301}, \href
  {https://ui.adsabs.harvard.edu/#abs/2001ApJ...546..681T} {546, 681}

\bibitem[\protect\citeauthoryear{{Tonry} et~al.,}{{Tonry}
  et~al.}{2012}]{tonry12}
{Tonry} J.~L.,  et~al., 2012, \mn@doi [\apj] {10.1088/0004-637X/750/2/99},
  \href {https://ui.adsabs.harvard.edu/#abs/2012ApJ...750...99T} {750, 99}

\bibitem[\protect\citeauthoryear{{Tonry} et~al.,}{{Tonry}
  et~al.}{2016}]{TNSTR583}
{Tonry} J.,  et~al., 2016, Transient Name Server Discovery Report, \href
  {https://ui.adsabs.harvard.edu/\#abs/2016TNSTR.583....1T} {2016-583, 1}

\bibitem[\protect\citeauthoryear{{Tonry} et~al.,}{{Tonry}
  et~al.}{2017a}]{TNSTR1371}
{Tonry} J.,  et~al., 2017a, Transient Name Server Discovery Report, \href
  {https://ui.adsabs.harvard.edu/#abs/2017TNSTR1371....1T} {2017-1371, 1}

\bibitem[\protect\citeauthoryear{{Tonry} et~al.,}{{Tonry}
  et~al.}{2017b}]{2017TNSTR15357}
{Tonry} J.,  et~al., 2017b, Transient Name Server Discovery Report, \href
  {https://ui.adsabs.harvard.edu/\#abs/2017TNSTR1431....1T} {2017-1431, 1}

\bibitem[\protect\citeauthoryear{{Tonry} et~al.,}{{Tonry}
  et~al.}{2017c}]{2017TNSTR361}
{Tonry} J.,  et~al., 2017c, Transient Name Server Discovery Report, \href
  {https://ui.adsabs.harvard.edu/\#abs/2017TNSTR.361....1T} {2017-361, 1}

\bibitem[\protect\citeauthoryear{{Tonry} et~al.,}{{Tonry}
  et~al.}{2017d}]{TNSTR860}
{Tonry} J.,  et~al., 2017d, Transient Name Server Discovery Report, \href
  {https://ui.adsabs.harvard.edu/\#abs/2017TNSTR.860....1T} {2017-860, 1}

\bibitem[\protect\citeauthoryear{{Toth} et~al.,}{{Toth} \&
  {Szab{\'o}}}{2000}]{toth00}
{Toth} I.,  et~al., 2000, \aap, \href
  {https://ui.adsabs.harvard.edu/#abs/2000A&A...361...63T} {361, 63}

\bibitem[\protect\citeauthoryear{{Treffers} et~al.,}{{Treffers}
  et~al.}{1993}]{IAUC5870}
{Treffers} R.~R.,  et~al., 1993, International Astronomical Union Circular,
  \href {https://ui.adsabs.harvard.edu/#abs/1993IAUC.5870....3T} {5870, 3}

\bibitem[\protect\citeauthoryear{{Treffers} et~al.,}{{Treffers}
  et~al.}{1994}]{IAUC5946}
{Treffers} R.~R.,  et~al., 1994, International Astronomical Union Circular,
  \href {https://ui.adsabs.harvard.edu/\#abs/1994IAUC.5946....2T} {5946, 2}

\bibitem[\protect\citeauthoryear{{Tsvetkov}}{{Tsvetkov}}{1982}]{tsvetkov82}
{Tsvetkov} D.~Y.,  1982, Soviet Astronomy Letters, \href
  {https://ui.adsabs.harvard.edu/#abs/1982SvAL....8..115T} {8, 115}

\bibitem[\protect\citeauthoryear{{Tsvetkov}}{{Tsvetkov}}{1986}]{tsvetkov86}
{Tsvetkov} D.~Y.,  1986, Peremennye Zvezdy, \href
  {https://ui.adsabs.harvard.edu/#abs/1986PZ.....22..279T} {22, 279}

\bibitem[\protect\citeauthoryear{{Tsvetkov} et~al.,}{{Tsvetkov} \&
  {Pavlyuk}}{1997}]{tsvetkov97}
{Tsvetkov} D.~Y.,  et~al., 1997, Astronomy Letters, \href
  {https://ui.adsabs.harvard.edu/#abs/1997AstL...23...26T} {23, 26}

\bibitem[\protect\citeauthoryear{{Tsvetkov} et~al.,}{{Tsvetkov}
  et~al.}{1990}]{tsvetkov90}
{Tsvetkov} D.~Y.,  et~al., 1990, \aap, \href
  {https://ui.adsabs.harvard.edu/\#abs/1990A&A...236..133T} {236, 133}

\bibitem[\protect\citeauthoryear{{Tsvetkov} et~al.,}{{Tsvetkov}
  et~al.}{2013}]{tsvetkov13}
{Tsvetkov} D.~Y.,  et~al., 2013, Contributions of the Astronomical Observatory
  Skalnate Pleso, \href
  {https://ui.adsabs.harvard.edu/#abs/2013CoSka..43...94T} {43, 94}

\bibitem[\protect\citeauthoryear{{Tsvetkov} et~al.,}{{Tsvetkov}
  et~al.}{2014}]{tsvetkov14}
{Tsvetkov} D.~Y.,  et~al., 2014, Contributions of the Astronomical Observatory
  Skalnate Pleso, \href
  {https://ui.adsabs.harvard.edu/#abs/2014CoSka..44...67T} {44, 67}

\bibitem[\protect\citeauthoryear{{Tucker} et~al.,}{{Tucker}
  et~al.}{2018}]{tucker18}
{Tucker} M.~A.,  et~al., 2018, arXiv e-prints, \href
  {https://ui.adsabs.harvard.edu/\#abs/2018arXiv181109635T} {p.
  arXiv:1811.09635}

\bibitem[\protect\citeauthoryear{{Tully} et~al.,}{{Tully}
  et~al.}{2008}]{tully08}
{Tully} R.~B.,  et~al., 2008, \mn@doi [\apj] {10.1086/527428}, \href
  {https://ui.adsabs.harvard.edu/\#abs/2008ApJ...676..184T} {676, 184}

\bibitem[\protect\citeauthoryear{{Tully} et~al.,}{{Tully}
  et~al.}{2013}]{tully13}
{Tully} R.~B.,  et~al., 2013, \mn@doi [\aj] {10.1088/0004-6256/146/4/86}, \href
  {https://ui.adsabs.harvard.edu/#abs/2013AJ....146...86T} {146, 86}

\bibitem[\protect\citeauthoryear{{Turatto} et~al.,}{{Turatto}
  et~al.}{1990}]{turatto90}
{Turatto} M.,  et~al., 1990, \mn@doi [\aj] {10.1086/115558}, \href
  {https://ui.adsabs.harvard.edu/#abs/1990AJ....100..771T} {100, 771}

\bibitem[\protect\citeauthoryear{{Turatto} et~al.,}{{Turatto}
  et~al.}{1996}]{turatto96}
{Turatto} M.,  et~al., 1996, \mn@doi [\mnras] {10.1093/mnras/283.1.1}, \href
  {https://ui.adsabs.harvard.edu/#abs/1996MNRAS.283....1T} {283, 1}

\bibitem[\protect\citeauthoryear{{Tutukov} et~al.,}{{Tutukov} \&
  {Fedorova}}{2007}]{tutukov07}
{Tutukov} A.~V.,  et~al., 2007, \mn@doi [Astronomy Reports]
  {10.1134/S1063772907040051}, \href
  {https://ui.adsabs.harvard.edu/\#abs/2007ARep...51..291T} {51, 291}

\bibitem[\protect\citeauthoryear{{Tutukov} et~al.,}{{Tutukov} \&
  {Yungelson}}{1979}]{tutukov79}
{Tutukov} A.~V.,  et~al., 1979, \actaa, \href
  {https://ui.adsabs.harvard.edu/#abs/1979AcA....29..665T} {29, 665}

\bibitem[\protect\citeauthoryear{{Tutukov} et~al.,}{{Tutukov} \&
  {Yungelson}}{1994}]{tutukov94}
{Tutukov} A.~V.,  et~al., 1994, \mn@doi [\mnras] {10.1093/mnras/268.4.871},
  \href {https://ui.adsabs.harvard.edu/abs/1994MNRAS.268..871T} {268, 871}

\bibitem[\protect\citeauthoryear{{Tutukov} et~al.,}{{Tutukov} \&
  {Yungelson}}{1996}]{tutukov96}
{Tutukov} A.,  et~al., 1996, \mn@doi [\mnras] {10.1093/mnras/280.4.1035}, \href
  {https://ui.adsabs.harvard.edu/\#abs/1996MNRAS.280.1035T} {280, 1035}

\bibitem[\protect\citeauthoryear{{Uddin} et~al.,}{{Uddin}
  et~al.}{2017}]{ATel10605}
{Uddin} S.,  et~al., 2017, The Astronomer's Telegram, \href
  {https://ui.adsabs.harvard.edu/#abs/2017ATel10605....1U} {10605, 1}

\bibitem[\protect\citeauthoryear{{Valenti} et~al.,}{{Valenti}
  et~al.}{2017}]{TCRN1020}
{Valenti} S.,  et~al., 2017, Transient Name Server Classification Report, \href
  {https://ui.adsabs.harvard.edu/#abs/2017TNSCR.613....1V} {2017-613, 1}

\bibitem[\protect\citeauthoryear{{Valentini} et~al.,}{{Valentini}
  et~al.}{2003}]{valentini03}
{Valentini} G.,  et~al., 2003, \mn@doi [\apj] {10.1086/377448}, \href
  {https://ui.adsabs.harvard.edu/\#abs/2003ApJ...595..779V} {595, 779}

\bibitem[\protect\citeauthoryear{{Vallely} et~al.,}{{Vallely}
  et~al.}{2019a}]{vallely}
{Vallely} P.~J.,  et~al., 2019a, arXiv e-prints, \href
  {https://ui.adsabs.harvard.edu/abs/2019arXiv190200037V} {p. arXiv:1902.00037}

\bibitem[\protect\citeauthoryear{{Vallely} et~al.,}{{Vallely}
  et~al.}{2019b}]{vallely18tb}
{Vallely} P.~J.,  et~al., 2019b, \mn@doi [\mnras] {10.1093/mnras/stz1445},
  \href {https://ui.adsabs.harvard.edu/abs/2019MNRAS.487.2372V} {487, 2372}

\bibitem[\protect\citeauthoryear{{Vanmunster} et~al.,}{{Vanmunster}
  et~al.}{1994}]{IAUC6115}
{Vanmunster} T.,  et~al., 1994, International Astronomical Union Circular,
  \href {https://ui.adsabs.harvard.edu/#abs/1994IAUC.6115....3V} {6115, 3}

\bibitem[\protect\citeauthoryear{{Verheijen} et~al.,}{{Verheijen} \&
  {Sancisi}}{2001}]{verheijen01}
{Verheijen} M.~A.~W.,  et~al., 2001, \mn@doi [\aap]
  {10.1051/0004-6361:20010090}, \href
  {https://ui.adsabs.harvard.edu/#abs/2001A&A...370..765V} {370, 765}

\bibitem[\protect\citeauthoryear{{Vernet} et~al.,}{{Vernet}
  et~al.}{2011}]{XSHOOTERref}
{Vernet} J.,  et~al., 2011, \mn@doi [\aap] {10.1051/0004-6361/201117752}, \href
  {https://ui.adsabs.harvard.edu/#abs/2011A&A...536A.105V} {536}

\bibitem[\protect\citeauthoryear{{Vettolani} et~al.,}{{Vettolani}
  et~al.}{1981}]{IAUC3584}
{Vettolani} G.,  et~al., 1981, International Astronomical Union Circular, \href
  {https://ui.adsabs.harvard.edu/#abs/1981IAUC.3584....1V} {3584, 1}

\bibitem[\protect\citeauthoryear{{Villegas} et~al.,}{{Villegas}
  et~al.}{2010}]{villegas10}
{Villegas} D.,  et~al., 2010, \mn@doi [\apj] {10.1088/0004-637X/717/2/603},
  \href {https://ui.adsabs.harvard.edu/#abs/2010ApJ...717..603V} {717, 603}

\bibitem[\protect\citeauthoryear{{Villi} et~al.,}{{Villi}
  et~al.}{1998}]{IAUC6899}
{Villi} M.,  et~al., 1998, International Astronomical Union Circular, \href
  {https://ui.adsabs.harvard.edu/#abs/1998IAUC.6899....1V} {6899, 1}

\bibitem[\protect\citeauthoryear{{Villi} et~al.,}{{Villi}
  et~al.}{2008}]{CBET1228}
{Villi} M.,  et~al., 2008, Central Bureau Electronic Telegrams, \href
  {https://ui.adsabs.harvard.edu/#abs/2008CBET.1228....1V} {1228, 1}

\bibitem[\protect\citeauthoryear{{Vink{\'o}} et~al.,}{{Vink{\'o}}
  et~al.}{2003}]{vinko03}
{Vink{\'o}} J.,  et~al., 2003, \mn@doi [\aap] {10.1051/0004-6361:20021469},
  \href {https://ui.adsabs.harvard.edu/\#abs/2003A&A...397..115V} {397, 115}

\bibitem[\protect\citeauthoryear{{Vladimirov} et~al.,}{{Vladimirov}
  et~al.}{2015}]{ATel7732}
{Vladimirov} V.,  et~al., 2015, The Astronomer's Telegram, \href
  {https://ui.adsabs.harvard.edu/\#abs/2015ATel.7732....1V} {7732, 1}

\bibitem[\protect\citeauthoryear{{Volkov}}{{Volkov}}{1991}]{volkov91}
{Volkov} I.~M.,  1991, Information Bulletin on Variable Stars, \href
  {https://ui.adsabs.harvard.edu/\#abs/1991IBVS.3581....1V} {3581, 1}

\bibitem[\protect\citeauthoryear{{Waagen} et~al.,}{{Waagen}
  et~al.}{1991}]{IAUC5239}
{Waagen} E.,  et~al., 1991, International Astronomical Union Circular, \href
  {https://ui.adsabs.harvard.edu/#abs/1991IAUC.5239....1W} {5239, 1}

\bibitem[\protect\citeauthoryear{{Walker} et~al.,}{{Walker}
  et~al.}{1994}]{IAUC5950}
{Walker} A.,  et~al., 1994, International Astronomical Union Circular, \href
  {https://ui.adsabs.harvard.edu/\#abs/1994IAUC.5950....1W} {5950, 1}

\bibitem[\protect\citeauthoryear{{Walker} et~al.,}{{Walker}
  et~al.}{2015}]{walker15}
{Walker} E.~S.,  et~al., 2015, \mn@doi [The Astrophysical Journal Supplement
  Series] {10.1088/0067-0049/219/1/13}, \href
  {https://ui.adsabs.harvard.edu/#abs/2015ApJS..219...13W} {219}

\bibitem[\protect\citeauthoryear{{Wang} et~al.,}{{Wang} et~al.}{2004}]{wang04}
{Wang} L.,  et~al., 2004, \mn@doi [\apj] {10.1086/383411}, \href
  {https://ui.adsabs.harvard.edu/\#abs/2004ApJ...604L..53W} {604, L53}

\bibitem[\protect\citeauthoryear{{Wang} et~al.,}{{Wang} et~al.}{2008}]{wang08}
{Wang} X.,  et~al., 2008, \mn@doi [\apj] {10.1086/526413}, \href
  {https://ui.adsabs.harvard.edu/\#abs/2008ApJ...675..626W} {675, 626}

\bibitem[\protect\citeauthoryear{{Webbink}}{{Webbink}}{1984}]{webbink84}
{Webbink} R.~F.,  1984, \mn@doi [\apj] {10.1086/161701}, \href
  {https://ui.adsabs.harvard.edu/#abs/1984ApJ...277..355W} {277, 355}

\bibitem[\protect\citeauthoryear{{Wells} et~al.,}{{Wells}
  et~al.}{1994}]{wells94}
{Wells} L.~A.,  et~al., 1994, \mn@doi [\aj] {10.1086/117236}, \href
  {https://ui.adsabs.harvard.edu/\#abs/1994AJ....108.2233W} {108, 2233}

\bibitem[\protect\citeauthoryear{{Weyant} et~al.,}{{Weyant}
  et~al.}{2018}]{weyant18}
{Weyant} A.,  et~al., 2018, \mn@doi [\aj] {10.3847/1538-3881/aab901}, \href
  {https://ui.adsabs.harvard.edu/#abs/2018AJ....155..201W} {155}

\bibitem[\protect\citeauthoryear{{Wheeler} et~al.,}{{Wheeler}
  et~al.}{1975}]{wheeler75}
{Wheeler} J.~C.,  et~al., 1975, \mn@doi [\apj] {10.1086/153771}, \href
  {https://ui.adsabs.harvard.edu/#abs/1975ApJ...200..145W} {200, 145}

\bibitem[\protect\citeauthoryear{{Whelan} et~al.,}{{Whelan} \&
  {Iben}}{1973}]{whelan73}
{Whelan} J.,  et~al., 1973, \mn@doi [\apj] {10.1086/152565}, \href
  {https://ui.adsabs.harvard.edu/#abs/1973ApJ...186.1007W} {186, 1007}

\bibitem[\protect\citeauthoryear{{Wiethoff} et~al.,}{{Wiethoff}
  et~al.}{2017}]{ATel10521}
{Wiethoff} W.,  et~al., 2017, The Astronomer's Telegram, \href
  {https://ui.adsabs.harvard.edu/#abs/2017ATel10521....1W} {10521, 1}

\bibitem[\protect\citeauthoryear{{Wills} et~al.,}{{Wills}
  et~al.}{1980}]{UVITSref}
{Wills} B.~J.,  et~al., 1980, \mn@doi [\apj] {10.1086/157871}, \href
  {https://ui.adsabs.harvard.edu/abs/1980ApJ...237..319W} {237, 319}

\bibitem[\protect\citeauthoryear{{Wood-Vasey} et~al.,}{{Wood-Vasey}
  et~al.}{2002a}]{IAUC7902}
{Wood-Vasey} W.~M.,  et~al., 2002a, International Astronomical Union Circular,
  \href {https://ui.adsabs.harvard.edu/#abs/2002IAUC.7902....3W} {7902, 3}

\bibitem[\protect\citeauthoryear{{Wood-Vasey} et~al.,}{{Wood-Vasey}
  et~al.}{2002b}]{IAUC7959}
{Wood-Vasey} W.~M.,  et~al., 2002b, International Astronomical Union Circular,
  \href {https://ui.adsabs.harvard.edu/#abs/2002IAUC.7959....1W} {7959, 1}

\bibitem[\protect\citeauthoryear{{Woods} et~al.,}{{Woods}
  et~al.}{2006}]{woods06}
{Woods} D.~F.,  et~al., 2006, \mn@doi [\aj] {10.1086/504834}, \href
  {https://ui.adsabs.harvard.edu/\#abs/2006AJ....132..197W} {132, 197}

\bibitem[\protect\citeauthoryear{{Woods} et~al.,}{{Woods}
  et~al.}{2018}]{woods18}
{Woods} T.~E.,  et~al., 2018, \mn@doi [\apj] {10.3847/1538-4357/aad1ee}, \href
  {https://ui.adsabs.harvard.edu/#abs/2018ApJ...863..120W} {863, 120}

\bibitem[\protect\citeauthoryear{{Woosley} et~al.,}{{Woosley} \&
  {Weaver}}{1994}]{woosley94}
{Woosley} S.~E.,  et~al., 1994, \mn@doi [\apj] {10.1086/173813}, \href
  {https://ui.adsabs.harvard.edu/\#abs/1994ApJ...423..371W} {423, 371}

\bibitem[\protect\citeauthoryear{{Wyrzykowski} et~al.,}{{Wyrzykowski}
  et~al.}{2015}]{ATel8484}
{Wyrzykowski} L.,  et~al., 2015, The Astronomer's Telegram, \href
  {https://ui.adsabs.harvard.edu/\#abs/2015ATel.8484....1W} {8484, 1}

\bibitem[\protect\citeauthoryear{{Yamanaka} et~al.,}{{Yamanaka}
  et~al.}{2015}]{yamanaka15}
{Yamanaka} M.,  et~al., 2015, \mn@doi [\apj] {10.1088/0004-637X/806/2/191},
  \href {https://ui.adsabs.harvard.edu/\#abs/2015ApJ...806..191Y} {806, 191}

\bibitem[\protect\citeauthoryear{{Yaron} et~al.,}{{Yaron} \&
  {Gal-Yam}}{2012}]{WRref}
{Yaron} O.,  et~al., 2012, \mn@doi [Publications of the Astronomical Society of
  the Pacific] {10.1086/666656}, \href
  {https://ui.adsabs.harvard.edu/#abs/2012PASP..124..668Y} {124, 668}

\bibitem[\protect\citeauthoryear{{Yoon} et~al.,}{{Yoon} \&
  {Langer}}{2003}]{yoon03}
{Yoon} S.~C.,  et~al., 2003, \mn@doi [\aap] {10.1051/0004-6361:20034607}, \href
  {https://ui.adsabs.harvard.edu/#abs/2003A&A...412L..53Y} {412, L53}

\bibitem[\protect\citeauthoryear{{Yoon} et~al.,}{{Yoon} \&
  {Langer}}{2005}]{yoon05}
{Yoon} S.~C.,  et~al., 2005, \mn@doi [\aap] {10.1051/0004-6361:20042542}, \href
  {https://ui.adsabs.harvard.edu/abs/2005A&A...435..967Y} {435, 967}

\bibitem[\protect\citeauthoryear{{Yu} et~al.,}{{Yu} et~al.}{2000}]{IAUC7458}
{Yu} C.,  et~al., 2000, International Astronomical Union Circular, \href
  {https://ui.adsabs.harvard.edu/#abs/2000IAUC.7458....1Y} {7458, 1}

\bibitem[\protect\citeauthoryear{{Zhai} et~al.,}{{Zhai} et~al.}{2016}]{zhai16}
{Zhai} Q.,  et~al., 2016, \mn@doi [\aj] {10.3847/0004-6256/151/5/125}, \href
  {https://ui.adsabs.harvard.edu/#abs/2016AJ....151..125Z} {151}

\bibitem[\protect\citeauthoryear{{Zhang} et~al.,}{{Zhang} \&
  {Wang}}{2014}]{ATel6813}
{Zhang} J.,  et~al., 2014, The Astronomer's Telegram, \href
  {https://ui.adsabs.harvard.edu/\#abs/2014ATel.6813....1Z} {6813, 1}

\bibitem[\protect\citeauthoryear{{Zhang} et~al.,}{{Zhang}
  et~al.}{2010}]{zhang10}
{Zhang} T.,  et~al., 2010, \mn@doi [\pasp] {10.1086/649851}, \href
  {https://ui.adsabs.harvard.edu/abs/2010PASP..122....1Z} {122, 1}

\bibitem[\protect\citeauthoryear{{Zhang} et~al.,}{{Zhang}
  et~al.}{2011}]{CBET2708}
{Zhang} T.,  et~al., 2011, Central Bureau Electronic Telegrams, \href
  {https://ui.adsabs.harvard.edu/#abs/2011CBET.2708....3Z} {2708, 3}

\bibitem[\protect\citeauthoryear{{Zhang} et~al.,}{{Zhang}
  et~al.}{2014}]{zhang14}
{Zhang} J.-J.,  et~al., 2014, \mn@doi [\aj] {10.1088/0004-6256/148/1/1}, \href
  {https://ui.adsabs.harvard.edu/#abs/2014AJ....148....1Z} {148}

\bibitem[\protect\citeauthoryear{{Zhang} et~al.,}{{Zhang}
  et~al.}{2018}]{2018TNSCR1755}
{Zhang} J.,  et~al., 2018, Transient Name Server Classification Report, \href
  {https://ui.adsabs.harvard.edu/\#abs/2018TNSCR.293....1Z} {2018-293, 1}

\bibitem[\protect\citeauthoryear{{Zheng} et~al.,}{{Zheng}
  et~al.}{2013}]{ATel5637}
{Zheng} W.,  et~al., 2013, The Astronomer's Telegram, \href
  {https://ui.adsabs.harvard.edu/\#abs/2013ATel.5637....1Z} {5637, 1}

\bibitem[\protect\citeauthoryear{{de Vaucouleurs} et~al.,}{{de Vaucouleurs}
  et~al.}{1991}]{devaucouleurs91}
{de Vaucouleurs} G.,  et~al., 1991, {Third Reference Catalogue of Bright
  Galaxies}

\bibitem[\protect\citeauthoryear{{van Driel} et~al.,}{{van Driel}
  et~al.}{2001}]{vandriel01}
{van Driel} W.,  et~al., 2001, \mn@doi [\aap] {10.1051/0004-6361:20011241},
  \href {https://ui.adsabs.harvard.edu/#abs/2001A&A...378..370V} {378, 370}

\bibitem[\protect\citeauthoryear{{van Dyk} et~al.,}{{van Dyk}
  et~al.}{1994}]{IAUC6105}
{van Dyk} S.~D.,  et~al., 1994, International Astronomical Union Circular,
  \href {https://ui.adsabs.harvard.edu/#abs/1994IAUC.6105....1V} {6105, 1}

\bibitem[\protect\citeauthoryear{{van Genderen}}{{van
  Genderen}}{1975}]{vangenderen75}
{van Genderen} A.~M.,  1975, \aap, \href
  {https://ui.adsabs.harvard.edu/\#abs/1975A&A....45..429V} {45, 429}

\makeatother
\end{thebibliography}
